\documentclass[aps,prb,reprint,groupedaddress]{revtex4-2}

\usepackage{graphicx}
\usepackage{amsmath,amssymb}

\begin{document}



\title{Effects of nonmagnetic impurities and subgap states on the kinetic inductance, complex conductivity, quality factor and depairing current density}


\author{Takayuki Kubo}
\email[]{kubotaka@post.kek.jp}
\altaffiliation[]{Stay-at-home dad on two-year paternity leave, New York, NY, USA}
\affiliation{High Energy Accelerator Research Organization (KEK), Tsukuba, Ibaraki 305-0801, Japan}
\affiliation{The Graduate University for Advanced Studies (Sokendai), Hayama, Kanagawa 240-0193, Japan}



\begin{abstract}
We investigate how a combination of a nonmagnetic-impurity scattering rate $\gamma$ and finite subgap states parametrized by Dynes $\Gamma$ affects various physical quantities relevant to to superconducting devices: kinetic inductance $L_k$, complex conductivity $\sigma$, surface resistance $R_s$, quality factor $Q$, and depairing current density $J_d$.
All the calculations are based on the Eilenberger formalism of the BCS theory. 
We assume the device materials are extreme type-II $s$-wave superconductors.
It is well known that the optimum impurity concentration ($\gamma/\Delta_0 \sim 1$) minimizes $R_s$. 
Here,  $\Delta_0$ is the pair potential for the idealized ($\Gamma\to 0$) superconductor for the temperature $T\to 0$. 
We find the optimum $\Gamma$ can also reduce $R_s$ by one order of magnitude for a clean superconductor ($\gamma/\Delta_0 < 1$) and a few tens $\%$ for a dirty superconductor ($\gamma/\Delta_0 > 1$). 
Also, we find a nearly-ideal ($\Gamma/\Delta_0 \ll 1$) clean-limit superconductor exhibits a frequency-independent $R_s$ for a broad range of frequency $\omega$, 
which can significantly improve $Q$ of a very compact cavity with a few tens of GHz frequency. 
As $\Gamma$ or $\gamma$ increases, the plateau disappears, and $R_s$ obeys the $\omega^2$ dependence. 
The subgap-state-induced residual surface resistance $R_{\rm res}$ is also studied, 
which can be detected by an SRF-grade high-$Q$ 3D resonator.  
We calculate $L_k(\gamma, \Gamma,T)$ and $J_d(\gamma, \Gamma,T)$, 
which are monotonic increasing and decreasing functions of $(\gamma, \Gamma,T)$, respectively.  
Measurements of $(\gamma, \Gamma)$ of device materials can give helpful information on engineering $(\gamma, \Gamma)$ via materials processing, 
by which it would be possible to improve $Q$, engineer $L_k$, and ameliorate $J_d$. 
\end{abstract}

\maketitle


\section{Introduction} \label{introduction}

The microscopic theory of superconductivity is the linchpin of research and development of various superconducting devices, 
such as superconducting radio-frequency (SRF) cavities for particle accelerators~\cite{2017_Padamsee, 2012_Gurevich_review, 2017_Gurevich_SUST}, superconducting magnets and cables~\cite{2021_Eley}, kinetic inductance detectors (KID)~\cite{2012_Zmuidzinas}, hot-electron bolometer~\cite{2017_Klapwijk_Semenov}, 
other superconducting instruments for astrophysics and cosmology~\cite{2004_Zmuidzinas}, 
superconducting single-photon detectors (SSPD)~\cite{2012_Natarajan, 2015_Engel}, 
and superconducting qubit for quantum information processing (QIP)~\cite{2017_Wendin, 2019_Q_report}. 
Models of superconducting devices based on the idealized BCS superconductor without any pair breakers usually provide good starting points to understand the operating principles of these technologies. 
However, such models are sometimes too naive to analyze the physics of real devices. 
The representative example is the quality factor $Q$ of superconducting resonators. 
The surface resistance and then $Q$ are sensitive to the detail of the quasiparticle density of states (DOS), 
which is affected by various pair breakers in the real world, such as the current and magnetic impurities~\cite{Tinkham, deGennes, Abrikosov, Maki, Kopnin}. 
Hence, an analysis of $Q$ based on the ideal BCS DOS, i.e., the Mattis-Bardeen (MB) theory~\cite{MB}, is qualitative and sometimes inadequate even for qualitative analyses. 
In fact, the MB theory cannot explain the rf-field dependent nonlinear $Q$, 
effects of material-treatment on $Q$, and saturation of $Q$ at $T\to 0$. 
Another example is the depairing current density $J_d$. 
The well-known Kupriyanov-Lukichev-Maki theory~\cite{1963_Maki, 1980_Kupriyanov} gives $J_d$ for the ideal BCS superconductor. 
However, pair-breakers in device materials (e.g., magnetic impurities) can degrade $J_d$ and thus reduce the ultimate limit of the current density in superconducting cables and that of the accelerating field of the SRF cavity~\cite{2012_Lin_Gurevich}. 
To understand the physics of superconducting devices and to discern the causes of unexpected performance limitations of real devices, 
we need to take various pair-breakers in real materials into account (e.g., current, magnetic impurities, metallic suboxides, hydride, and non-stoichiometric regions) and to assess the effects of such non-ideal features on device performances~\cite{2017_Gurevich_Kubo, 2019_Kubo_Gurevich}.

\begin{figure}[tb]
   \begin{center}
   \includegraphics[width=0.47\linewidth]{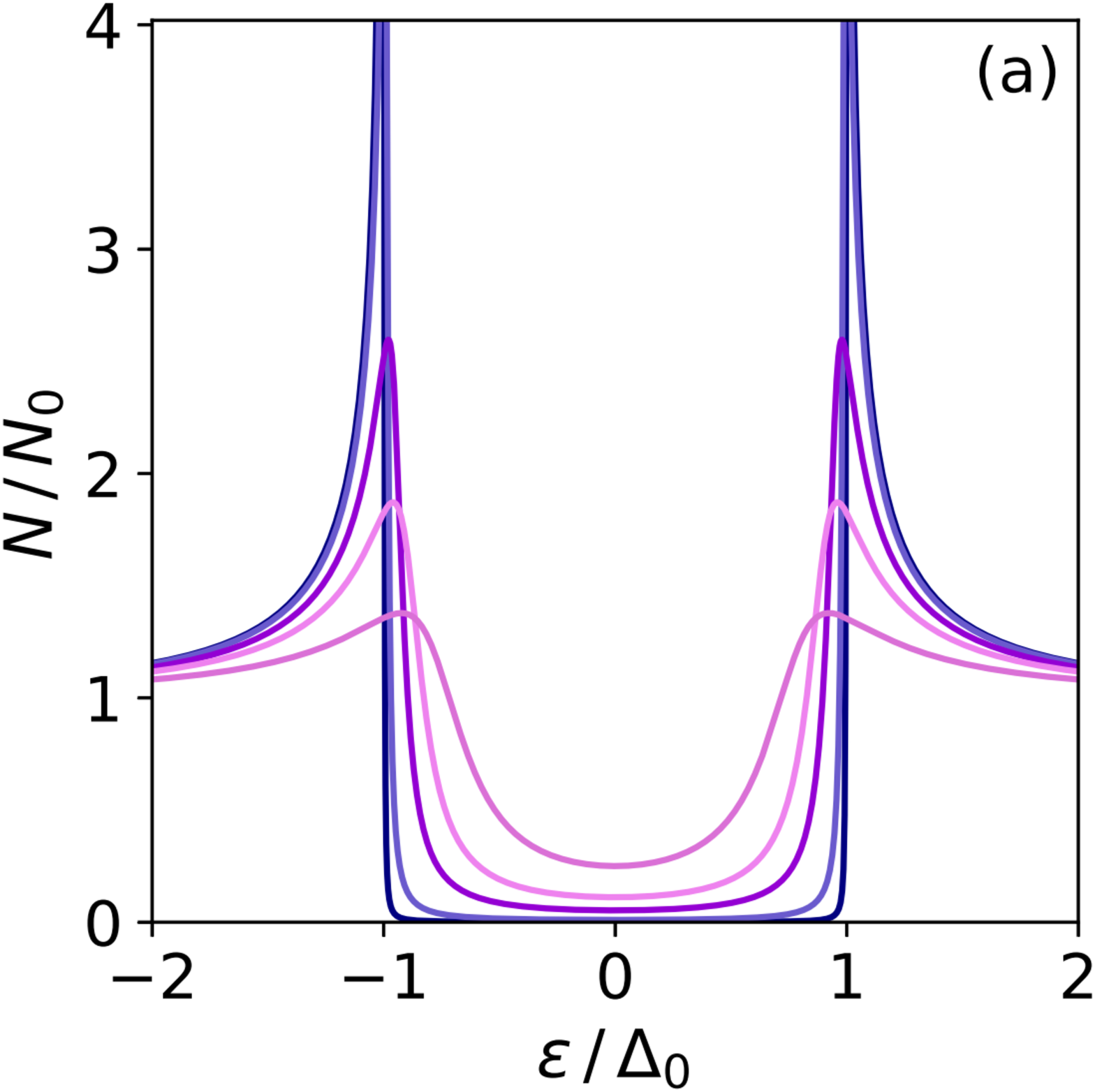}
   \includegraphics[width=0.51\linewidth]{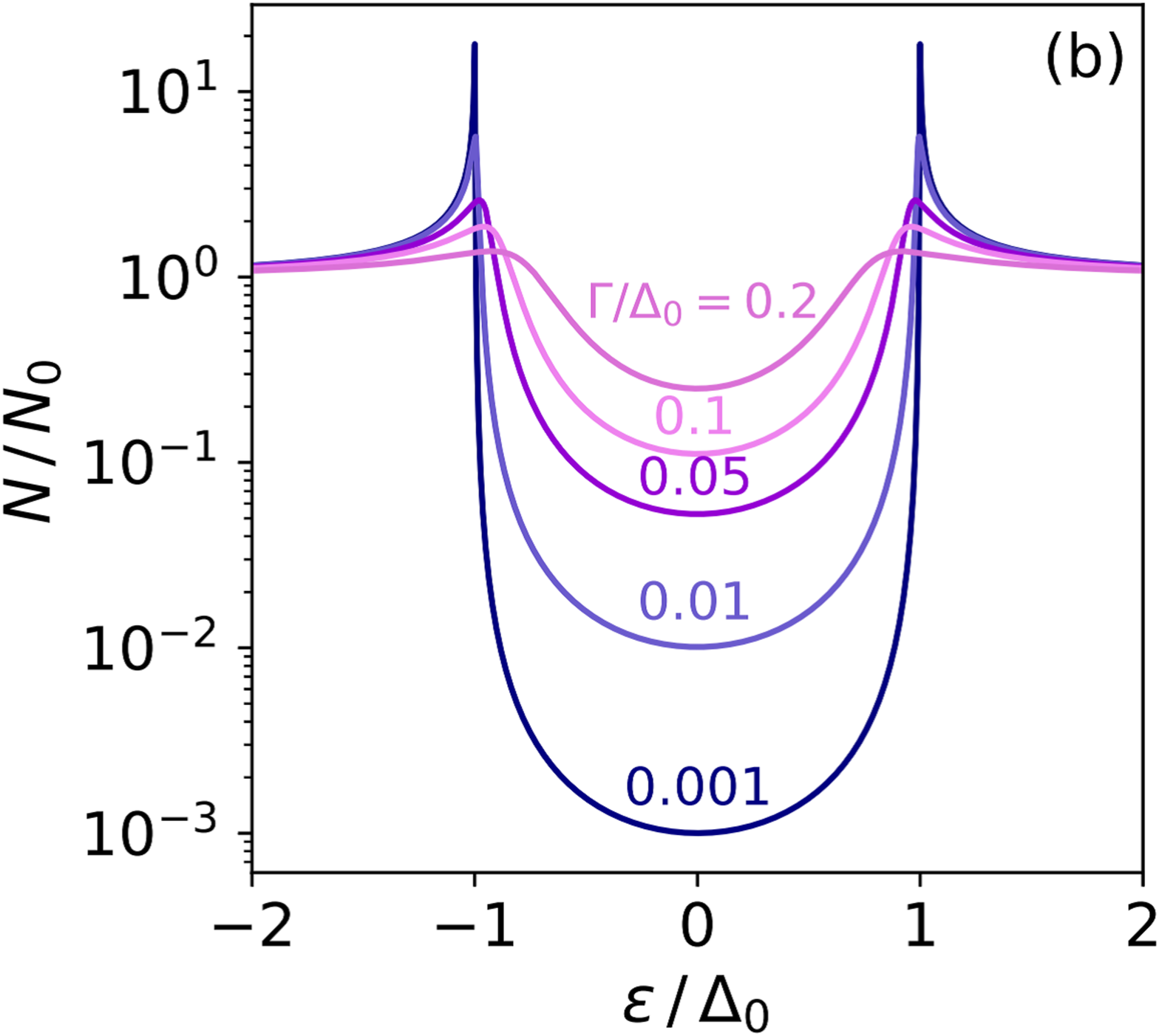}
   \end{center}\vspace{0 cm}
   \caption{
(a) Quasiparticle density of states (DOS) at $T\to 0$ calculated for $\Gamma/\Delta_0=0.2$, 0.1, 0.05, 0.01, and 0.001. 
(b) DOS in a logarithmic scale. 
Note DOS in the zero-current state is independent of a concentration of nonmagnatic impurities. 
   }\label{fig1}
\end{figure}

One of the non-ideal features observed in various superconducting materials is the broadening and subgap states in DOS~\cite{2003_Zasa, 2010_Kamlapure, 2013_Noat, 2013_Dhakal, 2014_Groll, 2015_Becker, 2020_Lechner}, 
in contrast to the sharp spectrum gap in the idealized BCS superconductor. 
Various mechanisms can contribute to the subgap states, e.g., inelastic scattering of quasiparticles on phonons~\cite{D}, Coulomb correlations~\cite{B}, anisotropy of the Fermi surface~\cite{Bennet}, local inhomogeneities of the BCS pairing constant~\cite{LO_inhomogeneous}, magnetic impurities~\cite{imp_review}, and effects of spatial correlations in impurity scattering~\cite{Simons}.
Such a broadened DOS is often described by using the phenomenological Dynes formula~\cite{1978_Dynes, 1984_Dynes}, 
$N(\epsilon)/N_0 = {\rm Re} [(\epsilon + i\Gamma)/\sqrt{ (\epsilon + i\Gamma)^2 - \Delta^2}]$ (see Fig.~\ref{fig1}). 
Here $\Delta$ is the pair potential, $N_0$ is the normal-state DOS at the Fermi level, 
and $\Gamma$ is the Dynes broadening parameter. 
We can reproduce the ideal BCS-DOS by taking $\Gamma \to 0$. 
Irrespective of microscopic models of the Dynes formula, 
we can incorporates $\Gamma$ into the quasiclassical formalism of the BCS theory (see e.g., Refs.~\cite{2017_Gurevich_Kubo, 2019_Kubo_Gurevich, 2020_Kubo_1, 2020_Kubo_2, 2017_Vischi, 2020_Tang}). 
Also, microscopic derivations of the Dynes formula have been investigated~\cite{1991_Mikhailovsky, 2016_Herman}.

The first way to calculate physical quantities based on realistic spectrum is to incorporate all the broadening mechanisms in materials into a device model. 
However, such a model should contain a bunch of free parameters and practically useless unless those parameters are determined from a number of experiments.  
Another way is to introduce the phenomenological Dynes $\Gamma$ into a device model. 
Here, a value of $\Gamma$ can be determined from experiments such as tunneling spectroscopy~\cite{2003_Zasa} and measurements of complex conductivity~\cite{2021_Bafia, 2021_Herman}. 
Even if we do not know what kind of pair breakers contribute to subgap states, 
such an experimentally determined $\Gamma$ makes it possible to calculate physical quantities based on the realistic quasiparticle spectrum. 
In the present paper, we pursue the second approach, 
which would give reliable results rather than those based on the idealized BCS-DOS.

According to the previous studies for the diffusive limit~\cite{2017_Gurevich_Kubo, 2019_Kubo_Gurevich, 2020_Kubo_1, 2020_Kubo_2}, 
influences of a finite $\Gamma$ is far-reaching: 
it affects, e.g.,the pair potential $\Delta$, the critical temperature $T_c$, the penetration depth $\lambda$, the kinetic inductivity $L_k$, the complex conductivity $\sigma$, the coherence peak of dissipative-conductivity $\sigma_1$, the surface impedance, $Q$ factor, and the depairing current density $J_d$. 
One of the nontrivial consequences is the fact that there exists the optimum $\Gamma$, 
which maximizes $Q$ at some temperature range~\cite{2017_Gurevich_SUST, 2017_Gurevich_Kubo, 2019_Kubo_Gurevich, 2020_Kubo_1}. 
This $\Gamma$-induced $Q$-rise comes from the same mechanism as the $Q$ rise due to current-induced DOS broadening~\cite{2014_Gurevich, 2014_Ciovati}. 
On the other hand, finite subgap states in the vicinity of the Fermi level induce a residual dissipative conductivity $\lim_{T\to 0}\sigma_1$~\cite{2012_Gurevich_review, 2017_Gurevich_SUST, 2017_Herman, 2017_Gurevich_Kubo, 2020_Kubo_1}.

The effects of a finite $\Gamma$ in a moderately clean superconductor have been studied less extensively except for some calculations of $\sigma$~\cite{2017_Herman, 2021_Herman}. 
In the present study, we investigate the effects of a combination of Dynes $\Gamma$ and nonmagnetic-impurity scattering rate $\gamma$ on the kinetic inductivity $L_k$, the surface resistance $R_s$, and the depairing current density $J_d$, 
which are the physical quantities relevant to modern superconducting devices. 
The kinetic inductance is known to limit the reset time after a detection event of SSPD~\cite{2006_Kerman} and also plays an essential role in the operating mechanism of KID, which detects a shift of resonant frequency $\delta f \propto -\delta L_k$ due to the arrival of pair-breaking photons~\cite{2012_Zmuidzinas}. 
The surface resistance $R_s$ is proportional to the dissipation at the inner surface of a 3D resonant cavity for SRF and QIP. 
Reduction of $R_s$ (or improvement of $Q$) has been the primary interest of researchers of resonant cavities over the last decades. 
Vortex-free Nb cavities~\cite{2014_Romanenko, 2016_Huang, 2016_Posen, 2016_Checchin, 2021_Ooi, 2021_Miyazaki} exhibit huge quality factor $Q \sim 10^{10}$-$10^{12}$ at $T< 2\,{\rm K}$~\cite{2020_Romanenko, 2020_Posen, 2021_Ito, 2021_He} even under the strong rf current~\cite{2007_Geng, 2014_Kubo_IPAC, 2017_Grassellino, 2018_Dhakal} close to the depairing current density. 
The depairing current density $J_d$ is related to the bias current of SSPD and is the maximum current that SRF cavities and superconducting cables can support. 
The screening current density on the inner surface of the cutting-edge Nb cavity reaches a current density close to $J_d$, and SRF researchers study next-generation cavities using alternative materials with higher $J_d$ theoretically~\cite{2006_Gurevich, 2014_Kubo, 2015_Gurevich, 2017_Kubo_SUST, 2017_Liarte_SUST, 2021_Kubo} and experimentally~\cite{2017_Anne-Marie, 2017_Posen, 2016_Tan, 2018_Junginger, 2019_Antoine, 2019_Keckert, 2019_Thoeng, 2020_Ito, 2021_Lin, 2021_Leith, 2021_Harshani}.

Our calculations are based on the well-established Eilenberger formalism of the BCS theory~\cite{1968_Eilenberger, Kopnin}, 
which can include an arbitrary impurity concentration.  
We assume that the penetration depth $\lambda$ is much larger than the coherence length $\xi$, 
then a superconductor obeys the local electrodynamics. 
Large-$\lambda/\xi$ superconductors include dirty Nb and also  NbN, NbTiN, and ${\rm Nb_3 Sn}$ for any impurity concentration from the clean limit to the dirty limit.

The paper is organized as follows. 
In Section~II, we briefly review the Eilenberger formalism of the BCS theory~\cite{1968_Eilenberger, Kopnin} and basic consequences of a finite $\Gamma$ in the zero-current state. 
In Sec.~III, we evaluate the penetration depth $\lambda(\gamma, \Gamma, T)$ and the kinetic inductivity $L_k (\gamma, \Gamma, T)$. 
Convenient formulas for $\lambda$ and $L_k$ are also summarized. 
In Sec.~IV, we summarize the complex conductivity formulas, evaluate the $T$ dependence and the coherence peak, and consider the low $T$ limit and moderately low $T$ regime ($\hbar\omega \ll k_B T \ll \Delta_0$). 
Then, combining the results of Sec.~III and Sec.~IV A - IV D, we study the effects of $(\gamma, \Gamma)$ on the surface resistance and $Q$ factor in the low $T$ limit and the moderately-low $T$ regime. 
In Sec. V, we solve the Eilenberger equation in the current-carrying state and evaluate the effects of $(\gamma, \Gamma)$ on the depairing current density $J_d$. 
In Sec. VI, we discuss the implications of our results.

\section{Theory}

\subsection{Eilenberger equation}

The normal ($g_m$) and anomalous ($f_m$) quasiclassical Matsubara Green's functions obey the Eilenberger equation~\cite{1968_Eilenberger, Kopnin}.
Since we consider the case $\lambda \gg \xi$, 
the current distribution varies slowly over the coherence length, 
and then the spatial-differentiation terms are negligible.  
The Eilenberger equation for the current-carrying state reduces to
\begin{eqnarray}
&&\biggl[i\frac{\pi s}{2} \cos \theta + (\hbar \omega_m + \Gamma) \biggr] f_m - \Delta g_m \nonumber \\
&&= \gamma \bigl\{ \langle f_m \rangle g_m -  \langle g_m \rangle f_m \bigr\}.\label{Eilenberger}
\end{eqnarray}
Here, 
$g_m^2+f_m^2=1$, 
$\hbar \omega_m=2\pi k_B T (m+1/2)$ is the Matsubara frequency, 
$\Gamma$ is the Dynes parameter, 
$\gamma = \hbar/2\tau_{\rm imp}$ is the nonmagnetic-impurity scattering rate, 
$\tau_{\rm imp}$ is the electron scattering time on nonmagnetic impurities,  
$s=\hbar q v_f/\pi = (q/q_0) \Delta_0$, 
$\hbar q$ is the superfluid momentum,  
$v_f$ is the Fermi velocity, 
$q_0 = 1/\xi_0 = \pi \Delta_0/\hbar v_f$ is the inverse of the BCS coherence length, 
$\Delta$ is the pair potential, 
$\Delta_0$ is the pair potential of the idealized ($\Gamma\to 0$) BCS superconductor in the zero-current state ($q=0$) at $T\to 0$, 
$\theta$ is the angle between the Fermi velocity and the current, 
and the bracket $\langle X \rangle$ is the angular averaging of a quantity $X$ over the Fermi surface. 
We assume a spherical Fermi surface: $\langle X \rangle = (1/2)\int_0^{\pi} X \sin\theta d\theta$. 
Note here that considering the average distance the electron travels between electron-nonmagnetic impurity collisions is given by $\ell_{\rm imp} = v_f \tau_{\rm imp}$, we obtain the relation $\ell_{\rm imp} /\xi_0 = \pi \Delta_0 /2 \gamma$.

In the zero-current state ($q= 0$), the $\theta$ dependences of $f_m$ and $g_m$ drop off: 
$\langle f_m \rangle = f_m$ and $\langle g_m \rangle =g_m$. 
Then, Eq.~(\ref{Eilenberger}) reduces to 
\begin{eqnarray}
(\hbar \omega_m + \Gamma) f_m - \Delta g_m  = 0 , \label{Eilenberger_zero}
\end{eqnarray}
which no longer includes the nonmagnetic-impurity scattering rate $\gamma$. 
Hence, in the zero-current state, $\Delta$ does not depend on a concentration of nonmagnetic impurities. 
This robustness of $s$-wave superconductor to nonmagnetic impurity is known as Anderson's theorem~\cite{Kopnin}.

The Eilenberger equation is compensated by the self-consistency equation
\begin{eqnarray}
\ln\frac{T_{c0}}{T} = 2\pi k_B T \sum_{\omega_m >0} \biggl( \frac{1}{\hbar \omega_m} - \frac{\langle f_m \rangle}{\Delta} \biggr) .
\label{self-consistency}
\end{eqnarray}
Here, $k_B T_{c0} = \Delta_0 \exp(\gamma_E)/\pi  \simeq \Delta_0/1.76$ is the critical temperature of the idealized ($\Gamma \to 0$) BCS superconductor, and $\gamma_E = 0.577$ is the Euler's constant. 
Solving Eqs.~(\ref{Eilenberger}) and (\ref{self-consistency}) for the current-carrying state or Eqs.~(\ref{Eilenberger_zero}) and (\ref{self-consistency}) for the zero-current state, 
we can calculate various physical quantities.

\subsection{Brief review of the consequences of a finite $\Gamma$ in the zero-current state} \label{brief_review}

\begin{figure}[tb]
   \begin{center}
   \includegraphics[width=0.49\linewidth]{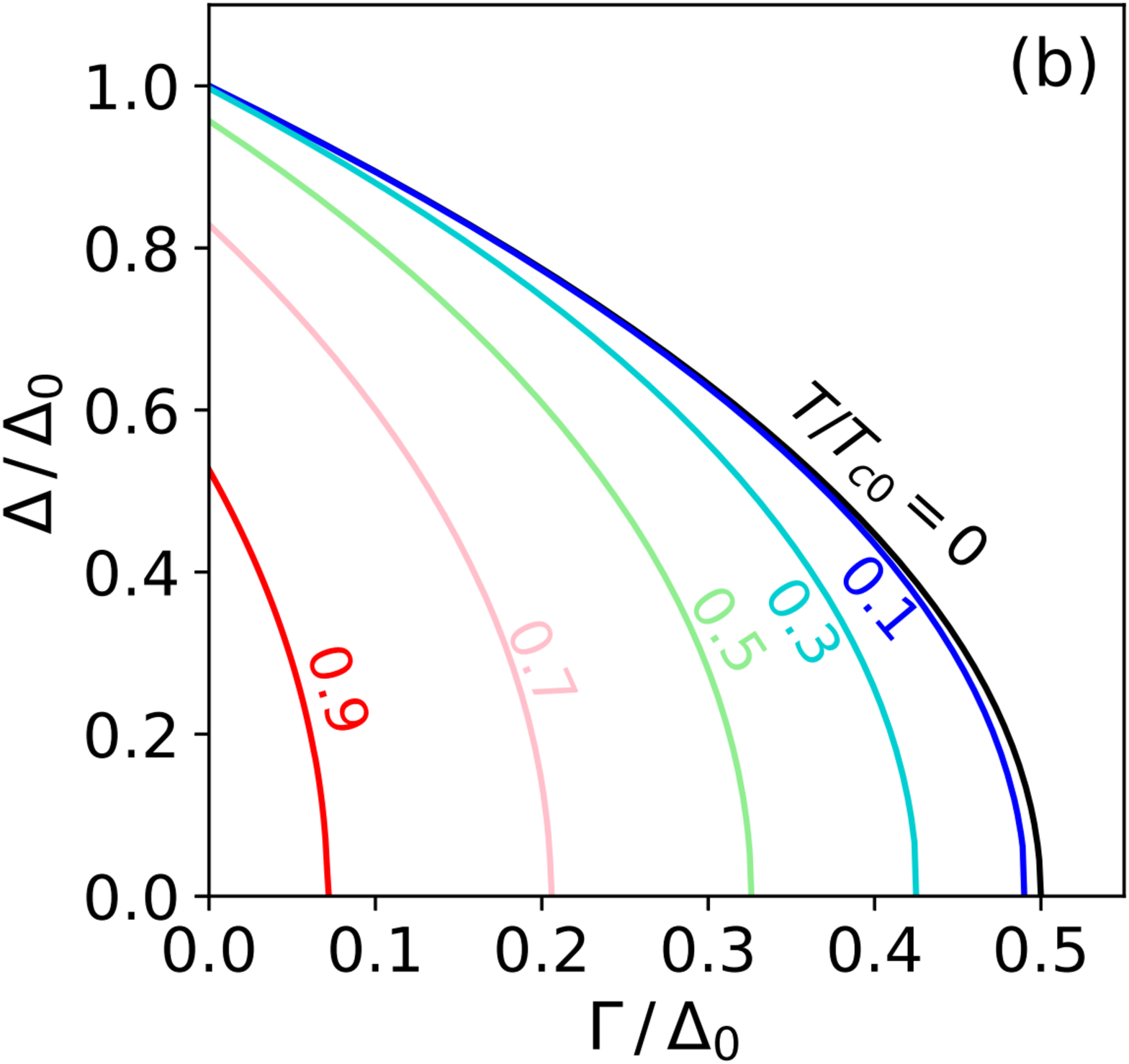}
   \includegraphics[width=0.49\linewidth]{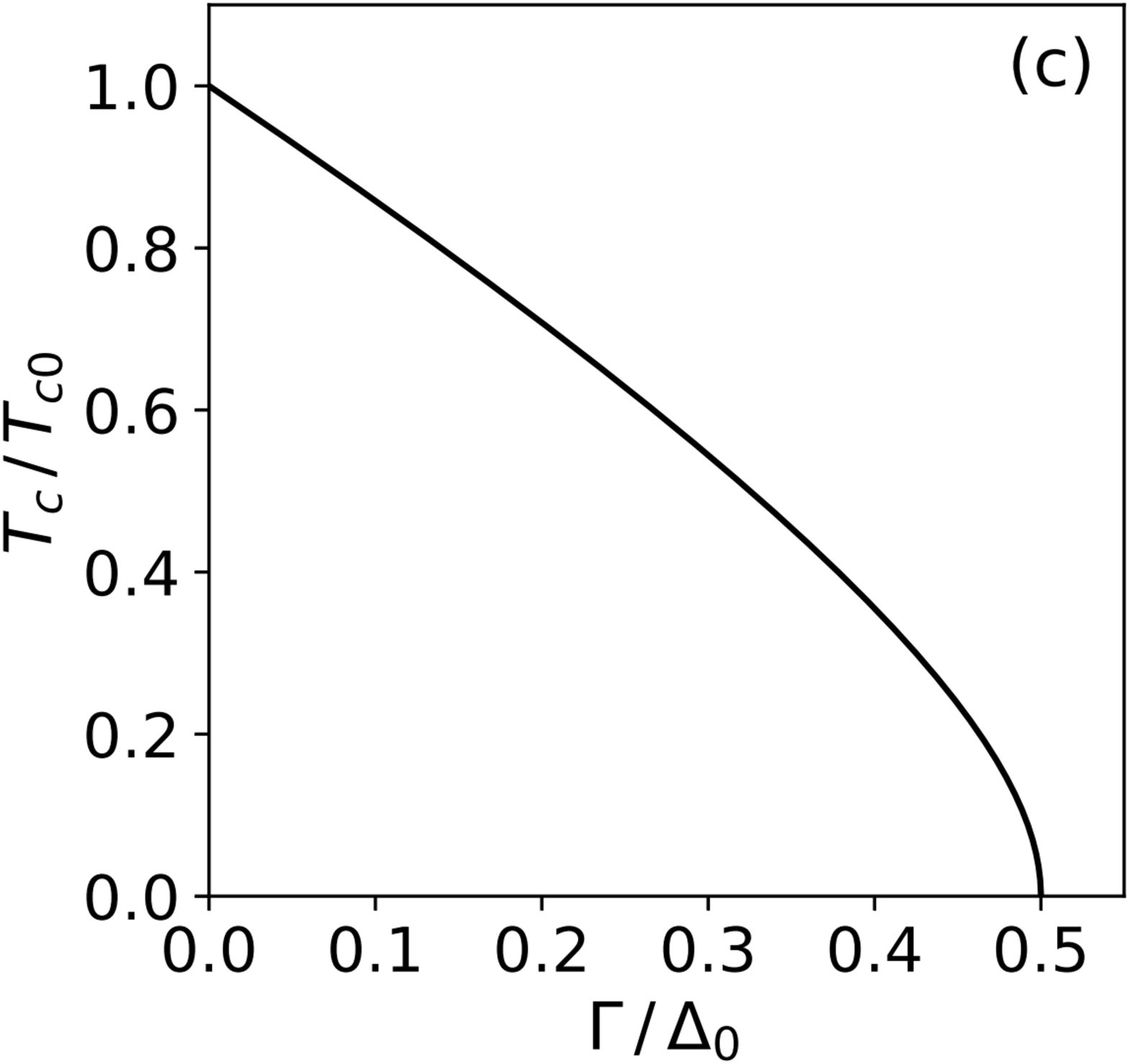}
   \end{center}\vspace{0 cm}
   \caption{
(a) Pair potential $\Delta$ as functions of $\Gamma$, 
calculated for $T/T_{c0} = 0$, 0.1, 0.3, 0.5, 0.7, and 0.9. 
(b) Critical temperature $T_c$ as a function of $\Gamma$. 
   }\label{fig2}
\end{figure}

Let us briefly review the effects of finite Dynes subgap states on $\Delta$, $T_c$, and DOS in the zero-current state. 
Solving Eq.~(\ref{Eilenberger_zero}), we get
\begin{eqnarray}
&&g_m = \frac{\hbar \omega_m+\Gamma}{\sqrt{(\hbar \omega_m+\Gamma)^2 + \Delta^2}} , \label{gm_zero_current} \\
&&f_m = \frac{\Delta}{\sqrt{(\hbar \omega_m+\Gamma)^2 + \Delta^2}} . \label{fm_zero_current} 
\end{eqnarray}
Here, $\Delta=\Delta(\Gamma,T)$ is calculated from Eq.~(\ref{self-consistency}).  
For $T\to 0$, it is easy to find~\cite{2017_Gurevich_Kubo, 2019_Kubo_Gurevich, 2020_Kubo_1, 2020_Kubo_2, 2016_Herman, 2017_Herman, 2021_Herman} 
\begin{eqnarray}
\Delta(\Gamma, T)|_{T\to 0} = \Delta_0 \sqrt{1-2\frac{\Gamma}{\Delta_0}} .\label{Delta_zeroT}
\end{eqnarray}
Hence, the zero-temperature pair-potential vanishes at the critical value $\Gamma = \Delta_0/2$. 
Eq.~(\ref{Delta_zeroT}) reduces to $\simeq \Delta_0-\Gamma$ when $\Gamma/\Delta_0 \ll 1$. 
A calculation for an arbitrary temperature is straightforward. 
Shown in Figure~\ref{fig2} (a) is $\Delta$ as functions of $\Gamma$ for different temperatures. 
The larger the temperature is, the smaller the critical value is.

The critical temperature $T_c=T_c(\Gamma)$ is obtained by substituting $T\to T_c$ and $\Delta \to 0$ into Eq.~(\ref{self-consistency}). 
We find~\cite{2017_Gurevich_Kubo, 2019_Kubo_Gurevich, 2020_Kubo_1, 2020_Kubo_2, 2016_Herman}
\begin{eqnarray}
\ln\frac{T_c}{T_{c0}} = \psi\biggl(\frac{1}{2} \biggr) - \psi\biggl(\frac{1}{2} + \frac{\Gamma}{2\pi k_B T_c} \biggr) , \label{Tc}
\end{eqnarray}
which reduces to $T_c \simeq T_{c0} -\pi\Gamma/4k_B$ for $\Gamma/\Delta_0 \ll 1$. 
It should be noted that Eq.~(\ref{Tc}) has the same form as the well-known formula of critical temperature for a superconductor with pair-breaking perturbations~\cite{Tinkham, deGennes, Abrikosov, Maki, Kopnin}.  
Shown in Figure~\ref{fig2} (b) is $T_c$ as a function of $\Gamma$, 
which is a monotonically decreasing function of $\Gamma$ and vanishes at $\Gamma=\Delta_0/2$.

In the real frequency representation ($\hbar\omega_m \to -i\epsilon$), we get
\begin{eqnarray}
&&g(\epsilon) = \frac{\epsilon+i\Gamma}{\sqrt{(\epsilon+i\Gamma)^2 - \Delta^2}} , \\
&&f(\epsilon) = \frac{\Delta}{\sqrt{(\epsilon+i\Gamma)^2 - \Delta^2}} . 
\end{eqnarray}
The quasiparticle DOS is given by $N(\epsilon)/N_0 = {\rm Re}\, [g]$, 
which corresponds with the Dynes formula shown in Section~\ref{introduction} (see also Fig~\ref{fig1}).

The Dynes $\Gamma$ is a phenomenological representation of some pair breakers in real materials. 
Introducing $\Gamma$ not only yields the broadened DOS (see Fig.~\ref{fig1}) but also results in various pair-breaking effects (e.g., suppression of $\Delta$ and $T_c$ shown in Fig.~\ref{fig2}). 
In the zero-current state, $\Delta$, $T_c$, and DOS are independent of a concentration of nonmagnetic impurities (Anderson theorem). 
Hence, the results shown in Sec.~\ref{brief_review} are coincident with those for the dirty limit discussed in the previous studies~\cite{2017_Gurevich_Kubo, 2019_Kubo_Gurevich, 2020_Kubo_1, 2020_Kubo_2}. 
In the following sections, we study physical quantities that depend on the Dynes $\Gamma$ and the nonmagnetic-impurity scattering rate $\gamma$.

\section{Penetration depth and kinetic inductance} \label{section_lambda_Lk}

\subsection{Penetration depth} \label{section_lambda}

First, consider the penetration depth in the zero-current limit. 
The penetration depth $\lambda(\gamma, \Gamma, T)$ is calculated from~\cite{Kopnin}
\begin{eqnarray}
\frac{1}{\lambda^{2}(\gamma, \Gamma, T)} 
= \frac{2\pi k_B T}{\lambda_0^2} \sum_{\omega_m>0} \frac{f_m^2}{d_m} . \label{penetration_depth}
\end{eqnarray}
Here, $\lambda_0^{-2}=\lambda^{-2}(0,0,0)=(2/3)\mu_0 e^2 N_0 v_f^2 = \mu_0 e^2 n/m$ is the idealized BCS penetration depth in the clean limit at $T\to 0$, 
$f_m$ is given by Eq.~(\ref{fm_zero_current}), 
and $d_m = \sqrt{(\hbar\omega_m + \Gamma)^2+\Delta^2} + \gamma$. 
Analytical expressions of $\lambda$ can be obtained for some special cases, 
e.g., a nearly-ideal ($\Gamma/\Delta_0 \ll 1$) moderately-clean ($\gamma/\Delta_0 \ll 1$) superconductor at $T\to 0$. The summation can be replaced with integral ($2\pi k_B T/\hbar \sum_{\omega_m>0} \to \int_0^{\infty}d\omega$) when $T\to 0$, and we find
\begin{eqnarray}
\frac{1}{\lambda^2(\gamma\ll \Delta_0, \Gamma\ll \Delta_0, 0)} = \frac{1}{\lambda_0^2} 
\biggl( 1 -\frac{\Gamma}{\Delta_0} -\frac{\pi\gamma}{4\Delta_0} \biggr) . 
\label{penetration_depth_zeroT_1}
\end{eqnarray}
Another example is the idealized ($\Gamma\to 0$) BCS superconductor at $T\to 0$ including an arbitrary concentration of nonmagnetic impurities. 
Replacing the summation with integral, Eq.~(\ref{penetration_depth}) reduces to the well-known formula (see e.g., Ref.~\cite{2003_Marsiglio, 2012_Gurevich_review, 2017_Gurevich_SUST}): 
\begin{equation} 
\frac{1}{\lambda^{2}(\gamma, 0, 0)} =
    \begin{cases}
        \frac{1}{\lambda_0^2 \gamma/\Delta_0}  \Bigl( \frac{\pi}{2} - \frac{\cos^{-1}(\gamma/\Delta_0)}{\sqrt{1-(\gamma/\Delta_0)^2}} \Bigr)    &   (\gamma/\Delta_0<1) \\
        \frac{1}{\lambda_0^2} \bigl( \frac{\pi}{2} -1 \bigr)   &   (\gamma/\Delta_0=1) \\
        \frac{1}{\lambda_0^2 \gamma/\Delta_0} \Bigl( \frac{\pi}{2} - \frac{\cosh^{-1}(\gamma/\Delta_0)}{\sqrt{(\gamma/\Delta_0)^2-1}} \Bigr)  &  (\gamma/\Delta_0 > 1)
    \end{cases}. \label{penetration_depth_zeroT_2}
\end{equation}
In the dirty limit ($\gamma/\Delta_0 = \pi \xi_0/2\ell_{\rm imp} \gg 1$), 
substituting $d_m \simeq \gamma$ into Eq.~(\ref{penetration_depth}), 
we reproduce the formula derived in the previous study~\cite{2017_Gurevich_Kubo}:  
\begin{eqnarray}
&&\frac{1}{\lambda^{2}(\gamma\gg \Delta_0, \Gamma, T)}
= \frac{4 k_B T}{\lambda_{0,{\rm dirty}}^{2}\Delta_0} \sum_{\omega_m>0} f_m^2 \nonumber \\
&&= \frac{2\Delta(\Gamma,T)}{\lambda_{0,{\rm dirty}}^{2} \pi \Delta_0} {\rm Im} \psi \biggl( \frac{1}{2} +\frac{\Gamma}{2\pi k_B T} + i \frac{\Delta(\Gamma, T)}{2\pi k_B T} \biggr) .
\label{penetration_depth_dirty}
\end{eqnarray}
Here, $\lambda_{0,{\rm dirty}}^{-2}=(\pi\Delta_0/2\gamma)\lambda_0^{-2}=\pi \mu_0 \Delta_0 \sigma_n/\hbar$ is the well-known BCS penetration depth in the dirty limit at $T\to 0$, 
and $\psi$ is the digamma function. 
For the idealized dirty-limit BCS superconductor, 
Eq.~(\ref{penetration_depth_dirty}) simplifies to~\cite{Tinkham, Kopnin}
\begin{eqnarray}
\frac{1}{\lambda^{2}(\gamma\gg \Delta_0, 0, T)}
= \frac{\Delta(0,T)}{\lambda_{0,{\rm dirty}}^{2} \Delta_0}\tanh\frac{\Delta(0,T)}{2k_B T}. 
\label{penetration_depth_dirty_idealBCS}
\end{eqnarray}
Eq.~(\ref{penetration_depth_dirty_idealBCS}) is widely used when analyzing experimental data of superconducting devices, but its range of applicability is somewhat limited.

\begin{figure}[tb]
   \begin{center}
   \includegraphics[width=0.5\linewidth]{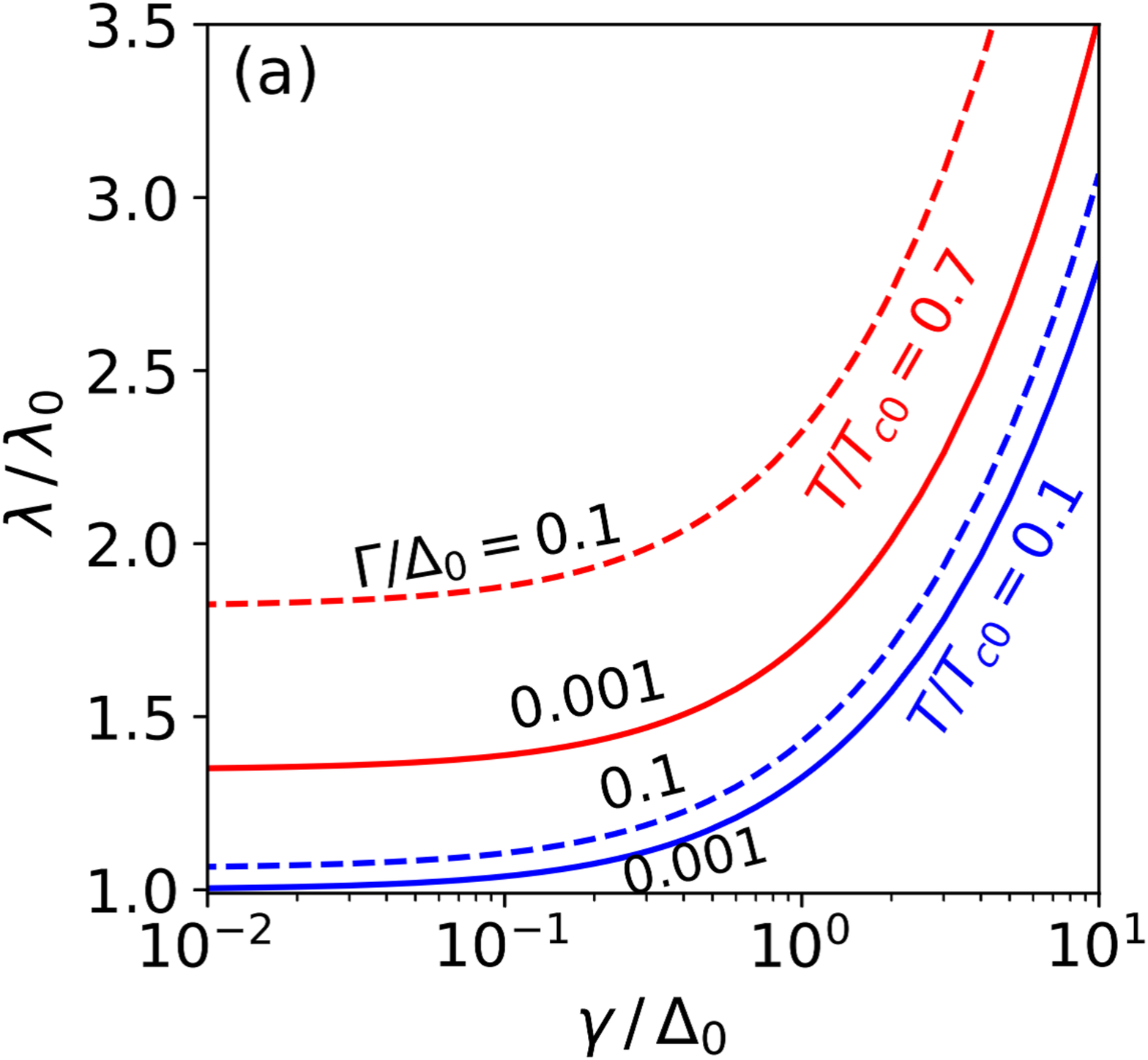}
   \includegraphics[width=0.48\linewidth]{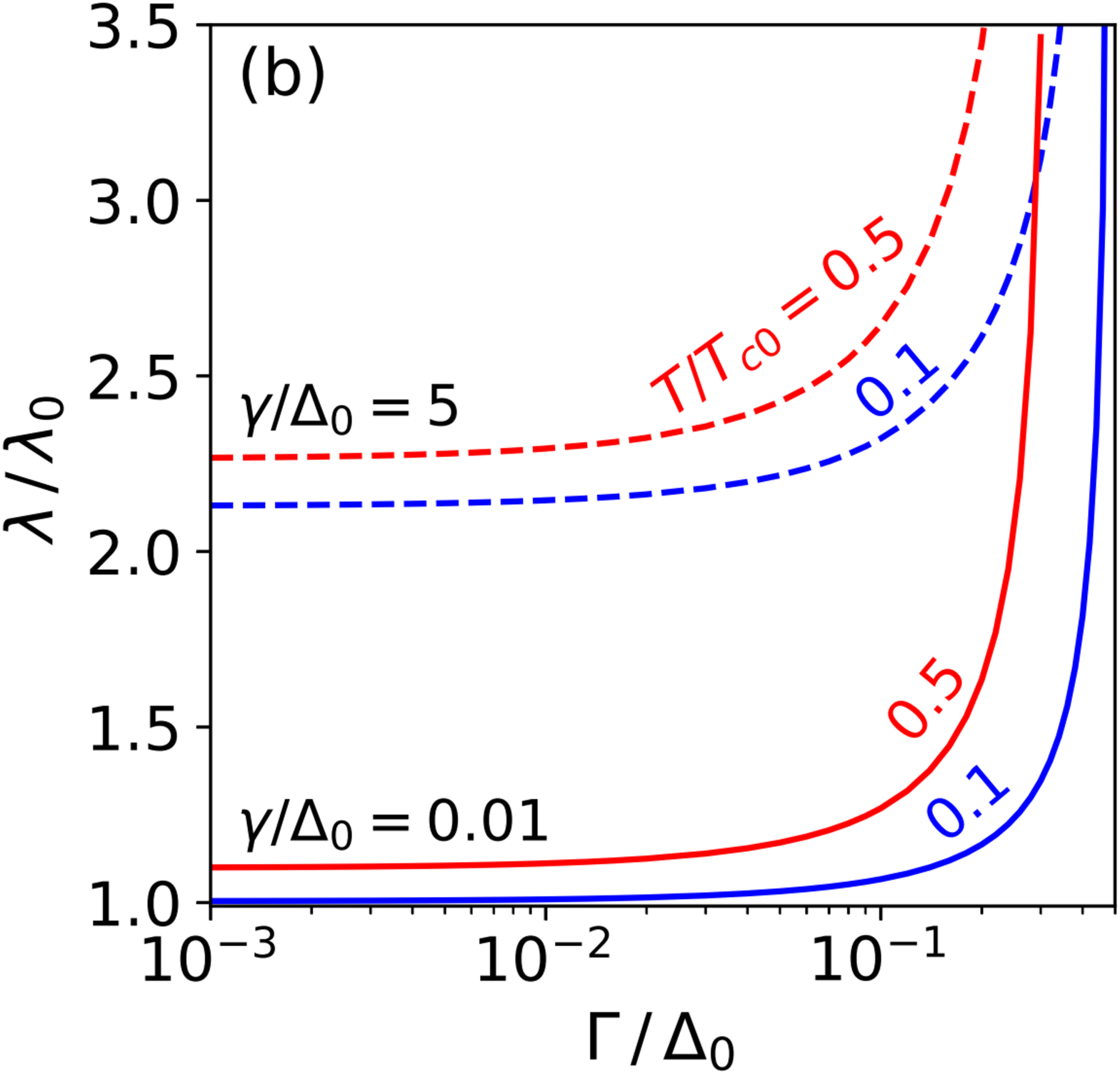}
   \end{center}\vspace{0 cm}
   \caption{
Penetration depth $\lambda(\gamma, \Gamma, T)$ as functions of (a) nonmagnetic-impurity scattering rate  $\gamma/\Delta_0=\pi\xi_0/2\ell_{\rm imp}$ and (b) Dynes parameter $\Gamma$. 
   }\label{fig3}
\end{figure}

To evaluate $\lambda$ for an arbitrary $(\gamma, \Gamma, T)$, 
we need to calculate Eq.~(\ref{penetration_depth}) numerically. 
Shown in Fig.~\ref{fig3} is $\lambda$ as functions of (a) $\gamma$ and (b) $\Gamma$ for different temperatures: 
$\lambda$ is a monotonic increasing function of $\gamma$, $\Gamma$, and $T$. 
The rapid increases of $\lambda$ at $\Gamma/\Delta_0 \gtrsim 0.1$ come from the pair-breaking effect of $\Gamma$ close to its critical value (see also Fig.~\ref{fig2} and Ref.~\cite{2020_Kubo_2}).

\subsection{Kinetic inductance}\label{section_Lk}

\begin{figure}[tb]
   \begin{center}
   \includegraphics[width=0.5\linewidth]{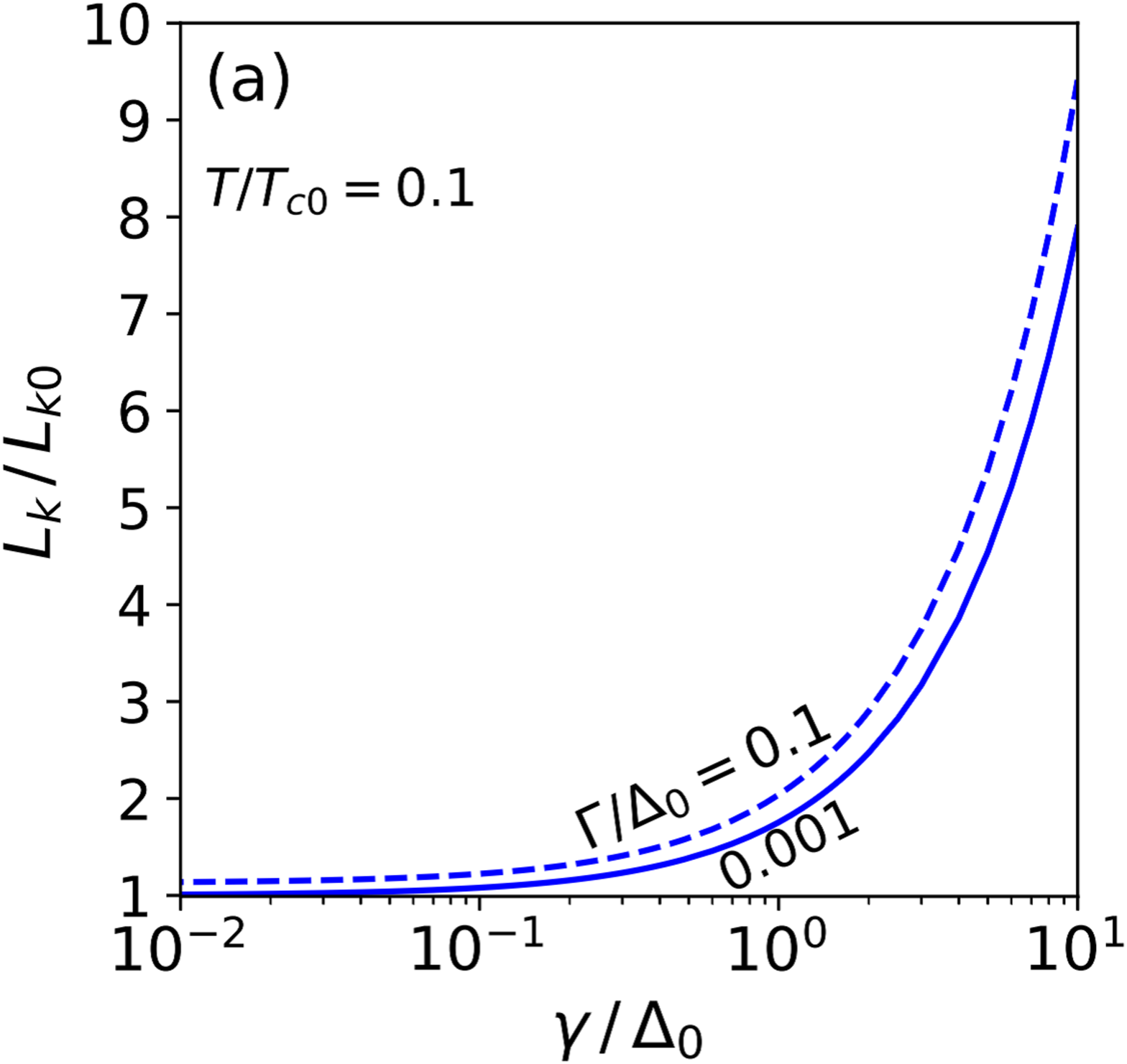}
   \includegraphics[width=0.48\linewidth]{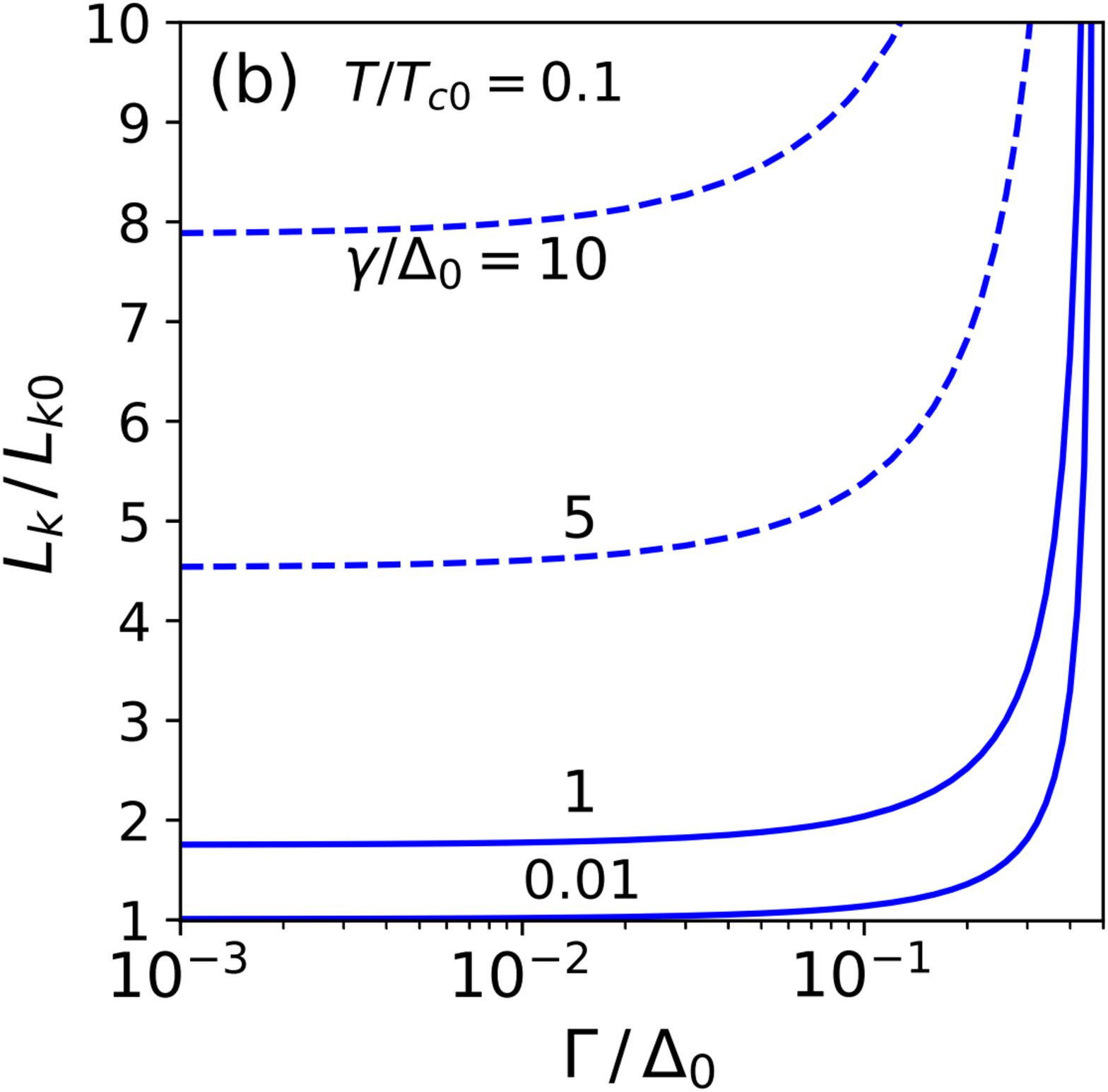}
   \end{center}\vspace{0 cm}
   \caption{
Kinetic inductivity at $T/T_{c0}=0.1$ as functions of (a) nonmagnetic-impurity scattering-rate $\gamma/\Delta_0=\pi\xi_0/2\ell_{\rm imp}$ and (b) Dynes $\Gamma$ parameter. 
   }\label{fig4}
\end{figure}

The direct application of the results of Sec.~\ref{section_lambda} is calculations of the kinetic inductance in the zero-current state,  
which is a crucial physical quantity for superconducting devices, such as KID and SSPD. 
The kinetic inductance of a thin and narrow film is given by $L_{\rm film} = [{\rm length}/({\rm width}\times {\rm thickness})] \times L_k$. 
Here, $L_k$ is the kinetic inductivity. 
In the zero-current limit, $L_k$ is given by $L_k = \mu_0 \lambda^2$ (see also Refs.~\cite{2010_Annunziata, 2012_Clem_Kogan, 2020_Kubo_2} for $L_k$ of a dirty-limit superconductor under the current). 
Since we have already calculated $\lambda$ in Sec.~\ref{section_lambda}, 
we have almost finished the calculations of $L_k$; see Fig.~\ref{fig4}.

For readers' convenience, let us summarize the analytical formulas of $L_k$ in the zero-current limit ($q\to 0$). 
For a superconductor including an arbitrary impurity concentration, 
$L_k$ is given by
\begin{eqnarray}
L_k(\gamma, \Gamma, T) 
= L_{k0} \frac{\lambda^2(\gamma, \Gamma, T)}{\lambda_0^2} .
\label{LK_general}
\end{eqnarray}
Here, $L_{k0}=L_k(0,0,0)=\mu_0 \lambda_0^2 = 3/2e^2 N_0 v_f^2$ is the kinetic inductivity for an idealized ($\Gamma\to 0$) clean-limit ($\gamma \propto \ell_{\rm imp}^{-1} \to 0$) BCS superconductor at $T\to 0$. 
In the dirty limit ($\gamma/\Delta_0 = \pi \xi_0/2\ell_{\rm imp} \gg 1$), 
taking $d_m \to \gamma$, we get 
\begin{eqnarray}
L_k(\gamma\gg \Delta_0, \Gamma, T) 
= L_{k0, {\rm dirty}} \frac{\lambda^2(\gamma\gg \Delta_0, \Gamma, T)}{\lambda_{0,{\rm dirty}}^2} .
\label{Lk_dirty}
\end{eqnarray}
Here, $L_{k0, {\rm dirty}} = \mu_0 \lambda_{0, {\rm dirty}}^2 = \hbar/\pi \Delta_0 \sigma_n$ is the inductivity of the idealized ($\Gamma\to 0$) dirty-limit BCS superconductor at $T\to 0$. 
Substituting Eqs.~(\ref{penetration_depth_zeroT_1}) or (\ref{penetration_depth_zeroT_2}) into Eq.~(\ref{LK_general}) or substituting Eqs.~(\ref{penetration_depth_dirty}) or (\ref{penetration_depth_dirty_idealBCS}) into Eq.~(\ref{Lk_dirty}), 
we can analytically evaluate $L_k$.

It should be noted that, Eqs.~(\ref{penetration_depth_dirty_idealBCS}) and (\ref{Lk_dirty}) yield the widely-used formula, $L_k= [\hbar \rho_n/\pi \Delta(0,T)] / \tanh[\Delta(0,T)/2k_B T]$, 
but it can apply only to the idealized ($\Gamma\to 0$) dirty-limit ($\gamma/\Delta_0 = \pi \xi_0/2\ell_{\rm imp} \gg 1$) BCS superconductor. 
The other formulas introduced in this section are more general and would be available for various situations.

\section{Complex conductivity, surface impedance, and $Q$ factor} 

\subsection{Complex-conductivity formulas} \label{section_sigma_formula}

The general formula for the complex conductivity $\sigma$ of a large-$\lambda/\xi$ superconductor, 
which can include an arbitrary concentration of nonmagnetic impurities, 
is given by~\cite{1967_Nam, 1991_Zimmermann, 1995_Sauls, 1991_Marsiglio, 2003_Marsiglio}
\begin{eqnarray}
&&\sigma (\gamma, \Gamma, T, \omega) =\sigma_1 + i \sigma_2 \nonumber \\
&&= \frac{-i\sigma_n}{4\omega \tau} \int_{-\infty}^{\infty}\!\!d\epsilon \bigl[ I_{1}\tanh\frac{\epsilon_{-}}{2kT} - I_{2}\tanh\frac{\epsilon_{+}}{2kT}  \bigr] , \label{complex_conductivity} \\
&& I_{1} = \frac{g_{+}g_{-} + f_{+}f_{-} -1  }{d_{+} + d_{-}} + 
\frac{g_{+}g_{-}^{*} + f_{+}f_{-}^{*}  +1  }{d_{+} - d^{*}_{-}} , \\
&& I_2 = \frac{g_{+}^{*}g_{-}^{*} + f_{+}^{*}f_{-}^{*} -1  }{-d_{+}^{*} - d_{-}^{*}} + 
\frac{g_{+}g_{-}^{*} + f_{+} f_{-}^{*}  +1  }{d_{+} - d_{-}^{*}} . 
\end{eqnarray}
Here, $\sigma$ is a function of  the impurity scattering rate $\gamma/\Delta_0 = \pi\xi_0/2\ell_{\rm imp}$, the Dynes parameter $\Gamma$, temperature $T$, and the photon frequency $\omega$;  
$\sigma_n=(2/3)e^2 N_0 v_f^2\tau$ is the dc conductivity in the normal state, 
$\tau$ is the electron relaxation-time, 
$\epsilon_{\pm} = \epsilon \pm \hbar\omega/2$, $g_{\pm}= g(\epsilon_{\pm})$, $f_{\pm}= f(\epsilon_{\pm})$, $d_{\pm}= d(\epsilon_{\pm})$, and
$d(\epsilon)=\sqrt{(\epsilon+i\Gamma)^2-\Delta^2}+i\gamma$.

In the normal state ($\Delta \to 0$), we have $g_{\pm}=1$, $f_{\pm}=0$, and $d_+ - d_{-}^*=\hbar \omega +2i (\Gamma + \gamma)$. 
Then, Eq.~(\ref{complex_conductivity}) reduces to the Drude model of the ac electrical conductivity of normal metal: 
$\sigma_{n1} = \sigma_n/[1+(\omega \tau)^2]$ and $\sigma_{n2} = \sigma_n \omega \tau/[1+(\omega \tau)^2]$ with the electron relaxation-time $\tau$, 
\begin{eqnarray}
\frac{1}{\tau} = \frac{1}{\tau_{\rm imp}} + \frac{1}{\tau_{\Gamma}} . \label{tau}
\end{eqnarray}
Here $\tau_{\rm imp} = \hbar/2\gamma$ and $\tau_{\Gamma} = \hbar/2\Gamma$. 
It is also possible to derive Eqs.~(\ref{tau}) considering a microscopic origin of $\Gamma$ such as the electron-phonon scattering~\cite{1991_Marsiglio} and electron-magnetic impurities scatterings~\cite{2016_Herman}.

In the low-frequency regime ($\hbar \omega \ll \gamma, \Gamma, k_B T, \Delta_0$), 
we can expand the integrand with respect to a tiny $\omega$.  
Then, Eq.~(\ref{complex_conductivity}) reduces to
\begin{eqnarray}
\sigma_1(\gamma, \Gamma, T, 0) &=& 
\frac{\sigma_n \hbar/\tau}{8k_B T} \int_{-\infty}^{\infty}\!\!d\epsilon 
\frac{1}{\cosh^2(\epsilon/2k_B T)} \biggl[ \frac{({\rm Re}\, g)^2}{{\rm Im} \,d} \nonumber \\
&& + \frac{ ({\rm Re}\,d \, {\rm Im}\,f - {\rm Im}\,d \,{\rm Re}\,f )^2  }{ \{ ({\rm Re}\,d)^2 + ({\rm Im}\,d)^2 \} {\rm Im}\, d} \biggr], \label{complex_conductivity_1_zero_omega} 
\end{eqnarray}
and
\begin{eqnarray}
\sigma_2 
&=& \frac{\sigma_n}{2\tau \omega} \int_{-\infty}^{\infty}\!\!d\epsilon \tanh\frac{\epsilon}{2k_B T} {\rm Re} \biggl[ \frac{f^2}{d} \biggr] \nonumber \\
&=& \frac{2e^2N_0 v_f^2}{3\mu_0 \omega } 2\pi k_B T \sum_{\omega_m>0} \frac{f_m^2}{d_m} = \frac{1}{\mu_0 \omega \lambda^2} , 
\label{complex_conductivity_2_zero_omega}
\end{eqnarray}
which are used in Ref.~\cite{2021_Herman} to study the effects of $\Gamma$ in the low-frequency regime.

In the dirty limit ($\gamma/\Delta_0=\pi\xi_0/2\ell_{\rm imp} \gg 1$), 
we can take $d_+ \simeq d_{-} \simeq i\gamma$. 
Then, Eq.~(\ref{complex_conductivity}) reduces to the well-known formulas~\cite{1967_Nam}: 
\begin{eqnarray}
&&\sigma_1 (\gamma\gg \Delta_0, \Gamma, T, \omega) \nonumber \\
&&= \frac{\sigma_n}{\hbar\omega} \int_{-\infty}^{\infty}\!\!d\epsilon [f_{\rm FD}(\epsilon)-f_{\rm FD}(\epsilon_{++})] M(\epsilon, \omega) , \label{complex_conductivity_1_dirty}  \\
&&M(\epsilon, \omega)={\rm Re}g(\epsilon){\rm Re}g(\epsilon_{++}) + {\rm Re}f(\epsilon){\rm Re}f(\epsilon_{++}) ,
\end{eqnarray}
and
\begin{eqnarray}
&&\sigma_2 (\gamma\gg \Delta_0, \Gamma, T, \omega) = \frac{\sigma_n}{\hbar\omega} \int_{-\infty}^{\infty}\!\!d\epsilon \tanh\frac{\epsilon}{2k_B T} L(\epsilon, \omega) ,  \label{complex_conductivity_2_dirty} \\
&&L(\epsilon, \omega)={\rm Re}g(\epsilon){\rm Im}g(\epsilon_{++}) + {\rm Re}f(\epsilon){\rm Im}f(\epsilon_{++}) .
\end{eqnarray}
Here, $\epsilon_{++} = \epsilon + \hbar \omega$, and $f_{\rm FD}=[1-\tanh(\epsilon/2k_B T)]/2$ is the Fermi-Dirac distribution. 
Eq.~(\ref{complex_conductivity_1_dirty}) is used in Refs.~\cite{2017_Gurevich_Kubo, 2019_Kubo_Gurevich, 2020_Kubo_1} to study the impact of imperfect surfaces on $Q$ factor. 
It should be noted that, for the idealized ($\Gamma\to 0$) dirty-limit ($\gamma/\Delta_0 \gg 1$) BCS superconductor, Eqs.~(\ref{complex_conductivity_1_dirty}) and (\ref{complex_conductivity_2_dirty}) reproduce the textbook formulas~\cite{2004_Zmuidzinas, Tinkham}:
\begin{eqnarray}
&&\sigma_1 (\gamma\gg \Delta_0, \Gamma\to 0, T, \omega) \nonumber \\
&&= \frac{2\sigma_n}{\hbar\omega} \int_{\Delta}^{\infty}\!\!d\epsilon \frac{[\epsilon(\epsilon+\hbar\omega)+\Delta^2] [f_{\rm FD}(\epsilon)-f_{\rm FD}(\epsilon_{++})]}{\sqrt{\epsilon^2-\Delta^2}\sqrt{\epsilon_{++}^2-\Delta^2}} , \label{MB_ideal_1} 
\end{eqnarray}
and
\begin{eqnarray}
&&\sigma_2 (\gamma\gg \Delta_0, \Gamma\to 0, T, \omega) \nonumber \\
&&= \frac{\sigma_n}{\hbar\omega} \int_{\Delta}^{\Delta+\hbar\omega}\!\!d\epsilon \tanh\frac{\epsilon}{2k_B T}
\frac{\epsilon(\epsilon-\hbar\omega)+\Delta^2}{\sqrt{\epsilon^2-\Delta^2}\sqrt{\Delta^2-\epsilon_{--}^2}} .\label{MB_ideal_2}
\end{eqnarray}
for $\hbar\omega < \Delta$. 
Here, $\epsilon_{--}=\epsilon-\hbar\omega$. 

Now, it would be evident that the famous handy-formulas, Eqs.~(\ref{MB_ideal_1}) and (\ref{MB_ideal_2}), can apply only to the special case and are usually not enough for quantitative understanding of dissipation in real devices, which generally include some pair breakers. 
The previous studies used Eqs.~(\ref{complex_conductivity_1_dirty}), (\ref{complex_conductivity_2_dirty}) and their non-equilibrium version and investigated the effects of Dynes subgap states, strong rf-current, magnetic impurities, and proximity-coupled normal layer at the surface~\cite{2014_Gurevich, 2017_Gurevich_Kubo, 2019_Kubo_Gurevich, 2020_Kubo_1}. 
In the following, we use Eq.~(\ref{complex_conductivity}), which can be applied to a superconductor with an arbitrary impurity concentration, to investigate how a combination of subgap states ($\Gamma$) and nonmagnetic impurities ($\gamma$) affects $\sigma$ and $Q$.

\subsection{Temperature dependence of $\sigma_1$ and coherence peak} \label{section_coherence_peak}

\begin{figure}[tb]
   \begin{center}
   \includegraphics[width=0.49\linewidth]{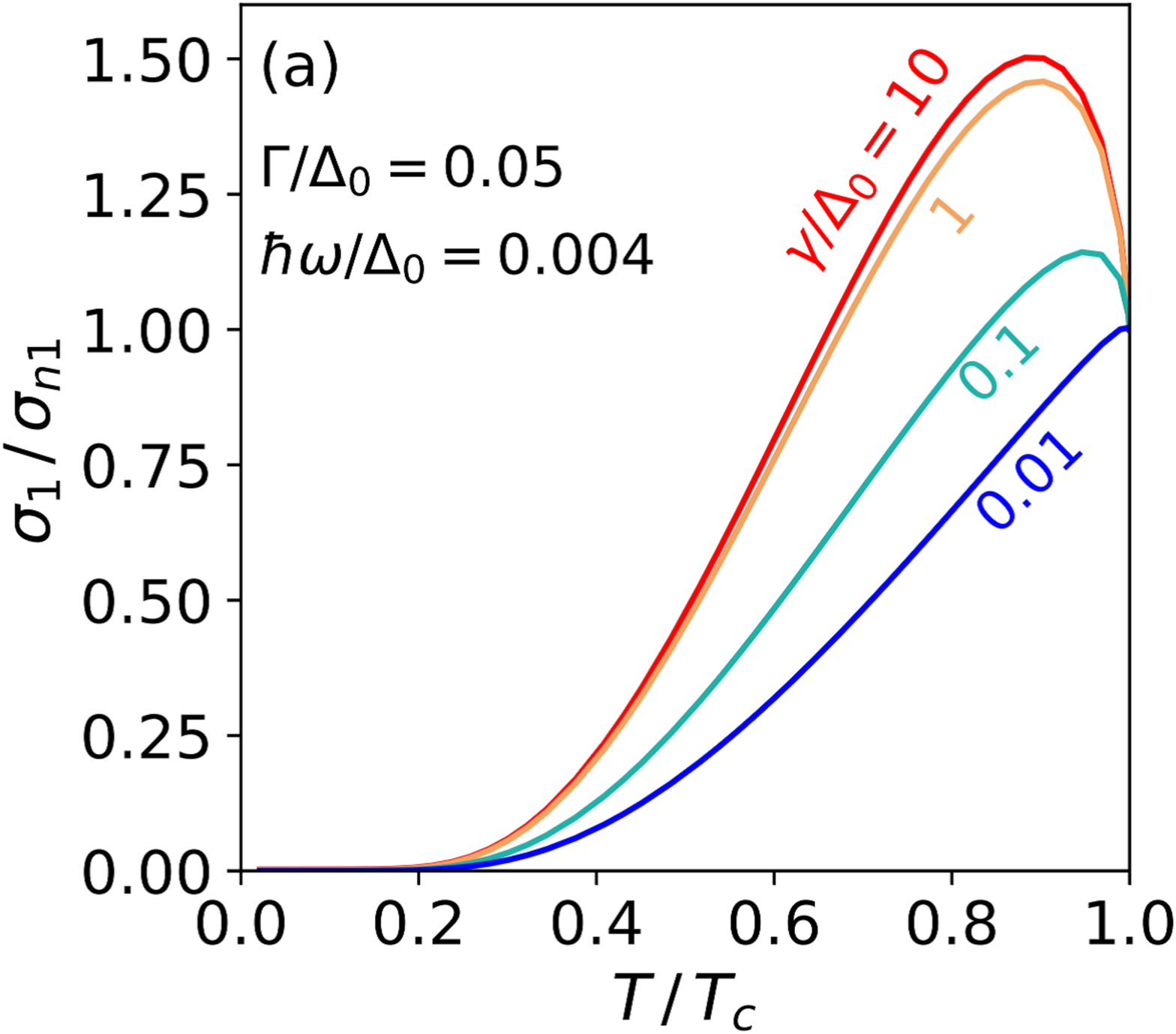}
   \includegraphics[width=0.49\linewidth]{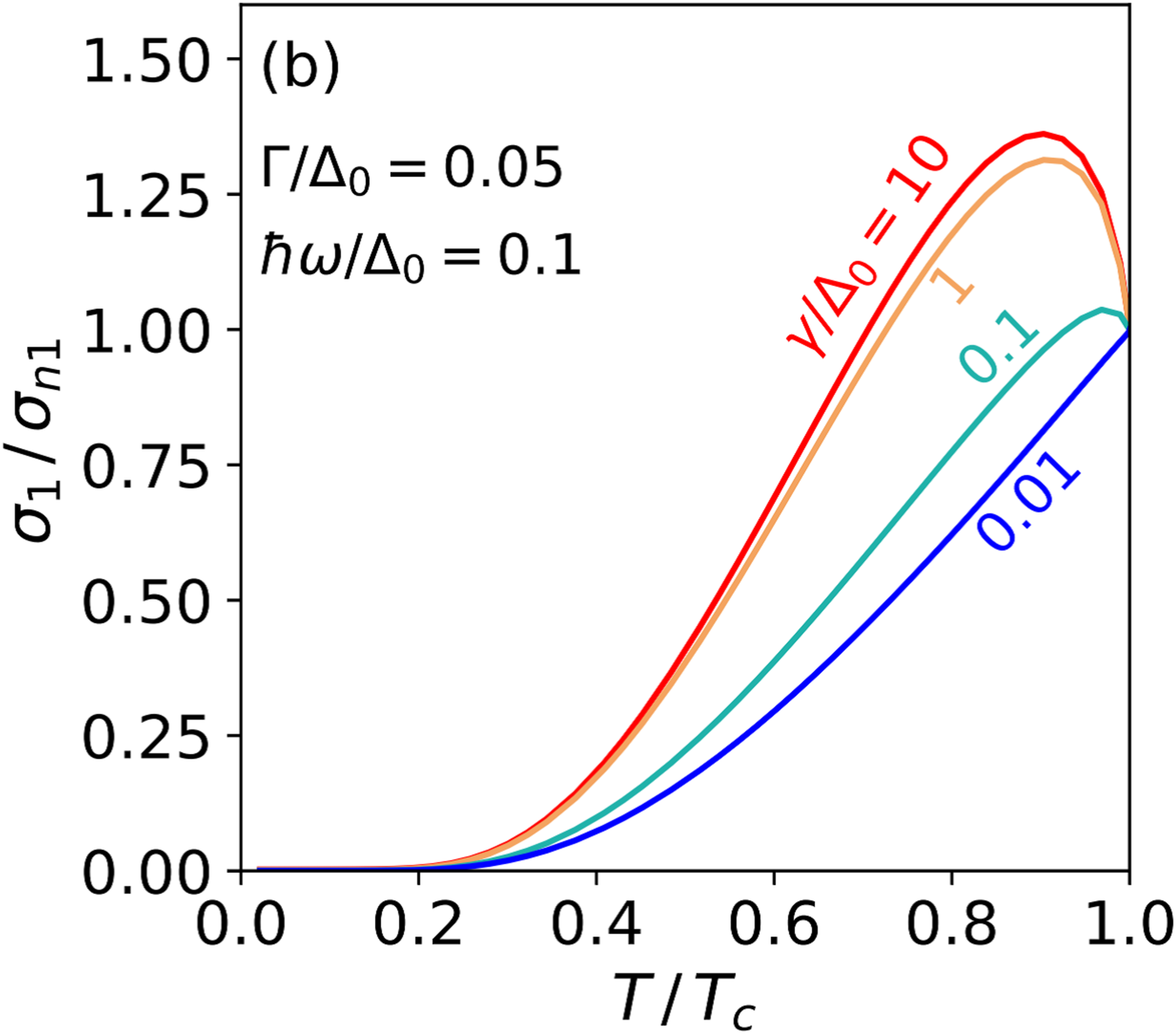}
   \includegraphics[width=0.49\linewidth]{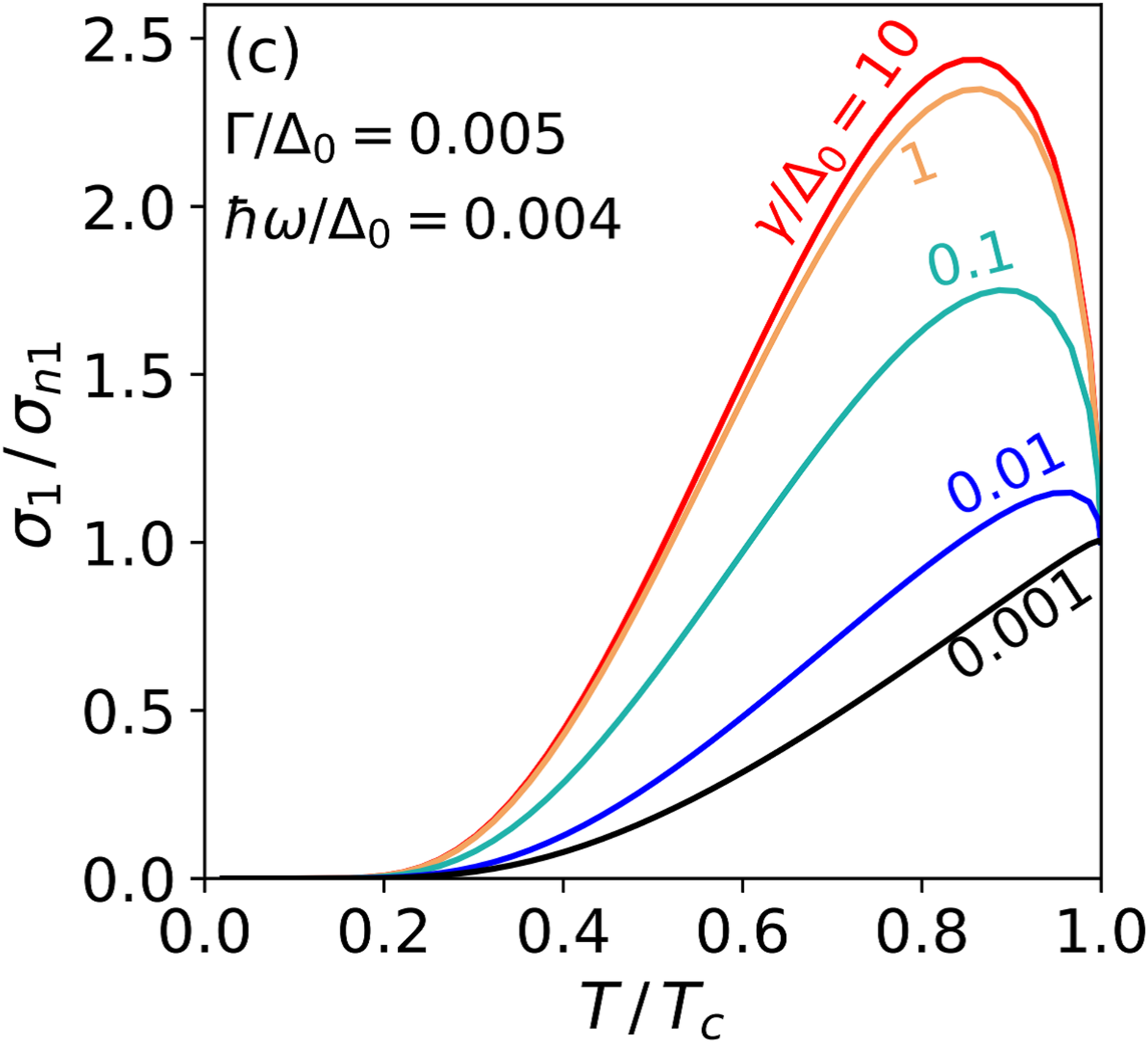}
   \includegraphics[width=0.49\linewidth]{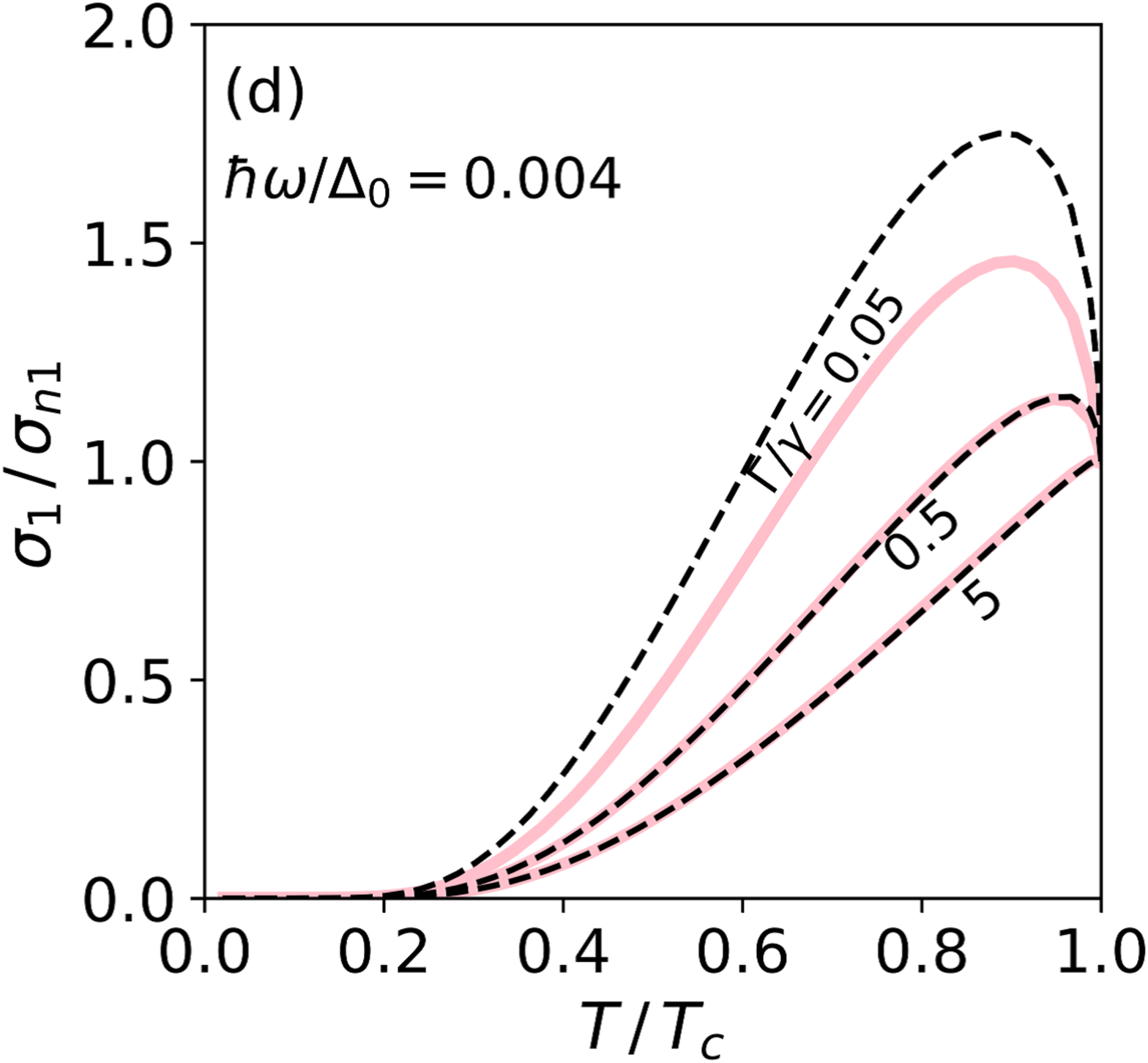}
   \end{center}\vspace{0 cm}
   \caption{
Temperature dependence of $\sigma_1(\gamma, \Gamma, T, \omega)$. 
(a) $\Gamma/\Delta_0=0.05$ and $\hbar\omega /\Delta_0 =0.004$. 
(b) a higher frequency: $\hbar\omega /\Delta_0 =0.1$. 
(c) a smaller Dynes $\Gamma$: $\Gamma/\Delta_0=0.005$. 
The impurity concentration is varied from $\gamma/\Delta_0=0.001$ (clean limit) to $\gamma/\Delta_0=10$ (dirty limit). 
(d) Comparison between (a) and (c). The solid-pink and dashed-black curves correspond to (a) and (c), respectively. The solid and dashed curves for $\Gamma/\gamma = 5$ and 0.5 almost overlap. 
See also the previous study~\cite{2021_Herman} for the results of $\omega \to 0$. 
   }\label{fig5}
\end{figure}

First, consider the effect of the Dynes $\Gamma$ and the nonmagnetic-impurity scattering rate $\gamma \propto \xi_0/\ell_{\rm imp}$ on the temperature dependence of $\sigma_1$ and the coherence peak. 
See also the recent study~\cite{2021_Herman}, in which they studied $\sigma_1(T)$ for $\omega \to 0$.

To examine $\sigma_1(T)$, we need $\Gamma(T)$ too. 
Here we assume $\Gamma$ is independent of $T$ for simplicity. 
Actually, the temperature dependence of $\Gamma$ depends on its microscopic origin (e.g., $T^3$ dependence~\cite{1991_Mikhailovsky} and constant $\Gamma$~\cite{2016_Herman}) and then can depend on materials and surface processing. 
However, according to experiments so far~\cite{2016_Szabo, 2017_Herman}, 
$\Gamma$ is nearly constant in some materials.  
Hence, the $T$-independent $\Gamma$ is a somewhat reasonable assumption and would be a good starting point for investigating the temperature dependence of $\sigma_1$.

Shown in Fig.~\ref{fig5} are the temperature dependences of $\sigma_1(\gamma, \Gamma, T, \omega)$ normalized by the real part of Drude's ac conductivity $\sigma_{n1}=\sigma_n/[1+(\omega\tau)^2]$. 
Fig.~\ref{fig5} (a) is calculated for $\Gamma/\Delta_0=0.05$, the photon frequencies $\hbar \omega/\Delta_0 =0.004$, and several different impurity concentrations. 
As the material gets clean (as $\gamma/\Delta_0 = \pi\xi_0/2\ell_{\rm imp}$ decreases), 
the height of the coherence peak decreases. 
This result may be counter-intuitive, but it is consistent with the previous theory~\cite{1991_Marsiglio, 2021_Herman} and supported by experiments~\cite{2008_Steinberg}. 
Fig.~\ref{fig5} (b) is calculated for the same $\Gamma$ as (a) but a higher frequency ($\hbar \omega/\Delta_0 =0.1$) and shows that $\omega$ reduces the height of the coherence peak. 
Fig.~\ref{fig5} (c) is calculated for the same frequency as (a) but for a smaller $\Gamma$, 
which shows $\Gamma$ decreases the height of the coherence peak.

Shown in Fig.~\ref{fig5} (d) is the comparison between (a) and (c). 
Interestingly, as first pointed out in Ref.~\cite{2021_Herman}, 
the temperature dependence $\sigma_1$ for a clean-limit superconductor does not depend on $\Gamma$ and $\gamma$ separately but on the ratio $\Gamma/\gamma$ (see the curves for $\Gamma/\gamma=5$ and 0.5). 
Note that this is not the case for a dirty superconductor (see the curve for $\Gamma/\gamma=0.05$).

\subsection{Residual $\sigma_1$ at $T\to 0$} \label{section_residual_sigma1}

\begin{figure}[tb]
   \begin{center}
   \includegraphics[width=0.49\linewidth]{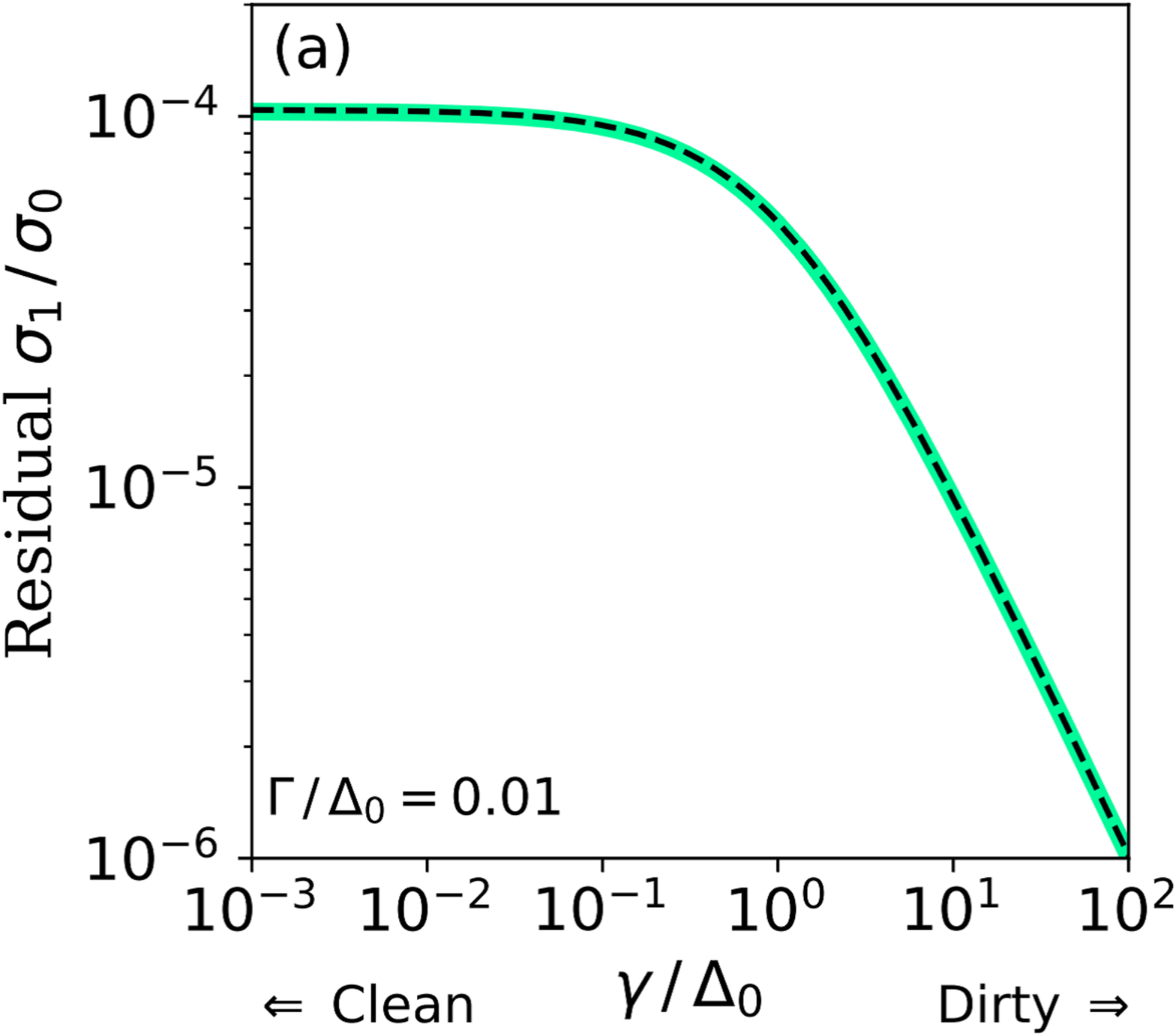}
   \includegraphics[width=0.49\linewidth]{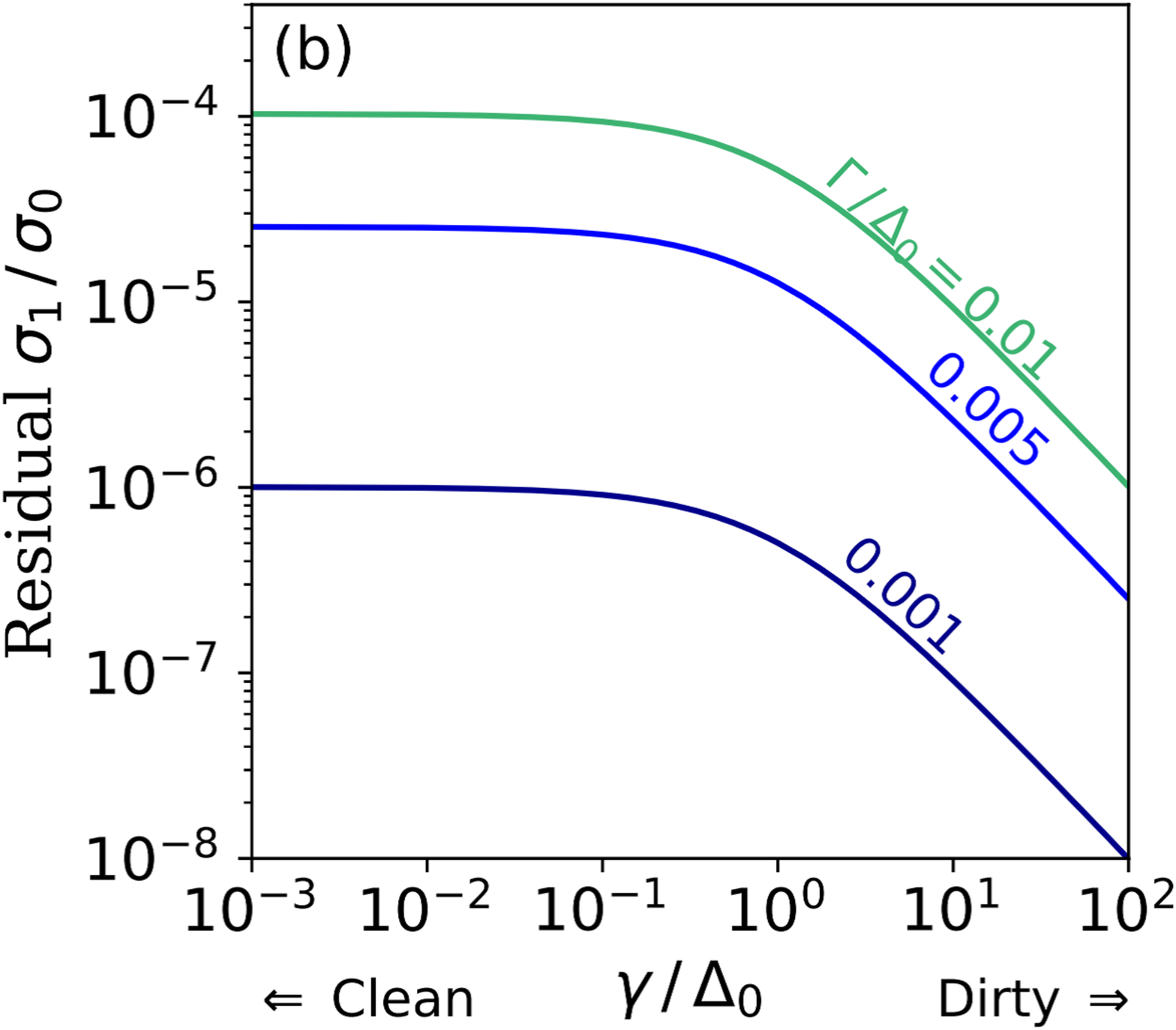}
   \end{center}\vspace{0 cm}
   \caption{
(a) Residual $\sigma_1$ at $T\to 0$ as functions of the nonmagnetic-impurity scattering rate calculated for $\Gamma/\Delta_0=0.01$. 
The dashed-black curve is the numerical result for $T/T_{c0}=0.001$ and $\hbar \omega/\Delta_0=0.1$. 
The solid-green curve, which overlaps the dashed-black curve, is calculated from Eq.~(\ref{sigma1_zeroT}). 
(b) Residual $\sigma_1$ at $T\to 0$ as functions of the nonmagnetic-impurity scattering rate for different $\Gamma$.
   }\label{fig6}
\end{figure}

At $T\to 0$, it is well-known that $\sigma_1$ of the idealized BCS superconductor ($\Gamma \to 0$) approaches zero. 
On the other hand, a realistic quasiparticle spectrum expressed with a finite $\Gamma$ results in a residual $\sigma_1$ due to a finite subgap states at the Fermi level~\cite{2012_Gurevich_review, 2017_Gurevich_SUST, 2017_Gurevich_Kubo, 2017_Herman, 2020_Kubo_1, 2021_Herman}. 
Suppose $\hbar\omega/\Delta_0 \ll 1$. 
Then, replacing $\tanh (\epsilon/2k_B T)$ with the Heaviside step function and using $g_{\pm}=\Gamma/\sqrt{\Gamma^2+\Delta^2}$,  $f_{\pm}=\Delta/i\sqrt{\Gamma^2+\Delta^2}$, and $d_{\pm}=i\sqrt{\Gamma^2+\Delta^2}+i\gamma$, 
Eq.~(\ref{complex_conductivity}) reduces to~\cite{2021_Herman} 
\begin{eqnarray}
\sigma_1 (\gamma, \Gamma, T, \omega)|_{T\to 0}
=\sigma_n \frac{\gamma+\Gamma}{\gamma+\sqrt{\Gamma^2+\Delta^2}} \frac{\Gamma^2}{\Gamma^2+\Delta^2}. \label{sigma1_zeroT}
\end{eqnarray}
Here, $\Delta=\Delta(\Gamma,T)|_{T\to0}=\Delta_0\sqrt{1-2\Gamma/\Delta_0}$ [see Eq.~(\ref{Delta_zeroT})]. 
In the dirty limit ($\gamma/\Delta_0 = \pi\xi_0/2\ell_{\rm imp} \gg 1$), 
we reproduce~\cite{2017_Gurevich_SUST, 2017_Gurevich_Kubo, 2019_Kubo_Gurevich, 2020_Kubo_1, 2017_Herman}
\begin{eqnarray}
\sigma_1 (\gamma, \Gamma, T, \omega)|_{T\to 0}
=\sigma_n \frac{\Gamma^2}{\Gamma^2+\Delta^2} . \label{sigma1_zeroT_dirty}
\end{eqnarray}
Note that $\sigma_n$ depends on electron relaxation time: $\sigma_n \propto \tau$.

Shown in Fig.~\ref{fig6} (a) is the residual dissipative conductivity $\lim_{T\to 0}\sigma_1$ as a function of the nonmagnetic-impurity scattering rate $\gamma/\Delta_0=\pi\xi_0/2\ell_{\rm imp}$. 
Note that $\sigma_1$ is normalized by $\sigma_0 = (2/3)e^2 N_0 v_f^2 \tau_0 = (\tau_0/\tau)\sigma_n$, 
independent of the electron mean free path of materials in contrast to $\sigma_n$. 
Here, $\tau_0$ is defined by $\tau_0 = \hbar/2\Delta_0$. 
The dashed-black curve is numerically calculated for $\hbar\omega=0.1$ and $T/T_{c0}=10^{-3}$. 
The solid-green curve, calculated from Eq.~(\ref{sigma1_zeroT}), overlaps the dashed-black curve and is a good approximation as long as $\hbar\omega \ll \Delta_0$. 
We find the residual $\sigma_1$ decreases as materials get dirty [see Eq.~(\ref{sigma1_zeroT_dirty}): $\sigma_1 \propto \sigma_n \propto 1/\gamma$]. 
Shown in Fig.~\ref{fig6} (b) is $\lim_{T\to 0}\sigma_1$ calculated for different $\Gamma$ using Eq.~(\ref{sigma1_zeroT}). 
The residual $\sigma_1$ increases as $\Gamma$ increases (i.e., as subgap states at Fermi level increase). 
Hence, the subgap-states-induced residual $\sigma_1$ can be significantly suppressed by using a nearly ideal ($\Gamma \ll \Delta_0$) dirty-limit ($\gamma/\Delta_0 =\pi \xi_0/2\ell_{\rm imp} \gg 1$) superconductor.

\subsection{Moderately low-temperature regime ($\hbar \omega \ll k_B T \ll \Delta_0$)} \label{lowF_lowT}

\begin{figure}[tb]
   \begin{center}
   \includegraphics[width=0.49\linewidth]{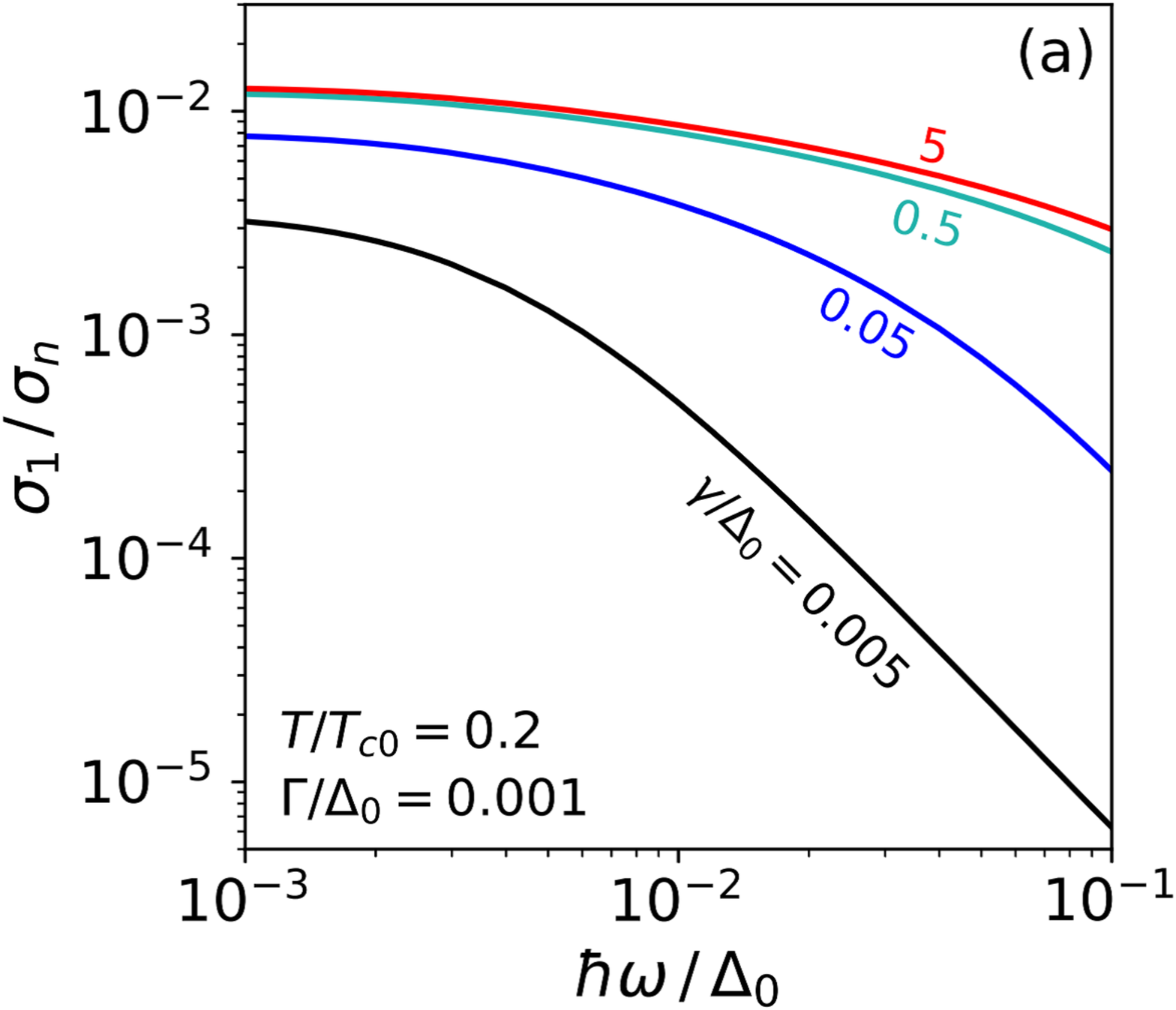}
   \includegraphics[width=0.49\linewidth]{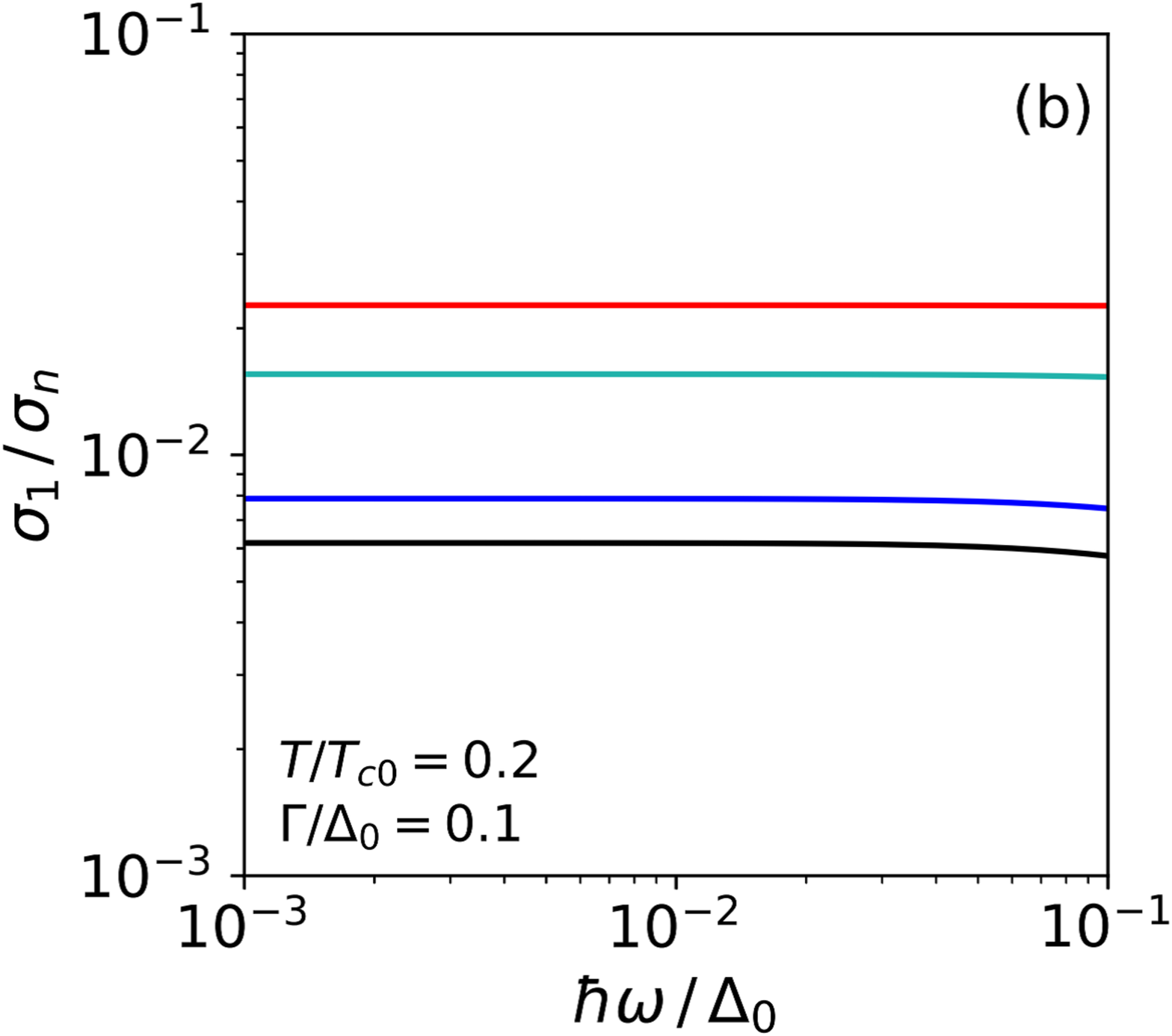}
   \includegraphics[width=0.49\linewidth]{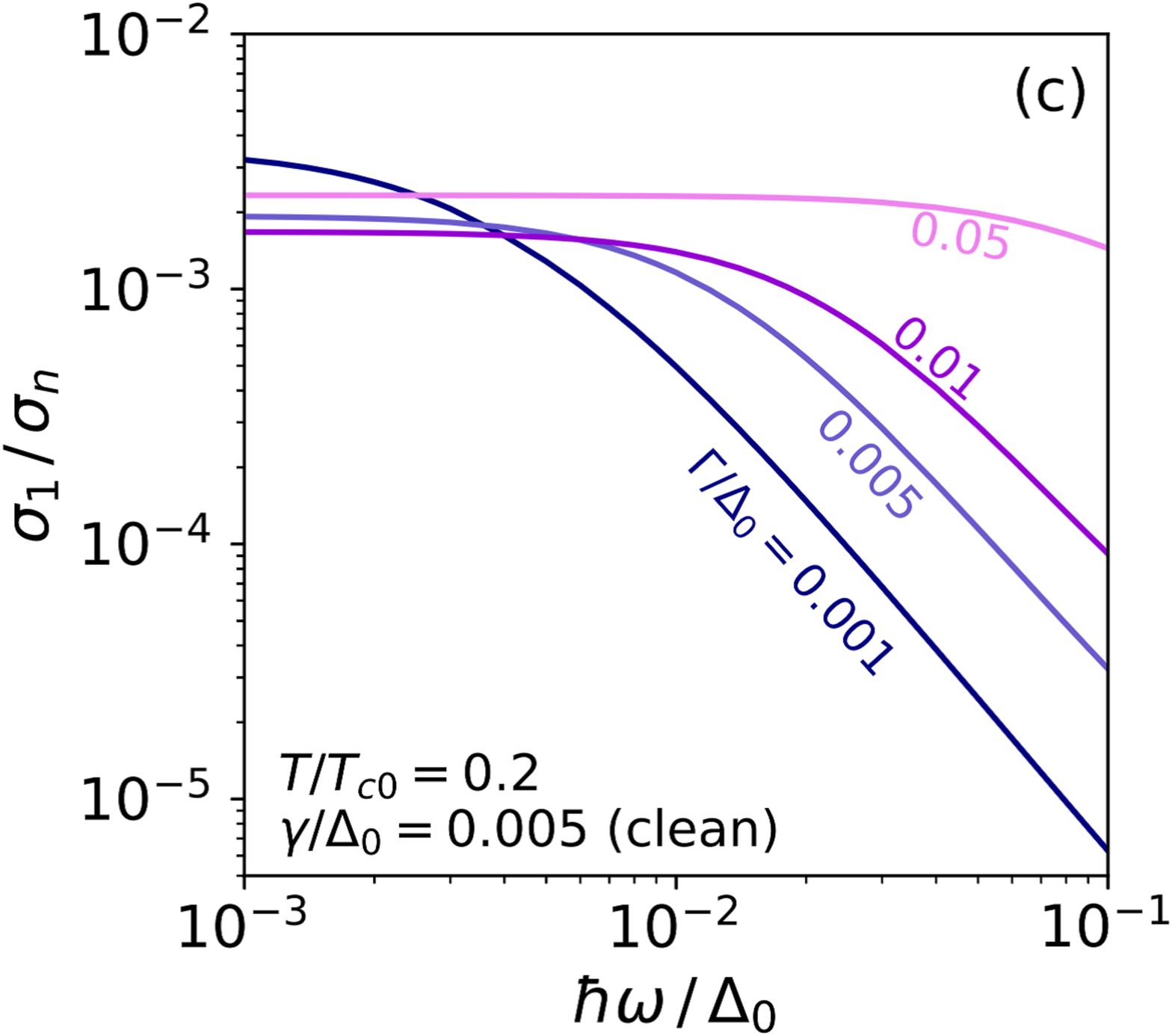}
   \includegraphics[width=0.49\linewidth]{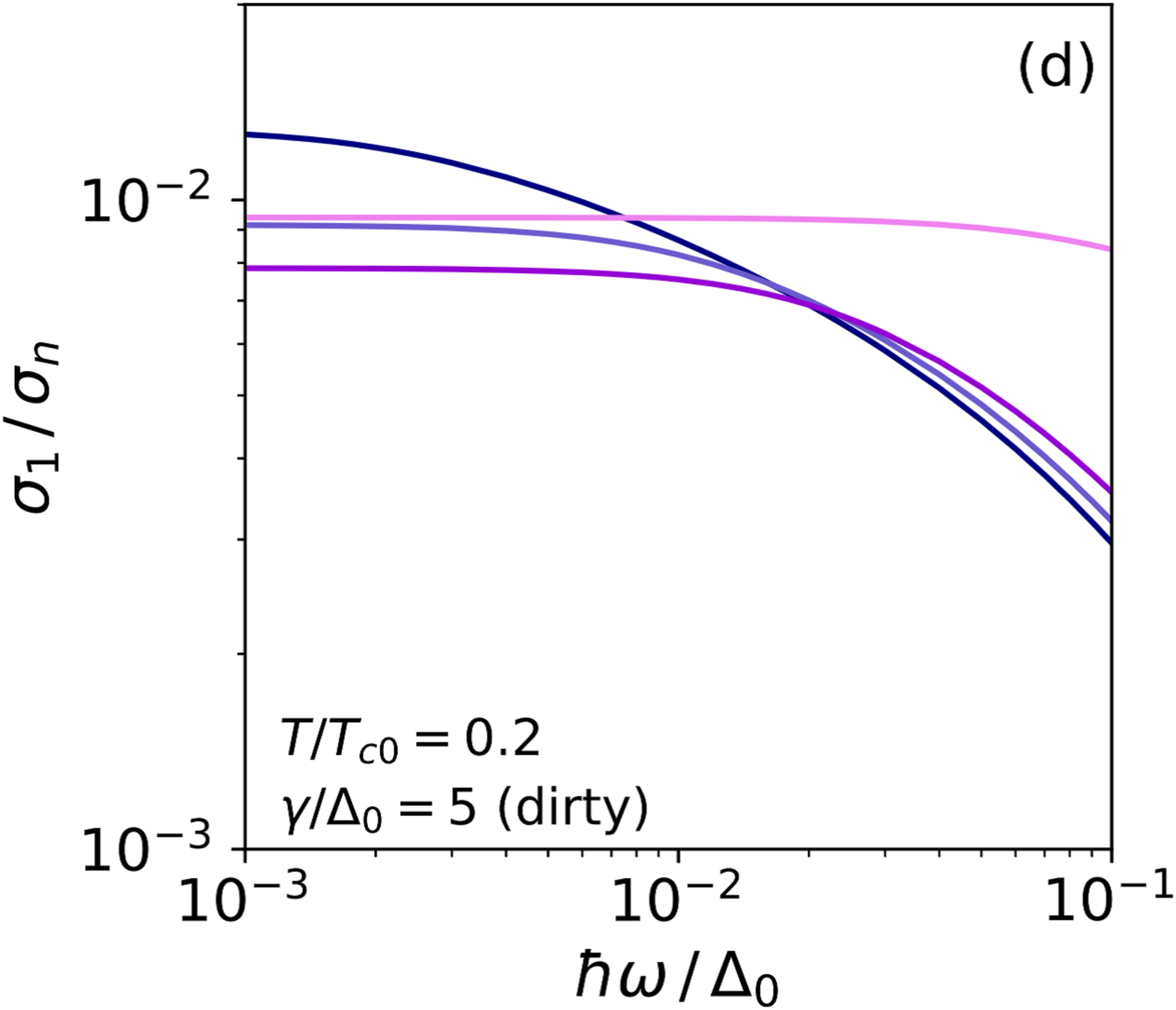}
   \end{center}\vspace{0 cm}
   \caption{
The dissipative conductivity $\sigma_1$ as functions of the photon frequency $\omega$. 
(a) $\sigma_1$ of a nearly-ideal ($\Gamma/\Delta_0=0.001$) superconductor calculated for different nonmagnetic-impurity scattering rate $\gamma/\Delta_0=\pi\xi_0/2\ell_{\rm imp}$ from the clean limit to the dirty limit.
(b) $\sigma_1$ of a superconductor with strongly-smeared gap-peaks ($\Gamma/\Delta_0=0.1$) calculated for different $\gamma$.
(c) $\sigma_1$ of a clean-limit ($\gamma/\Delta_0=0.005$) superconductor calculated for different Dynes $\Gamma$.
(d) $\sigma_1$ of a dirty-limit ($\gamma/\Delta_0=5$) superconductor calculated for different Dynes $\Gamma$.
   }\label{fig7}
\end{figure}

Type-II superconductors used for superconducting devices typically have the gap frequencies $\Delta_0/2\pi\hbar \gtrsim $ a few hundred GHz. 
Operating frequencies of devices are around $\sim 1$-$10\,{\rm GHz}$ and satisfy $\hbar \omega \ll \Delta_0$. 
The operating temperature depends on the type of device, 
but almost all devices satisfy $k_B T < 0.2 k_B T_c \simeq 0.1 \Delta_0 \ll \Delta_0$. 
Hence, we may consider that a typical operating condition of superconducting microwave devices falls into the category of the low-frequency and low-temperature regime: $\hbar \omega \ll \Delta_0$ and $k_B T \ll \Delta_0$. 
In particular, when $\hbar \omega \ll k_B T \ll \Delta_0$, 
$\sigma_1$ is mainly determined by thermally activated quasiparticles and is sensitive to the quasiparticle spectrum as shown below.

First, consider the idealized ($\Gamma \to 0$) BCS superconductor in the dirty limit ($\gamma/\Delta_0=\pi\xi_0/2\ell_{\rm imp} \gg 1$). 
When $\hbar \omega \ll \Delta_0$ and $k_B T \ll \Delta_0$, the main contribution to the integral in Eq.~(\ref{MB_ideal_1}) comes from a narrow strip of energy range $0 < \bar{\epsilon} \lesssim k_B T (\ll \Delta)$. 
Here, we defined $\bar{\epsilon} = \epsilon-\Delta$. 
Then, the factor $\epsilon^2 -\Delta^2$ reduces to $\simeq 2\bar{\epsilon} \Delta$, 
the distribution function reduces to $f_{\rm FD}(\epsilon) - f_{\rm FD}(\epsilon + \hbar\omega) \simeq e^{-\Delta/k_B T} e^{-\bar{\epsilon}/k_B T} (1 - e^{-\hbar\omega/k_B T})$, 
and then Eq.~(\ref{MB_ideal_1}) results in the famous handy-formula~\cite{2012_Zmuidzinas, 2017_Gurevich_SUST}: 
$\sigma_1^{\rm dirty}/\sigma_n = (4\Delta/\hbar\omega) e^{-\Delta/k_B T} \sinh(\hbar\omega/2k_B T) K_0( \hbar\omega/2k_B T)$. 
Here, the modified Bessel function $K_0( \hbar\omega/2k_B T)$ reduces to $\simeq \ln (4k_B T/e^{\gamma_E} \hbar\omega)$ when $\hbar \omega \ll k_B T$. 
Hence, we find $\sigma_1^{\rm dirty} \propto \ln (1/\omega)$. 
However, as materials get clean, $\sigma_1$ exhibits a much different behavior from the dirty limit. 
Applying the similar calculation as above to Eq.~(\ref{complex_conductivity}), 
we find $\sigma_1^{\rm clean} \propto \omega^{-2}$ when $\hbar\omega \ll k_B T \ll \Delta_0$. 
Fig.~\ref{fig7} (a) shows $\sigma_1$ of a nearly-ideal ($\Gamma/\Delta_0=0.001$) superconductor as functions of $\omega$ for different impurity concentrations from the clean limit (black) to the dirty limit (red). 
A dirty-limit superconductor (red) exhibits $\ln(1/\omega)$ dependence (see the red curve), 
while a clean-limit superconductor (black) exhibits $\omega^{-2}$ dependence at $\hbar\omega/\Delta \gtrsim 0.01$. 
The black curve deviates from $\omega^{-2}$ dependence at $\hbar \omega \lesssim \gamma$, 
where the clean-limit assumption is no longer appropriate.

Finite subgap states can drastically change the $\omega$ dependence of $\sigma_1$. 
Fig.~\ref{fig7} (b) is an example calculated for significantly smeared DOS peaks ($\Gamma/\Delta_0=0.1$, see also Fig.~\ref{fig1}), 
in which $\sigma_1$ is almost independent of $\omega$. 
Rather than such a large $\Gamma$ ($\sim k_B T$), 
a relatively small $\Gamma$ ($\hbar\omega \lesssim \Gamma \ll k_B T \ll \Delta_0$) induces nontrivial effects. 
Fig.~\ref{fig7} (c) and (d) are $\sigma_1$ for a clean and a dirty superconductor, respectively, calculated for different $\Gamma$. 
In both cases, $\sigma_1$ for low-frequency regions ($\hbar\omega \ll k_B T$) decreases as $\Gamma$ increases, takes the minimum when $\Gamma/\Delta_0 \sim 0.01$-$0.1$, and then increases with $\Gamma$.

\begin{figure}[tb]
   \begin{center}
   \includegraphics[width=0.47\linewidth]{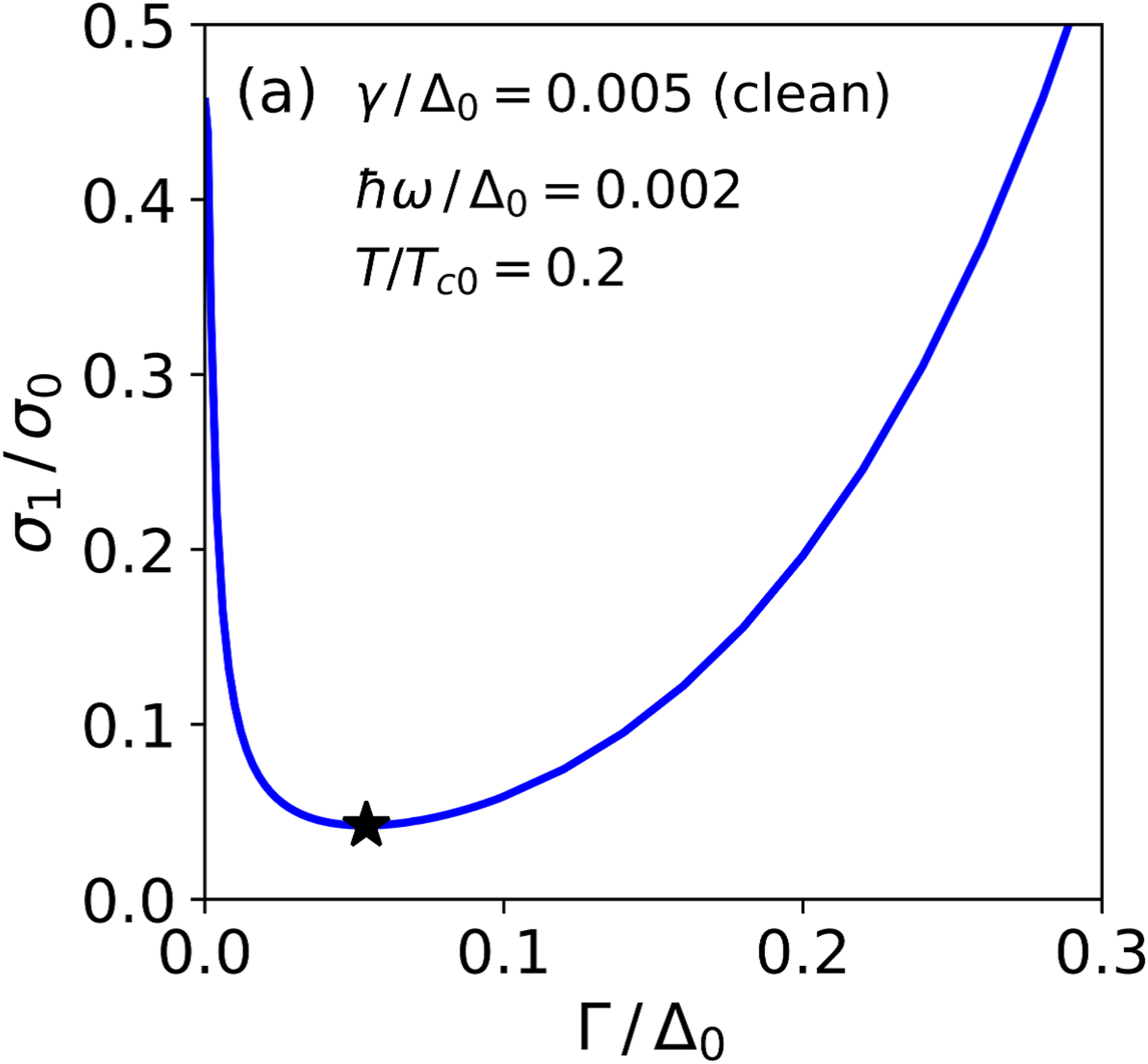}
   \includegraphics[width=0.51\linewidth]{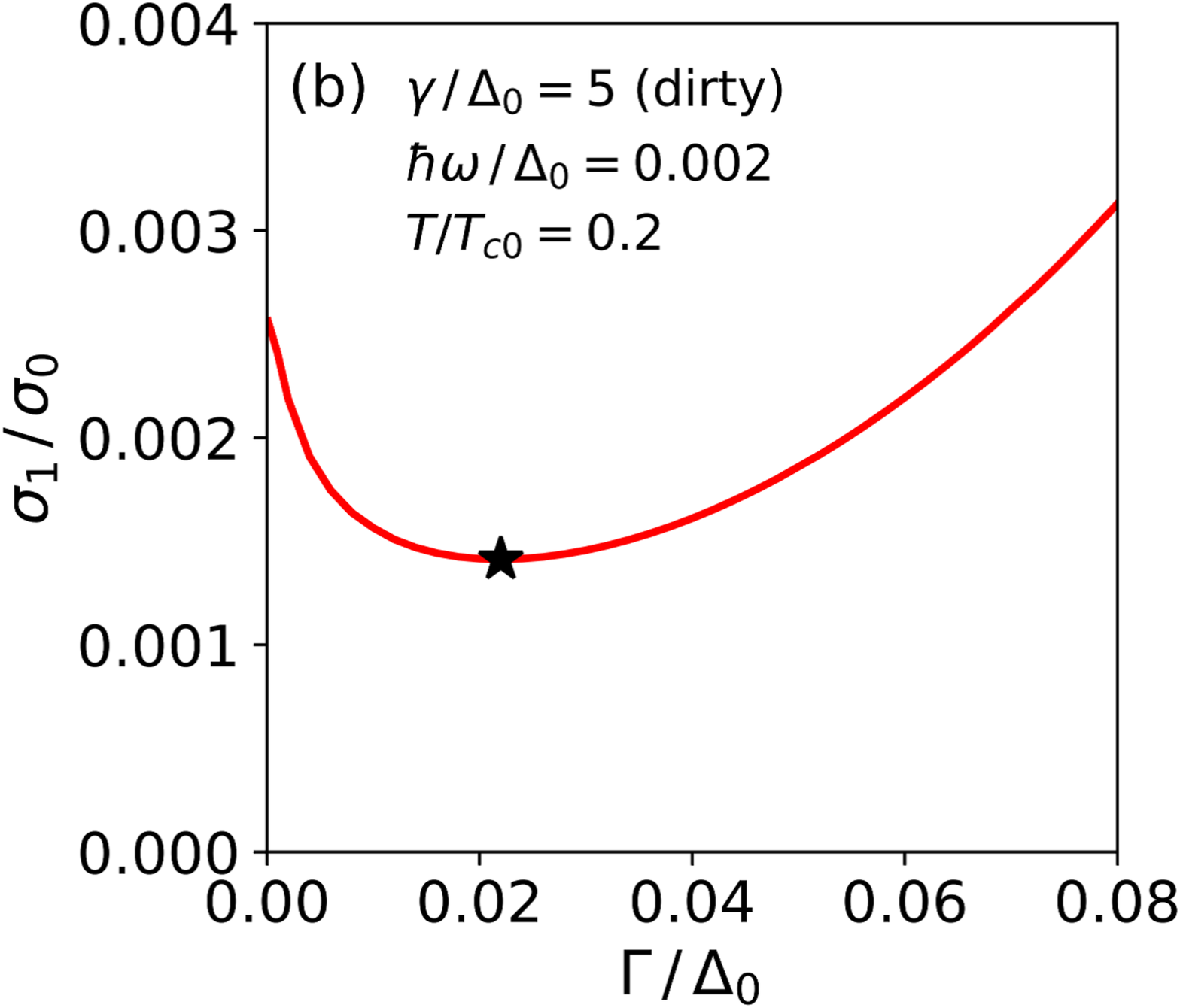}
   \end{center}\vspace{0 cm}
   \caption{
The dissipative conductivity $\sigma_1$ as functions of Dynes $\Gamma$. 
(a) $\sigma_1$ of a clean-limit superconductor ($\gamma/\Delta_0=\pi\xi_0/2\ell_{\rm imp}=0.005$) and (b) a dirty-limit superconductor ($\gamma/\Delta_0=5$). 
The black stars are the minimums. 
   }\label{fig8}
\end{figure}

Figs.~\ref{fig8} (a) and (b) show $\sigma_1$ as functions of Dynes $\Gamma$. 
In both the clean- and dirty-limit cases, the optimum value ($\Gamma_*$) minimizes $\sigma_1$. 
In the clean (dirty) limit,  $\sigma_1(\Gamma_*)$ is $\simeq 90\%$ ($40\%$) smaller than that of the idealized ($\Gamma\to 0$) BCS superconductor. 
These results can be qualitatively explained as follows.  
At moderately low temperatures, $\hbar\omega \ll k_B T \ll \Delta_0$, 
$\sigma_1$ is mainly determined by thermally activated quasiparticles and sensitive to the the broadening of the DOS. 
For the dirty limit, the integral of the broadened quasiparticle-spectrum with width $\sim \Gamma > \hbar\omega$ yields $\sigma \propto \log (1/\Gamma)$~\cite{2017_Gurevich_SUST, 2014_Gurevich, 2017_Gurevich_Kubo, 2020_Kubo_1}.  
As $\Gamma$ increases, $\sigma_1$ logarithmically decreases [see the slope at $\Gamma \lesssim 0.01$ in Fig.~\ref{fig8} (b)]. 
In the clean limit, in addition to the broadened quasiparticle-spectrum, 
the $d$ factor contributes to the integral, 
resulting in a sharper drop of $\sigma_1 \propto \Gamma^{-1/2} \ln (1/\Gamma)$ [see the slope at $\Gamma \lesssim 0.05$ in Fig.~\ref{fig8} (a)].

According to Eq.~(\ref{complex_conductivity_2_zero_omega}), we have
\begin{eqnarray}
\sigma_2 (\gamma, \Gamma, T, \omega) = \frac{1}{\omega \mu_0 \lambda^2(\gamma, \Gamma, T) }, 
\end{eqnarray}
for $\hbar\omega \ll \Delta_0$. 
Hence the results are already shown in Figs.~\ref{fig3} and \ref{fig4}.
Here we do not repeat the calculations of $\sigma_2$. 
See also Refs.~\cite{2017_Herman, 2021_Herman}.

\subsection{Quality factor and surface resistance} \label{section_Rs}

The electromagnetic response of a superconductor is generally nonlocal~\cite{MB}. 
However, for an extreme type-II superconductor ($\lambda\gg \xi$), 
a calculation reduces to a local problem, 
and the surface impedance $Z_s$ is expressed with the complex conductivity. 
When $\sigma_1/\sigma_2 \ll 1$ and the thickness of the material is larger than $\lambda$, 
we have
\begin{eqnarray}
&& Z_s = R_s -i X_s = R_s + j X_s , \\
&& R_s  = \frac{1}{2} \mu_0^2 \omega^2 \lambda^3 \sigma_1, \label{Rs} \\
&& X_s = \sqrt{\frac{\mu_0 \omega}{\sigma_2}} = \mu_0 \omega \lambda .
\end{eqnarray}
The quality factor $Q$ can be expressed with these quantities. 
Remind $Q$ is defined by $Q=\omega U/P$. 
Here, $P=(1/2)\int R_s |{\bf H}|^2 dS$ and $U=(\mu_0/2) \int |{\bf H}|^2 dV$ are the dissipation and the stored energy in the cavity, and $dS$ and $dV$ represent the area element and the volume element, respectively. 
Then, we have $Q=\mu_0 \omega \int |{\bf H}|^2 dV/\int R_s |{\bf H}|^2 dS$, 
which simplifies to 
\begin{eqnarray}
&&Q = \frac{G}{R_s} , \\
&&G = \frac{\mu_0 \omega \int |{\bf H}|^2 dV}{\int |{\bf H}|^2 dS} , 
\end{eqnarray}
when $R_s$ is uniform on the inner surface of the cavity. 
We can calculate $R_s$ using the results of Sec.\ref{section_lambda_Lk}-\ref{lowF_lowT} and Eq.~(\ref{Rs}).

\begin{figure}[tb]
   \begin{center}
   \includegraphics[width=0.49\linewidth]{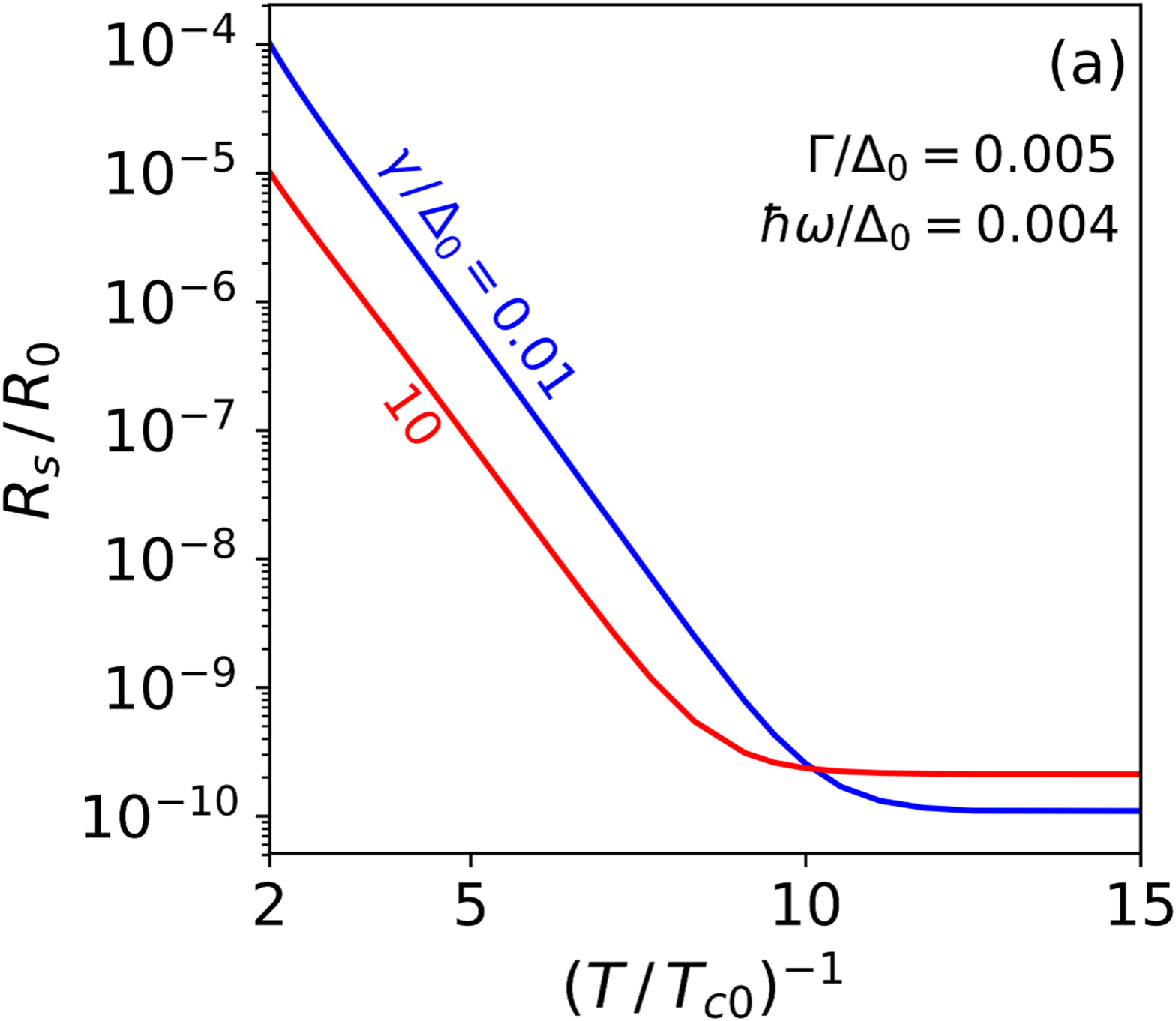}
   \includegraphics[width=0.49\linewidth]{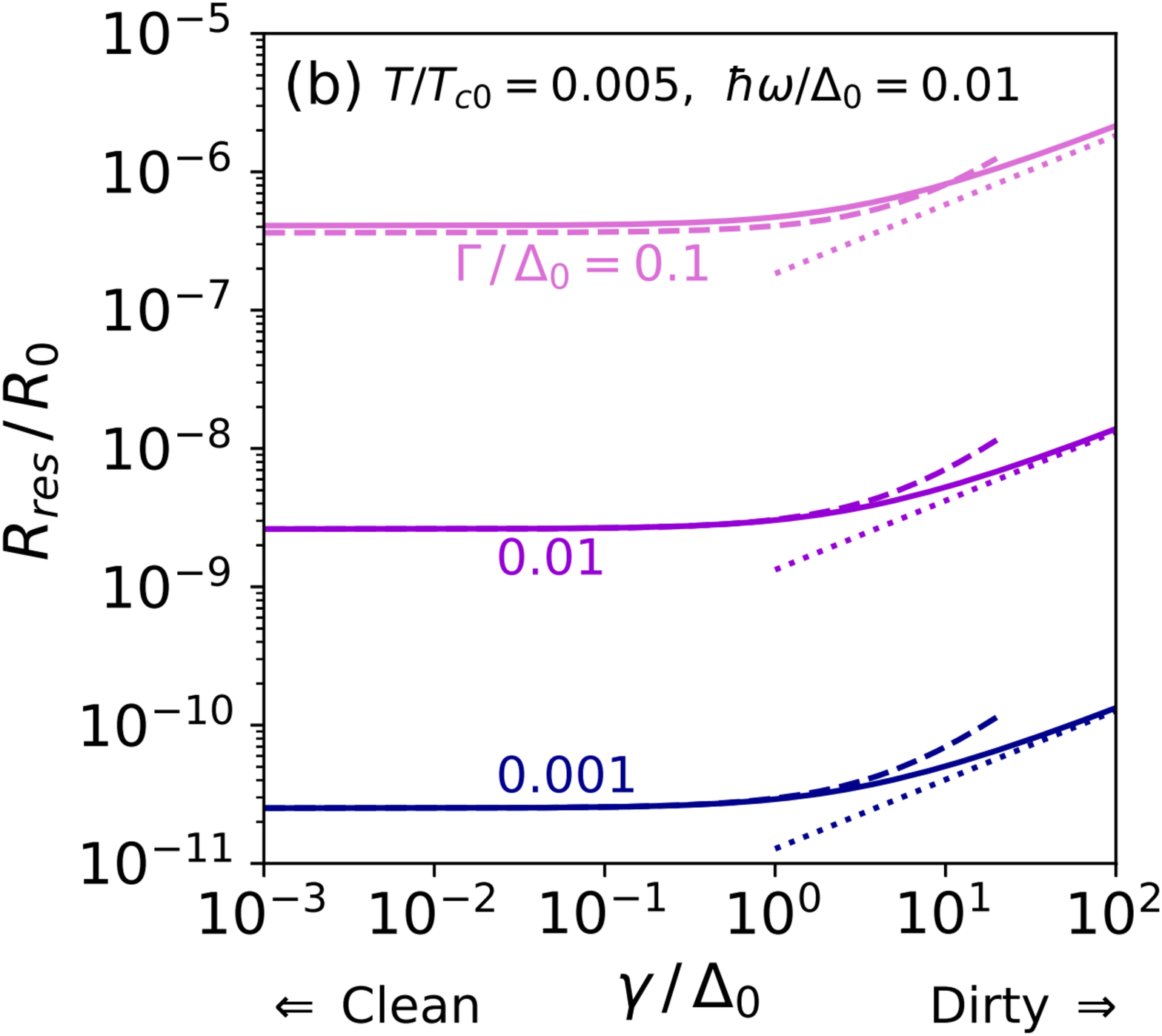}
   \end{center}\vspace{0 cm}
   \caption{
(a) Surface resistance $R_s$ as functions of $1/T$. Here, Dynes $\Gamma$ is assumed independent of $T$ for simplicity (see also Section~\ref{section_coherence_peak}). 
(b) Subgap-state-induced residual surface-resistance $R_{\rm res}= \lim_{T\to 0} R_s$ as functions of nonmagnetic-impurity scattering rate $\gamma/\Delta_0 = \pi\xi_0/2\ell_{\rm imp}$ calculated for different Dynes $\Gamma$. 
The solid curves are numerically calculated for $T/T_{c0}=0.005$ and $\hbar\omega/\Delta_0 =0.01$.  
The dashed and dotted curves are calculated from Eqs.~(\ref{Rres_approx}) and (\ref{Rres_approx_dirty}), respectively. 
   }\label{fig9}
\end{figure}

Shown in Fig.~\ref{fig9} (a) is $R_s$ as functions of $1/T$ for different impurity concentrations. 
Here, the normalization factor $R_0$ is given by 
\begin{eqnarray}
R_0 = \frac{1}{2}\mu_0^2\tau_0^{-2} \lambda_0^3 \sigma_0 = \frac{\mu_0 \Delta_0 \lambda_0}{\hbar} .
\end{eqnarray}
For instances, $R_0 \simeq 0.5\,{\rm \Omega}$ and $\simeq 0.8\,{\rm \Omega}$ for ${\rm Nb_3 Sn}$ and NbTiN, respectively. 
At higher temperatures, a dirty superconductor (red) exhibits smaller $R_s$. 
At lower temperatures, however, $R_s$ of  a clean superconductor (blue) becomes smaller than the red curve. 
Although the residual $\sigma_1$ at $T\to 0$ decreases as material gets dirtier (see Fig.~\ref{fig6}), 
the contribution from $\lambda^3$ significantly pushes up $R_s$ of a dirty superconductor. 
Shown in Fig.~\ref{fig9} (b) is the subgap-state-induced residual surface resistance $R_{\rm res} = \lim_{T\to 0}R_s$ as functions of the nonmagnetic-impurity scattering rate $\gamma/\Delta_0 = \pi\xi_0/2\ell_{\rm imp}$ (solid curves). 
$R_{\rm res}$ is a monotonic decreasing function of $\gamma$ and $\Gamma$. 
For $\Gamma/\Delta_0 \ll 1$ and $\gamma/\Delta_0 \ll 1$ (clean limit), 
we find an analytical formula of $R_{\rm res}$ for a clean superconductor: 
\begin{eqnarray}
R_{\rm res}^{\rm clean} = \frac{R_0}{4} \biggl( \frac{\hbar\omega}{\Delta_0}\biggr)^2 \biggl(\frac{\Gamma}{\Delta_0}\biggr)^2
\biggl[ 1 + \frac{9\Gamma}{2\Delta_0} + \biggl(\frac{3\pi}{8} -1\biggr)\frac{\gamma}{\Delta_0} \biggr]. \nonumber \\
\label{Rres_approx}
\end{eqnarray}
Here, Eqs.~(\ref{Delta_zeroT}), (\ref{penetration_depth_zeroT_1}), and (\ref{sigma1_zeroT}) are used. 
The dashed curves are calculated from Eq.~(\ref{Rres_approx}), 
which agree well with the numerical results (solid curves) at $\gamma/\Delta_0 \lesssim 1$. 
For $\Gamma/\Delta_0 \ll 1$ and $\gamma/\Delta_0 \gg 1$ (dirty limit), 
using Eqs.~(\ref{Delta_zeroT}), (\ref{penetration_depth_dirty}), and (\ref{sigma1_zeroT_dirty}), 
we find 
\begin{eqnarray}
R_{\rm res}^{\rm dirty} = \frac{R_0}{\pi\sqrt{2\pi}} \biggl( \frac{\hbar\omega}{\Delta_0}\biggr)^2 \biggl(\frac{\Gamma}{\Delta_0}\biggr)^2 \sqrt{\frac{\gamma}{\Delta_0}}
\biggl[ 1 +  \biggl(\frac{7}{2} +\frac{3}{\pi}\biggr)\frac{\Gamma}{\Delta_0} \biggr].  \nonumber \\
\label{Rres_approx_dirty}
\end{eqnarray}
The dotted curves are calculated from Eq.~(\ref{Rres_approx_dirty}) and agree with the numerical results (solid curves) at $\gamma/\Delta_0 \gtrsim 10$. 
Here, we have focused on the subgap-state-induced $R_{\rm res}$, 
which can be translated into the quality factor $Q_{\rm \Gamma}=G/R_{\rm res}$. 
Note that other factors [e.g., the two-level-system (TLS) defect~\cite{TLS}, nonequilibrium quasiparticles~\cite{noneq_qp}, and trapped vortices~\cite{2017_Gurevich_SUST}], can also contribute to the total $Q$: 
\begin{eqnarray}
Q^{-1} = Q^{-1}_{\rm \Gamma} + Q^{-1}_{\rm TLS} + \dots . \label{Q_Gamma_TLS}
\end{eqnarray}
Suppose $\gamma/\Delta_0 \sim 1$, $R_0\sim 0.1\,{\rm \Omega}$, and $G \sim 100\,{\rm \Omega}$. 
Then, $\Gamma/\Delta_0 \sim 0.01$ and $0.1$ yield $Q^{-1}_{\rm \Gamma} \sim 10^{-12}$ and $10^{-10}$, respectively. 
All other contributions need to be suppressed to a level below these values to observe $Q^{-1}_{\rm \Gamma}$.

\begin{figure}[tb]
   \begin{center}
   \includegraphics[width=0.49\linewidth]{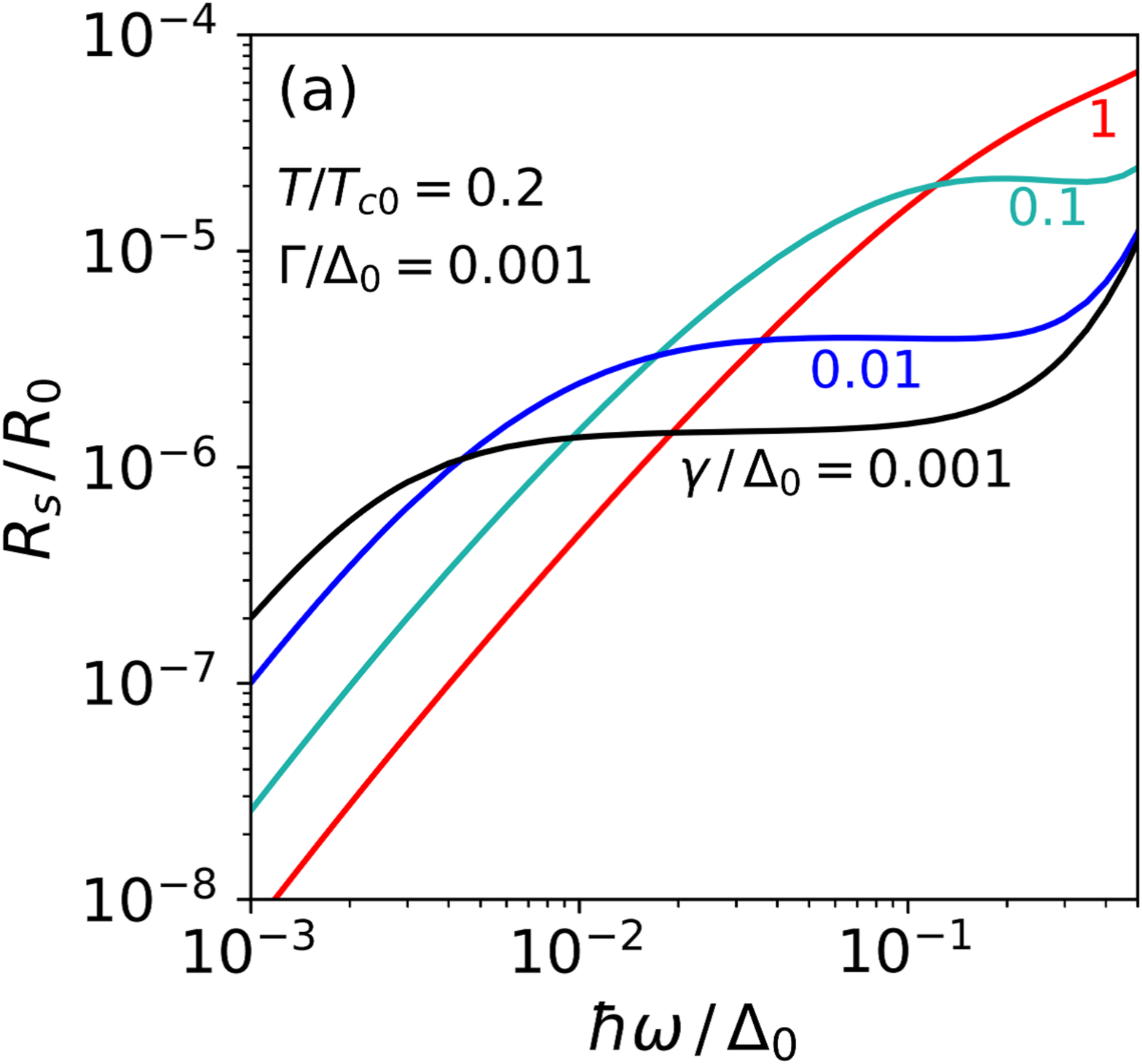}
   \includegraphics[width=0.49\linewidth]{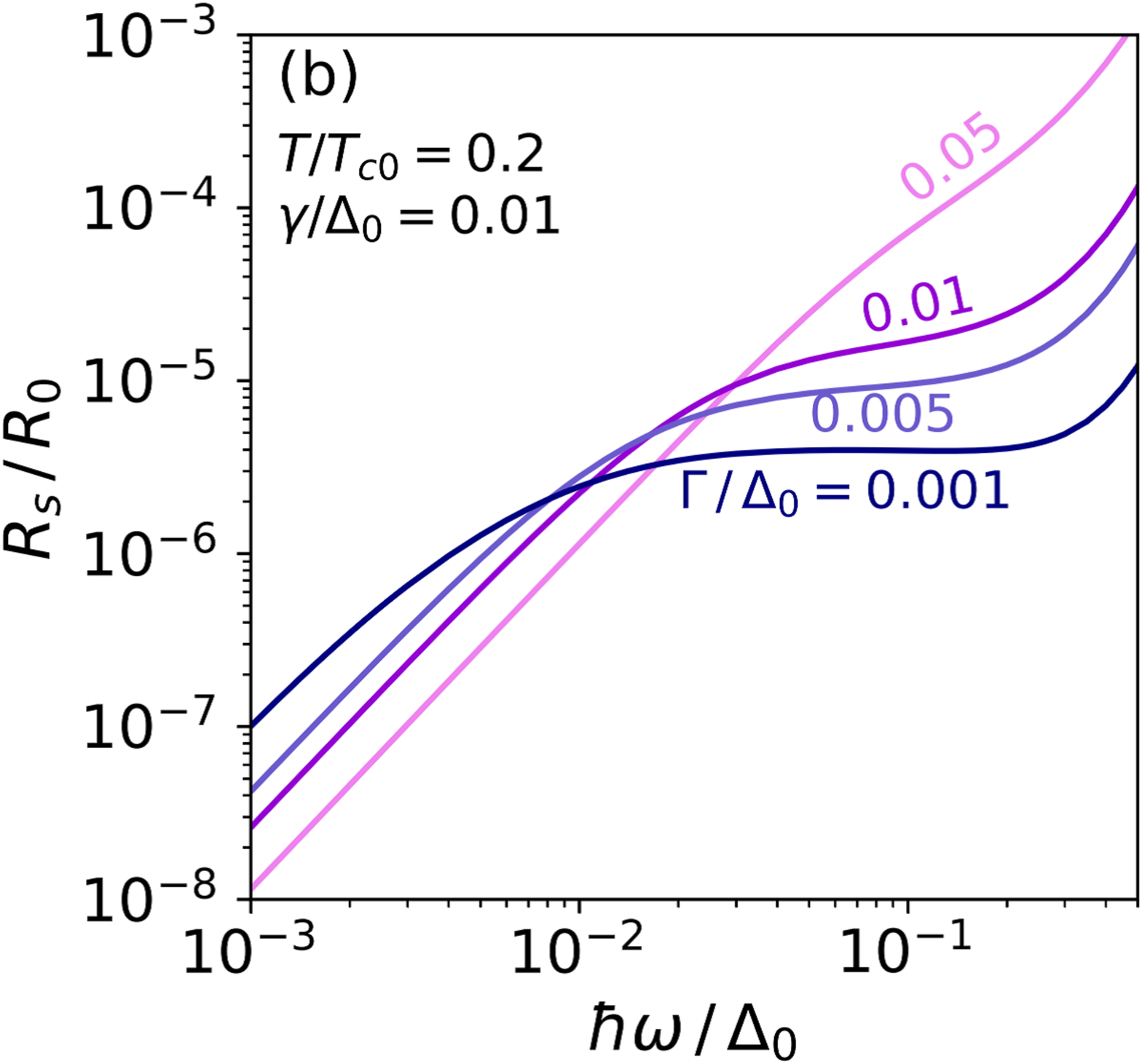}
   \end{center}\vspace{0 cm}
   \caption{
Frequency dependences of the surface resistance $R_s$ (a) calculated for different nonmagnetic-impurity scattering rate $\gamma$ and (b) calculated for different Dynes $\Gamma$. 
   }\label{fig10}
\end{figure}

Next, consider $R_s$ in the moderately-low-temperature regime ($\hbar\omega \ll k_B T\ll \Delta_0$), 
which corresponds to an operating condition of some types of microwave resonators (e.g., SRF and KID).  
Fig.~\ref{fig10} (a) shows $R_s$ of a nearly ideal BCS superconductor ($\Gamma/\Delta_0 =0.001\ll 1$) as functions of $\omega$ calculated for different impurity concentrations. 
The red curve corresponds to a dirty superconductor, 
exhibiting the famous $\omega^2$ dependence well below the gap frequency (see, e.g., Ref.~\cite{1991_Halbritter}). 
However, as the material gets clean (i.e., as $\gamma$ decreases), a plateau appears. 
When $\gamma/\Delta_0=\pi\xi_0/2\ell_{\rm imp}=0.001$ (black curve), 
$R_s$ is almost independent of $\omega$ for a wide frequency range, $0.005 \lesssim \hbar\omega/\Delta_0 \lesssim 0.1$, 
corresponding to $3.7\,{\rm GHz} < f < 75\,{\rm GHz}$ and $3.4\,{\rm GHz} < f < 68\,{\rm GHz}$ for ${\rm Nb_3 Sn}$ and NbTiN, respectively. 
This frequency-independent $R_s$ comes from the $\omega^{-2}$ dependence of $\sigma_1$ in the clean limit (see Fig.~\ref{fig7}) canceling out the $\omega$ dependence of $R_s \propto \omega^2 \sigma_1$. 
Shown in Fig.~\ref{fig10} (b) is calculated for different Dynes $\Gamma$. 
As $\Gamma$ increases, the plateau shrinks and disappears at $\Gamma/\Delta_0 \sim 0.05$. 
Only nearly ideal ($\Gamma/\Delta_0\ll 1$) clean-limit ($\gamma/\Delta_0\ll 1$) superconductors exhibit the plateau. 
It should be noted that this result is applicable to extreme type-II superconductors ($\lambda\gg \xi$), 
while pure Al and pure Nb, which are popular resonator-materials, do not fall into this category.

\begin{figure}[tb]
   \begin{center}
   \includegraphics[width=0.487\linewidth]{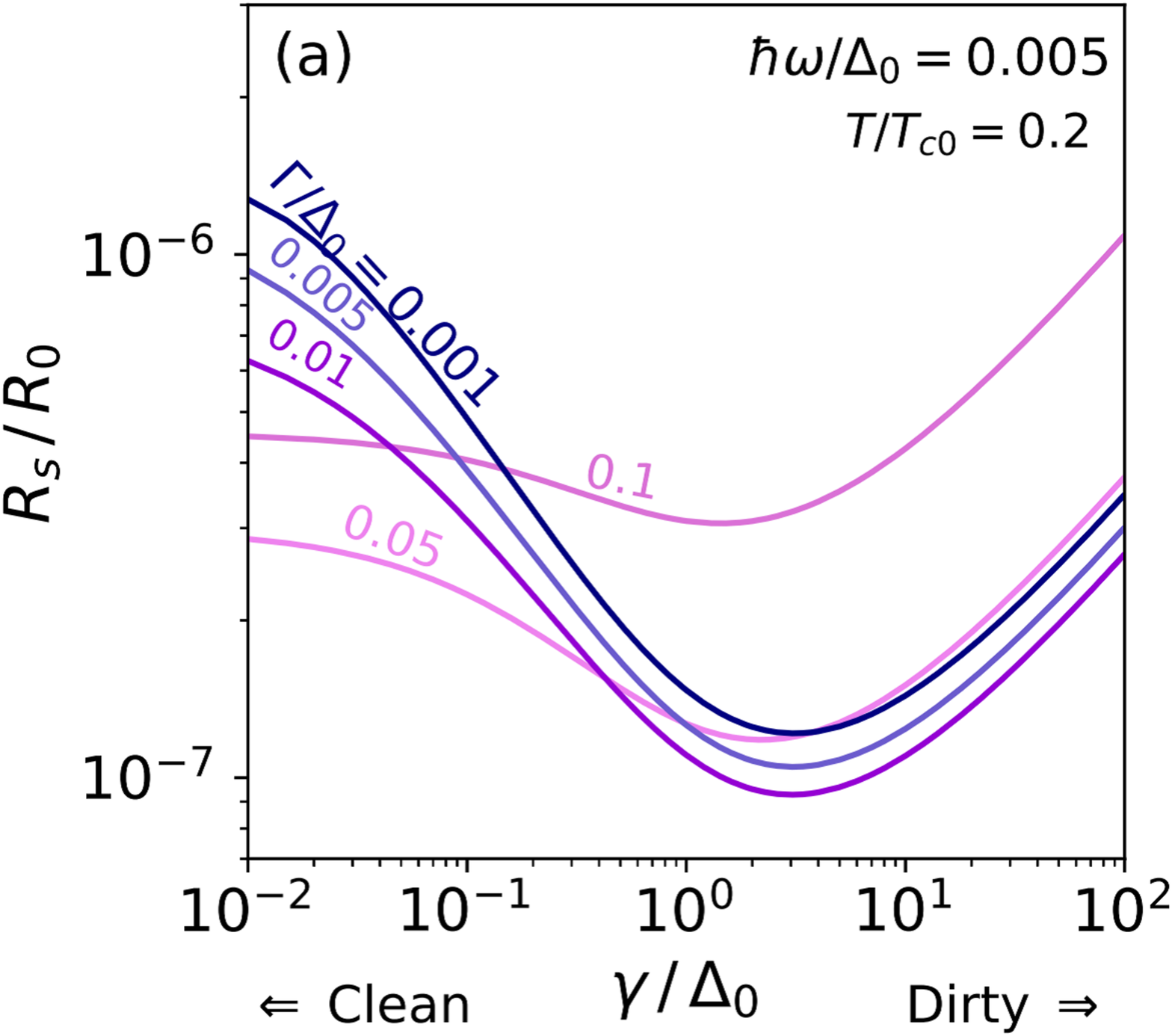}
   \includegraphics[width=0.493\linewidth]{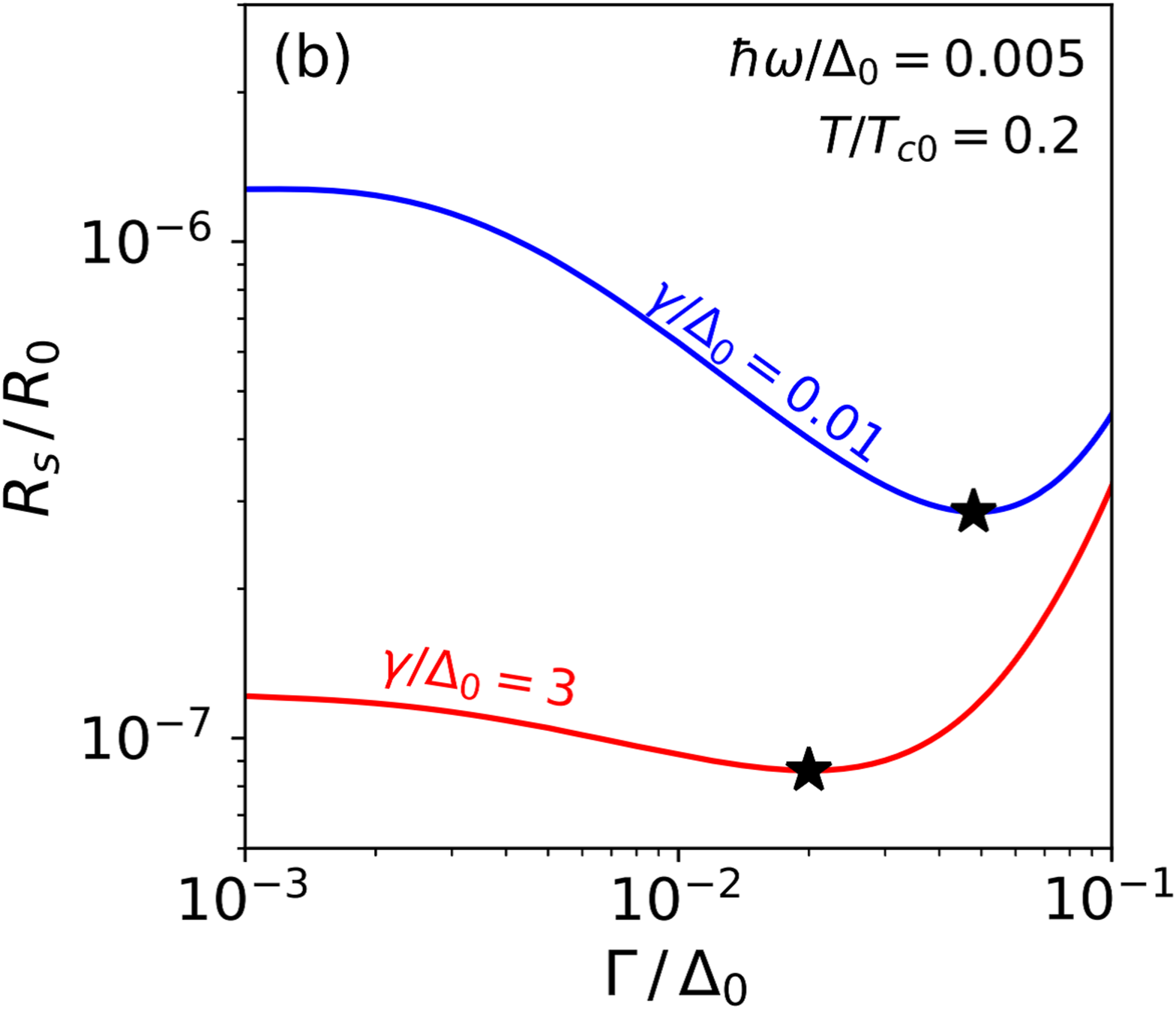}
   \end{center}\vspace{0 cm}
   \caption{
(a) $R_s$ as functions of nonmagnetic-impurity scattering-rate $\gamma/\Delta_0=\pi\xi_0/2\ell_{\rm  imp}$ calculated for different $\Gamma$. 
(b) $R_s$ as functions of $\Gamma$ calculated for $\gamma/\Delta_0 = 3$ (red) and  $\gamma/\Delta_0 = 0.01$ (blue). 
The black stars are the minimums.
   }\label{fig11}
\end{figure}

Now we investigate how to minimize $R_s$ (i.e., maximize $Q$). 
Shown in Fig.~\ref{fig11} (a) is $R_s$ as functions of nonmagnetic-impurity scattering rate $\gamma/\Delta_0=\pi\xi_0/2\ell_{\rm  imp}$. 
The navy curve is calculated for a nearly ideal BCS superconductor ($\Gamma/\Delta_0=0.001$), 
whose minimum locates at $\gamma/\Delta_0\simeq 3$, namely, $\ell_{\rm imp}/\xi_0 \simeq 0.5$. 
The existence of the optimum mean-free path is very well-known (see, e.g., Ref.~\cite{1991_Halbritter});
its origin is the interplay of $\lambda^3$ and $\sigma_1$, which are increasing and decreasing functions of $\gamma$, respectively. 
As $\Gamma$ increases, the minimum further decreases (see the dark-violet curve) and then increases at $\Gamma/\Delta_0 \gtrsim 0.01$. 
Fig.~\ref{fig11} (b) shows this non-monotonic behavior as functions of $\Gamma$. 
The red curve is calculated for a optimally dirty superconductor ($\gamma/\Delta_0= 3$), 
reaching the minimum at $\Gamma/\Delta_0 \simeq 0.02$, 
where $R_s$ is $\simeq 30\%$ smaller than that of the ideal superconductor ($\Gamma \to 0$)~\cite{2014_Gurevich, 2017_Gurevich_Kubo, 2019_Kubo_Gurevich, 2020_Kubo_1}.
The blue curve is calculated for a clean superconductor ($\gamma/\Delta_0= 0.01$). 
Its minimum locates at $\Gamma/\Delta_0 \simeq 0.05$, 
where $R_s$ is $\simeq 80\%$ smaller than that of the ideal ($\Gamma \to 0$) superconductor. 
The reduction of $R_s$ due to a finite $\Gamma$ comes from $\sigma_1(\Gamma)$, 
which is discussed in detail in Sec.~\ref{lowF_lowT} [see also Fig.~\ref{fig8}].

Remind that we have found that $R_s$ of a clean ($\gamma/\Delta_0\ll 1$) and nearly-ideal ($\Gamma/\Delta_0 \ll 1$) superconductor can be frequency independent for a wide frequency range (see Fig.~\ref{fig10}). 
For instance, the condition $\gamma/\Delta_0=\Gamma/\Delta_0 = 0.001$ yields the frequency plateau with $R_s/R_0 \sim 1\times 10^{-6}$, which is one order of magnitude larger than the global minimum for $\hbar\omega/\Delta_0=0.005$ [see the star on the red curve in Fig.~\ref{fig11} (b)].  
However, $R_s$ of such a dirty superconductor increases in proportion to $\omega^2$ and exceeds $R_s/R_0 \sim 10^{-5}$ at $\hbar\omega/\Delta_0 =0.1$, 
while a clean and nearly-ideal superconductor on the frequency plateau remains $R_s/R_0 \sim 10^{-6}$ even at $\hbar\omega/\Delta_0 =0.1$. 
Moderately dirty ($\ell_{\rm imp} \sim \xi_0$) superconductors are widely believed to minimize $R_s$, 
while the frequency plateau of a nearly-ideal clean superconductor ($\gamma/\Delta_0\simeq \Gamma/\Delta_0 \simeq 0.001$) can yield much smaller $R_s$ for moderately-high-frequency regions ($\hbar\omega \gtrsim 0.01$) for an extreme type-II superconductor.

\section{Depairing current density} \label{section_Jd}

The depairing current density $J_d$ is the maximum value of the supercurrent density. 
To obtain $J_d$, we need to solve Eq.~(\ref{Eilenberger}) and calculate the current density. 
The solution for $|q|> 0$ is given by~\cite{2012_Lin_Gurevich}
\begin{eqnarray}
g_m &=& \frac{a+i\cos\theta}{\sqrt{(a+i\cos\theta)^2 + b^2}} ,\\
f_m &=& \frac{b}{\sqrt{(a+i\cos\theta)^2 + b^2}} . 
\end{eqnarray}
Here $a$, $b$, $\langle g_m \rangle$, and $\langle f_m \rangle$ are given by 
\begin{eqnarray}
&& \hbar\omega_m + \Gamma + \langle g_m \rangle \gamma - \frac{\pi s}{2} a =0, \\
&& \Delta + \langle f_m \rangle \gamma -  \frac{\pi s}{2} b =0, \\
&& \langle g_m \rangle^4 + (a^2 +b^2 -1) \langle g_m \rangle^2 -a^2=0, \\
&& \frac{\langle f_m \rangle}{b} = \tan^{-1}\frac{\langle g_m \rangle}{a} ,
\end{eqnarray}
and $\Delta = \Delta(s,\gamma,\Gamma,T)$ satisfies Eq.~(\ref{self-consistency}). 
The current density can be calculated from~\cite{Kopnin} ${\bf J} = -4\pi k_B T e N_0 {\rm Im} \sum_{\omega_m>0} \langle {\bf v}_{f} g_m \rangle$ or 
\begin{eqnarray}
\frac{J}{J_0} = -\frac{\sqrt{6}\pi k_B T}{\Delta_0} \sum_{\omega_m>0}  \int_{-1}^{1} \!\!dc \bigl\{ c \sin u(c) \sin v(c) \bigr\}. 
\end{eqnarray}
Here, $u+iv =\tan^{-1}[b/(a+ic)]$, $c=\cos\theta$, $J_0=H_{c0}/\lambda_0$, 
$H_{c0}=\Delta_0\sqrt{N_0/\mu_0}$ is the thermodynamic critical field of the idealized BCS superconductor ($\Gamma \to 0$) in the zero-current state, 
and $\lambda_0=\lambda(0,0,0)$ is the zero-temperature ($T\to 0$) penetration depth of the idealized ($\Gamma \to 0$) clean-limit ($\gamma \to 0$) BCS superconductor in the zero-current state (see also Sec.~\ref{section_lambda}).

\begin{figure}[tb]
   \begin{center}
   \includegraphics[width=0.49\linewidth]{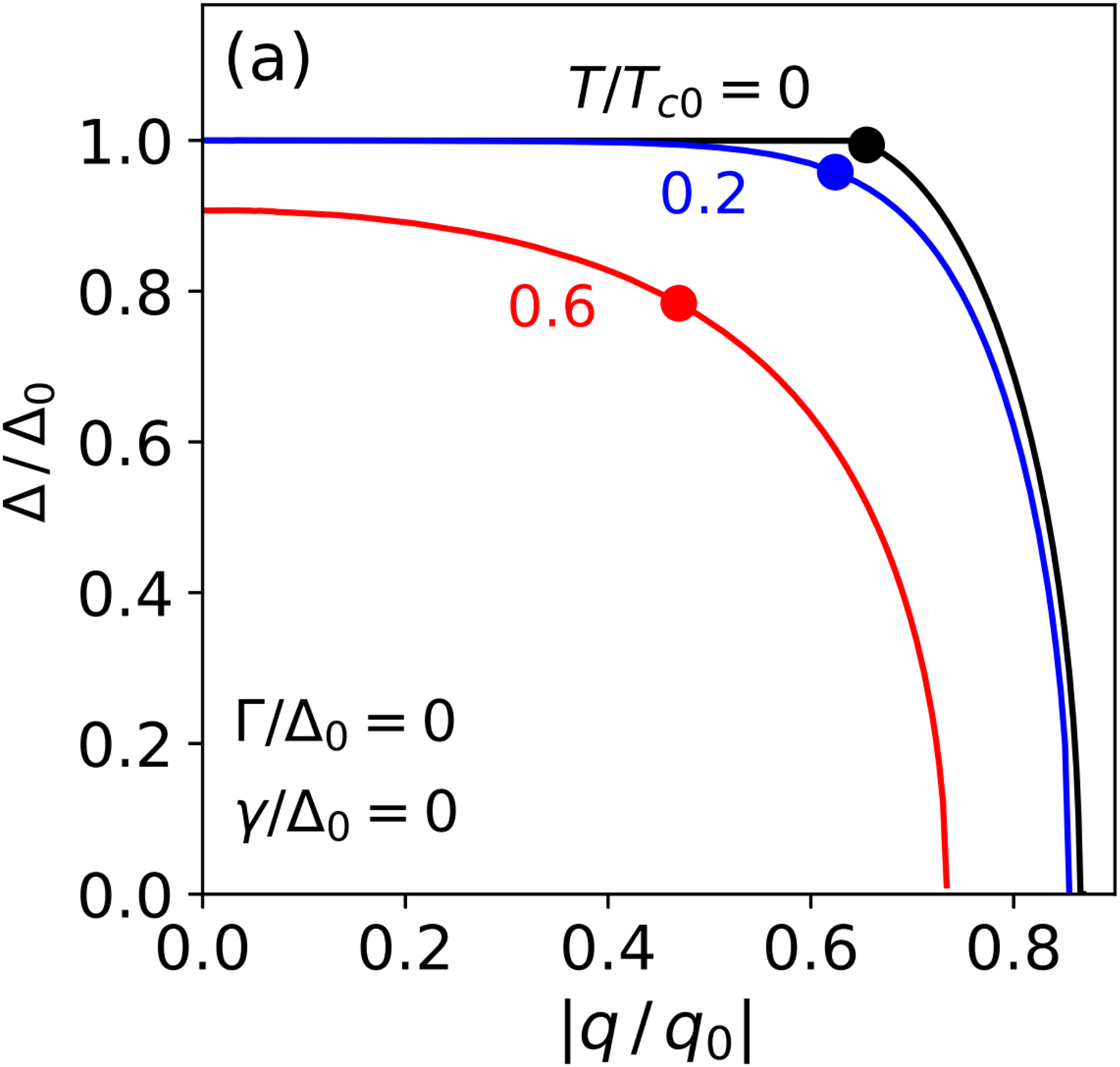}
   \includegraphics[width=0.49\linewidth]{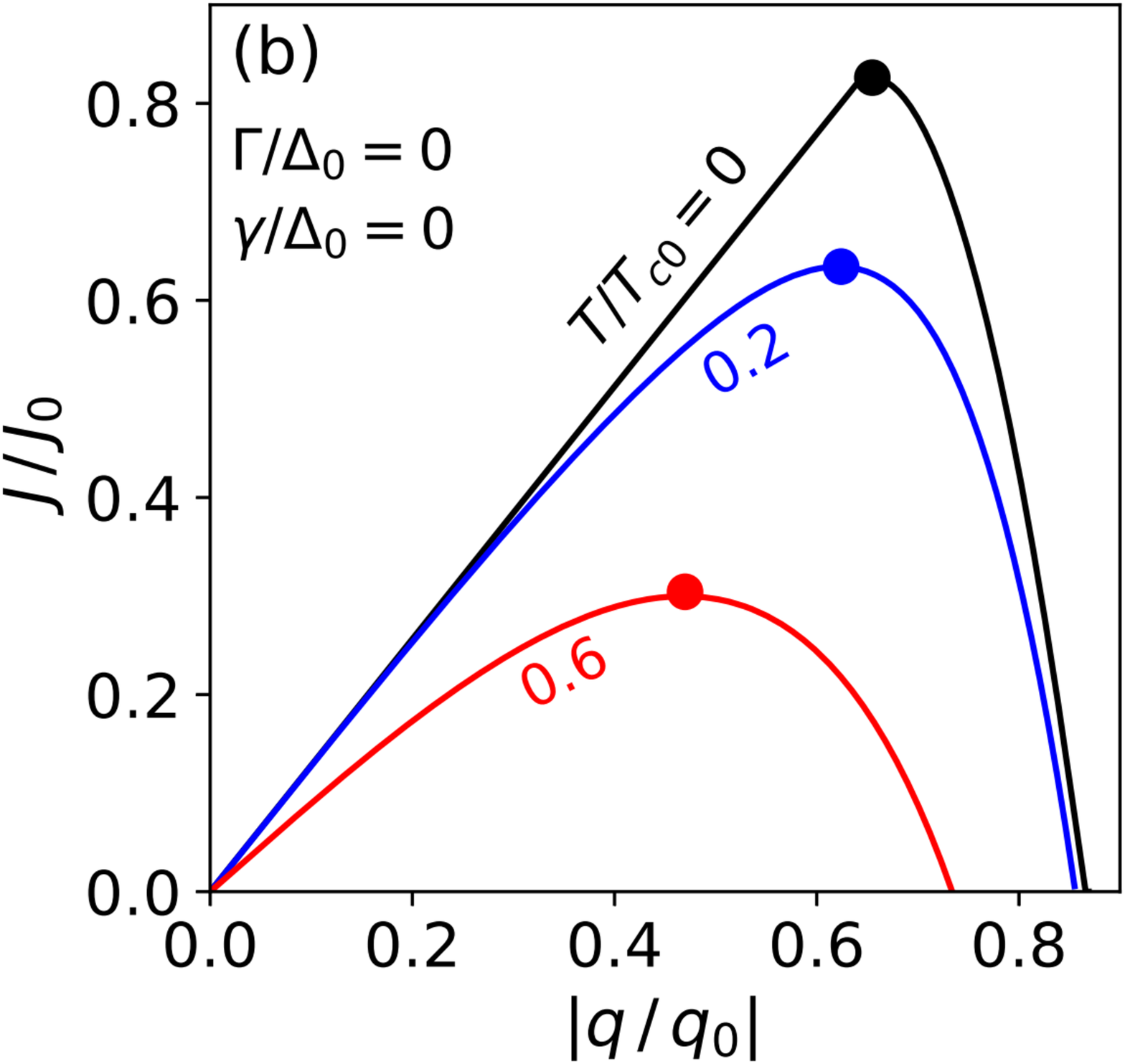}
   \end{center}\vspace{0 cm}
   \caption{
(a) $\Delta$ and (b) $J$ for the idealized BCS superconductor ($\Gamma\to 0$) in the clean limit ($\gamma=0$) calculated for for $T/T_{c0}=0$, 0.2, and 0.6. 
The blobs correspond to the depairing current densities. 
   }\label{fig12}
\end{figure}

\begin{figure}[tb]
   \begin{center}
   \includegraphics[width=0.49\linewidth]{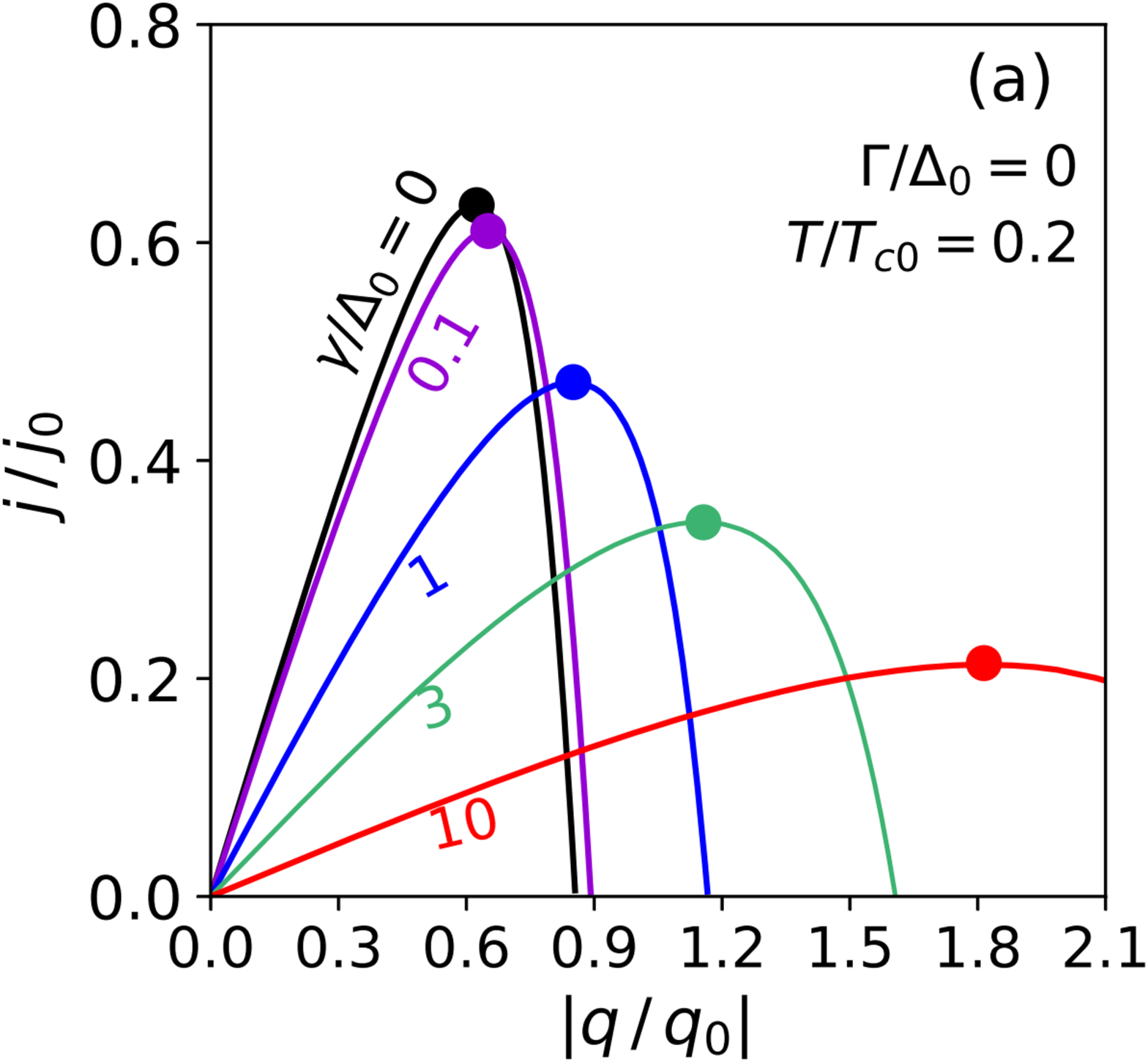}
   \includegraphics[width=0.46\linewidth]{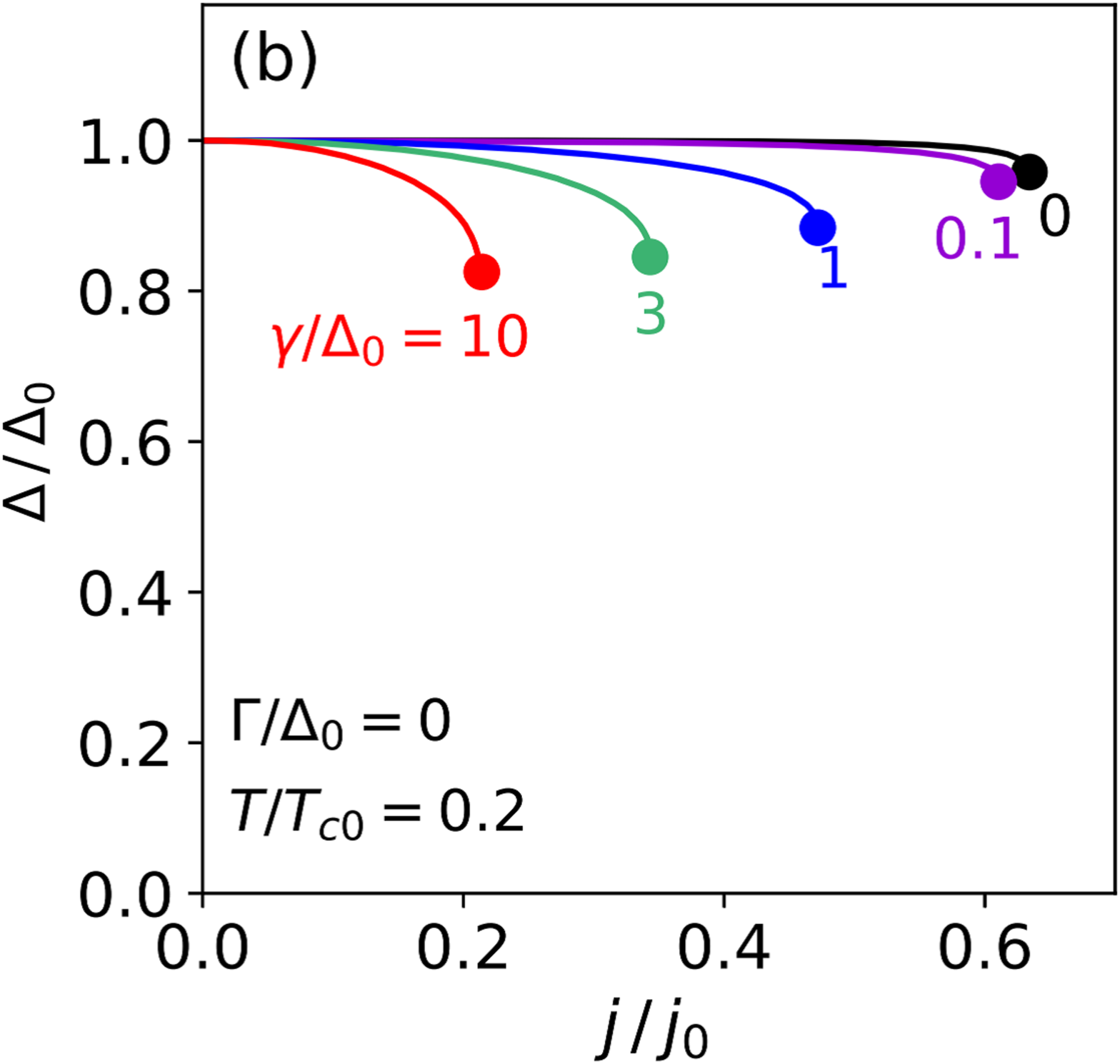}
   \end{center}\vspace{0 cm}
   \caption{
Effects of impurity scattering-rate $\gamma$ (See also the previous studies~\cite{1980_Kupriyanov, 2012_Lin_Gurevich}). 
(a) $J$ as functions of $q$ for the idealized BCS superconductor ($\Gamma\to 0$) calculated for $\gamma/\Delta_0=\pi\xi_0/2\ell_{\rm imp}=0$, 0.1, 1, 3, and 10. 
(b) $\Delta$ as functions of $J$ calculated for $J \le J_d$. 
The blobs represent the depairing current densities. 
   }\label{fig13}
\end{figure}

Let us briefly review the depairing current density for the idealized ($\Gamma\to 0$) BCS superconductor. 
Fig.~\ref{fig12} shows the pair potential $\Delta$ and the current density $J$ as functions of the superfluid momentum $q/q_0=s/\Delta_0$ for the idealized BCS superconductor ($\Gamma\to 0$) in the clean limit ($\gamma\to 0$) calculated for different temperatures.  
When $|q|$ is small, the current pair-breaking effect is limited [Fig~\ref{fig12} (a)], 
and $J$ linearly increases with $|q|$ [Fig~\ref{fig12} (b)]. 
However, as $|q|$ increases, the reduction of $\Delta$ becomes rapid [Fig~\ref{fig12} (a)], 
and $J$ ceases to increase [Fig~\ref{fig12} (b)]. 
The maximum value of $J$ (see the blob) is the depairing current density $J_d(\gamma,\Gamma,T)$. 
Interestingly, when $T/T_{c0}\ll 1$, a finite $|q|$ little affects $\Delta$ up to a large momentum close to the depairing momentum.  
For $\gamma=\Gamma=T=0$ (i.e., the black curve), the solution is well-known~\cite{Maki, 1980_Kupriyanov, 2012_Lin_Gurevich}.
The pair potential $\Delta=\Delta(s,0,0,0)$ is given by 
\begin{eqnarray}
&&\ln\frac{\Delta}{\Delta_0} \nonumber \\
&&=
\begin{cases}
0 & (s/\Delta \le 2/\pi) \\
-\cosh^{-1} \frac{\pi s}{2\Delta} + \sqrt{1-(2\Delta/\pi s)^2} & (s/\Delta > 2/\pi)
\end{cases} \label{delta_s000}, 
\end{eqnarray}
and the current density $J=J(s,0,0,0)$ is given by
\begin{eqnarray}
\frac{J}{J_0} 
=\frac{\pi s}{\sqrt{6}\Delta_0}
\begin{cases}
1 &  (s/\Delta \le 2/\pi)  \\
1- \{ 1-(2\Delta/\pi s)^2 \}^{3/2} & (s/\Delta > 2/\pi)
\end{cases} .
\end{eqnarray}
which reaches the depairing current density, 
\begin{eqnarray}
J_d(0,0,0) = 0.826 \frac{H_{c0}}{\lambda_0} ,
\end{eqnarray}
at the depairing momentum $q_d/q_0=s_d/\Delta_0=0.655$. 
According to Eq.~(\ref{delta_s000}), 
$\Delta$ is unaffected by the superfluid flow for $q/q_0=s/\Delta_0 \le 0.637$, 
which is very close to the depairing momentum. 
The nonexistence of current pair-breaking for a broad range of $q$ is the well-known anomalous feature of a clean-limit superconductor at $T/T_{c0}\ll 1$. 
It should be noted that this feature disappears as $T$ increases (see the red curve); 
also impure superconductors do not exhibit such an anomalous feature even at $T\to 0$ (see the dirty-limit results~\cite{1963_Maki, 2012_Clem_Kogan, 2020_Kubo_1, 2020_Kubo_2} and Fig.~\ref{fig13}). 
Shown in Fig.~\ref{fig13} are $J(q)$ and $\Delta(J)$ of the idealized BCS superconductor ($\Gamma\to 0$) calculated for different impurity concentrations.
As $\gamma$ increases (as materials get dirty), $J_d$ and $\Delta(J)|_{J\sim J_d}$ decrease: 
nonmagnetic impurities become pair breakers and significantly impact $\Delta$ in the current-carrying state, in contrast to their pair-conserving nature in the zero-current state (see Sec.~\ref{brief_review}).

\begin{figure}[tb]
   \begin{center}
   \includegraphics[width=0.48\linewidth]{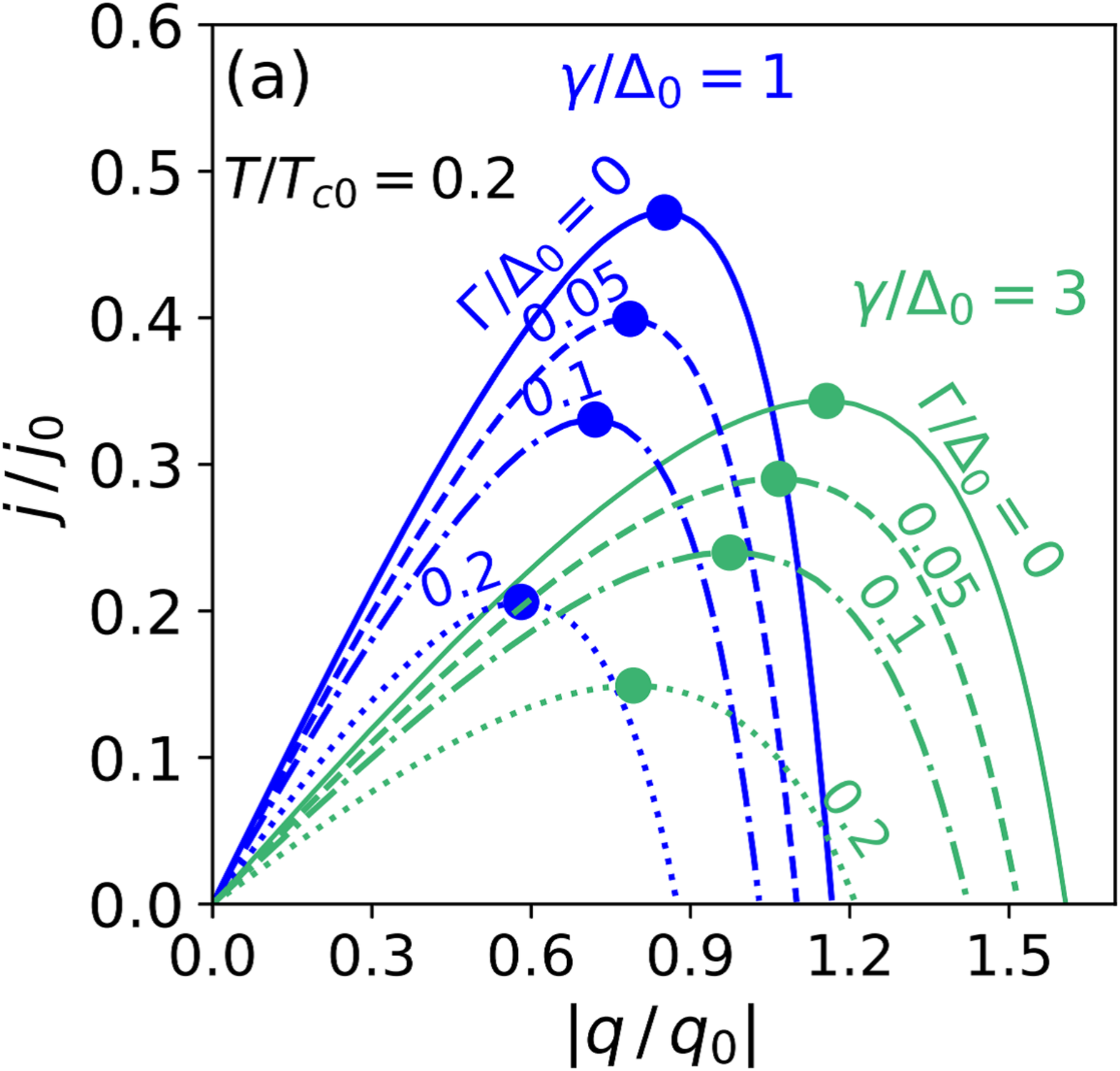}
   \includegraphics[width=0.49\linewidth]{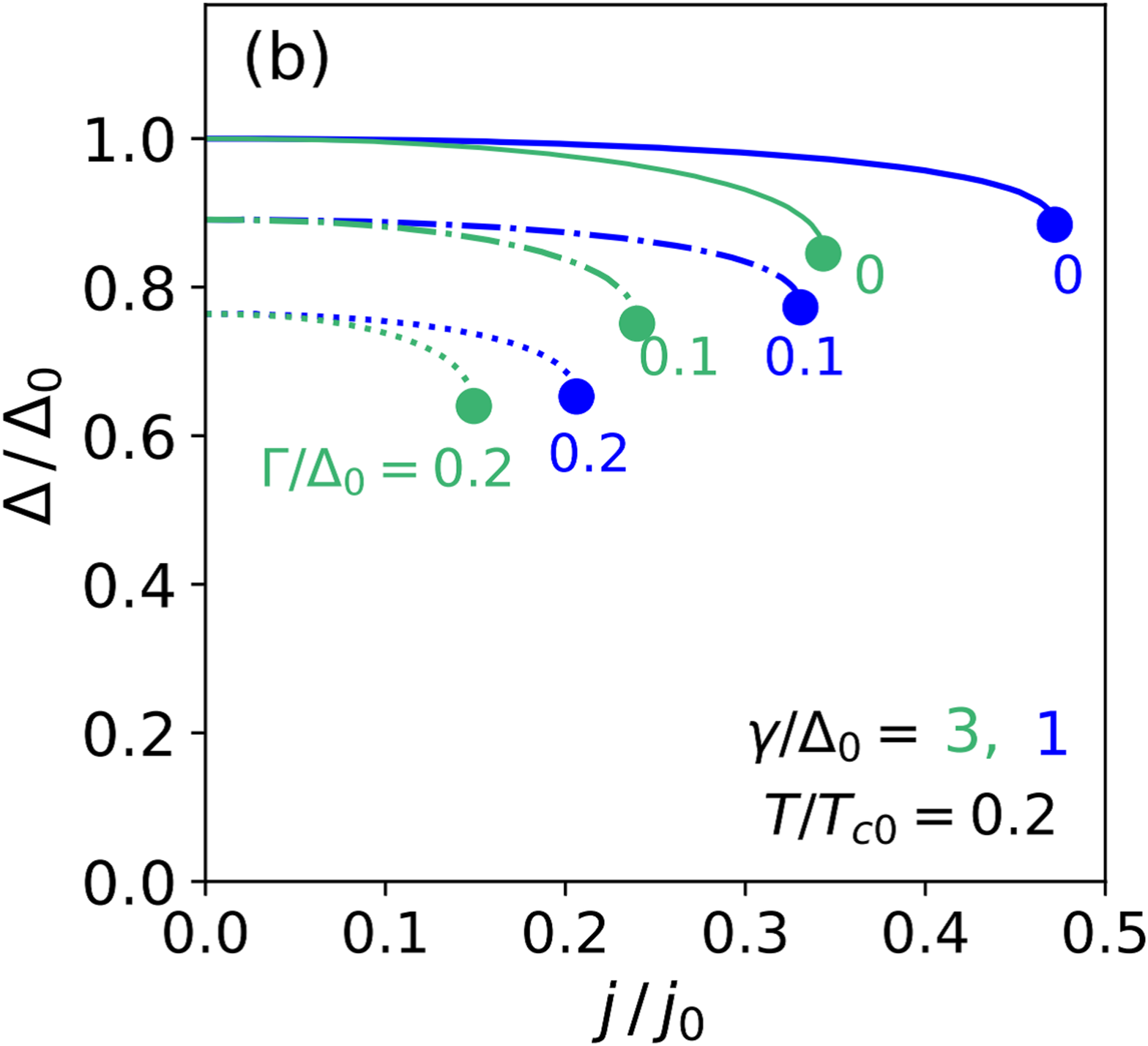}
   \end{center}\vspace{0 cm}
   \caption{
Effects of Dynes $\Gamma$ on $J(q)$ and $\Delta(J)$. 
(a) $J$ as functions of the superfluid momentum $q$ calculated for $\gamma/\Delta_0=1$ and 3 and $\Gamma/\Delta_0=0$, 0.05, 0.1, and 0.2. 
The blobs correspond to the depairing current densities. 
(b) $\Delta$ as functions of $J$ calculated for $J \le J_d$. 
   }\label{fig14}
\end{figure}

Shown in Fig.~\ref{fig14} are the effects of Dynes $\Gamma$ on $J(q)$ and $\Delta(J)$. 
As $\Gamma$ increases, $J_d$ and $\Delta$ decrease. 
In contrast to nonmagnetic impurities, Dynes $\Gamma$ is always pairbreaking even in the zero-current state as shown in Sec.~\ref{brief_review}; compare Fig.~\ref{fig13} (b) with Fig.~\ref{fig14} (b).

\begin{figure}[tb]
   \begin{center}
   \includegraphics[width=0.49\linewidth]{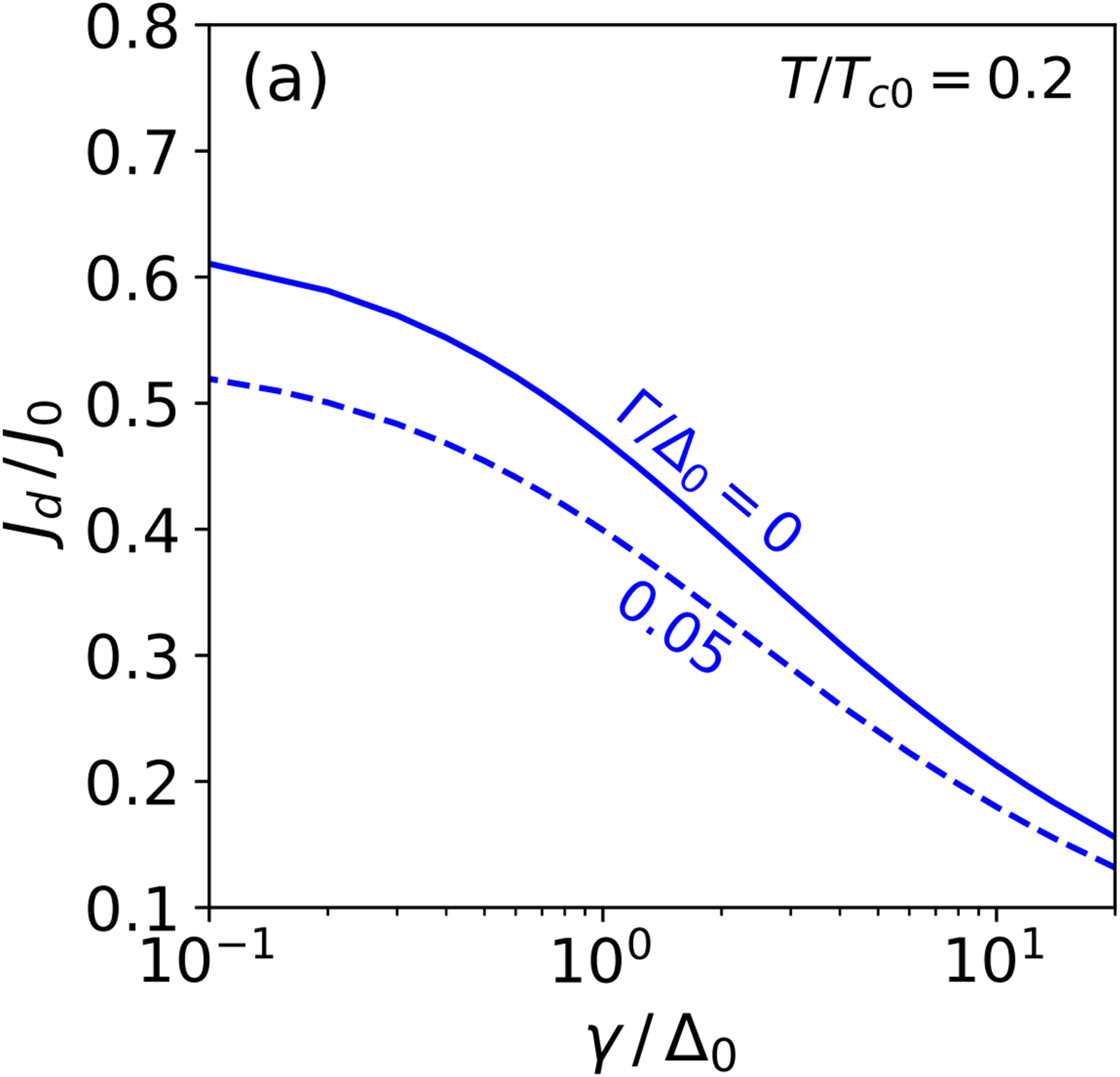}
   \includegraphics[width=0.49\linewidth]{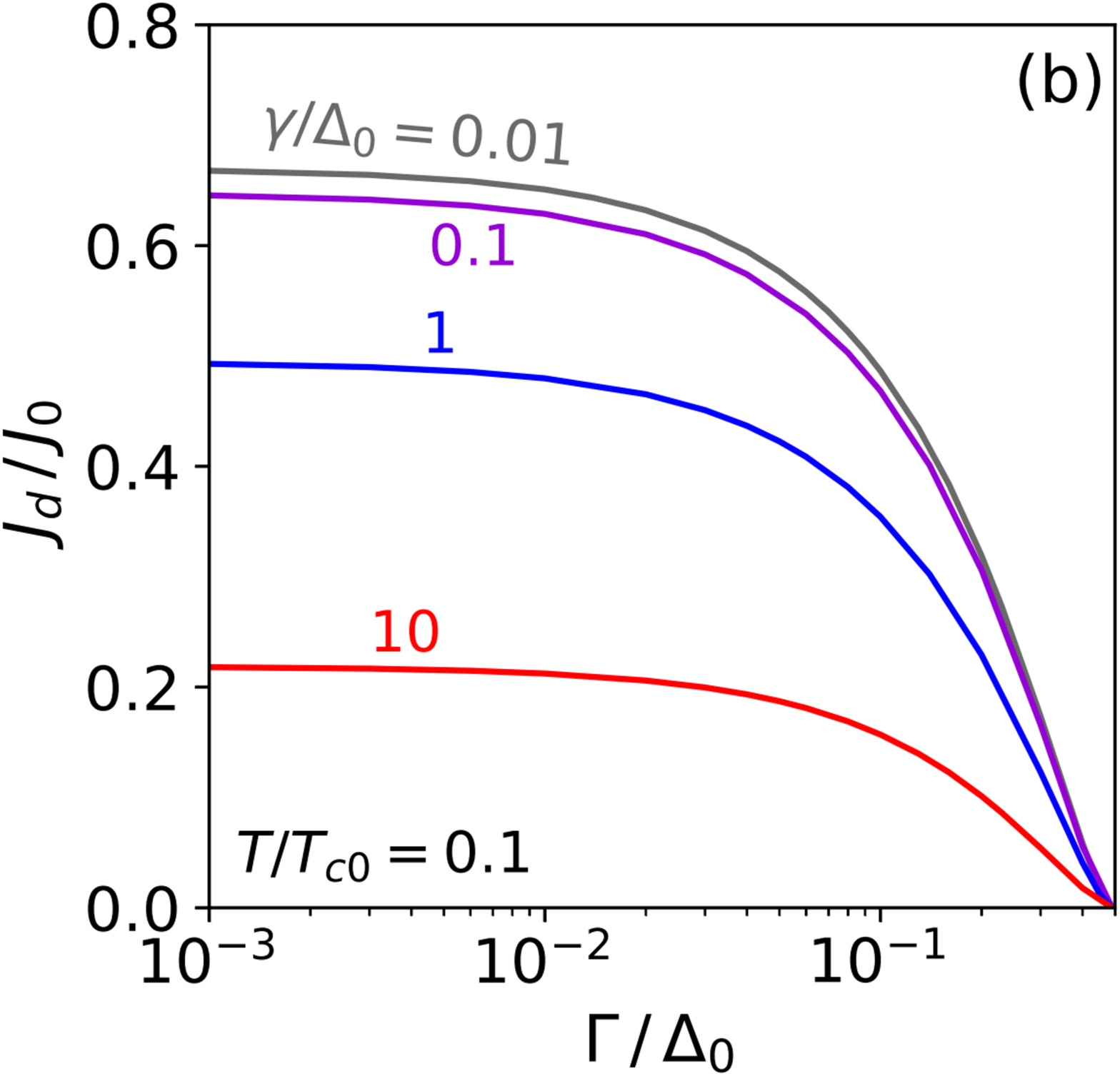}
   \end{center}\vspace{0 cm}
   \caption{
The depairing current density $J_d$ as functions of (a) the nonmagnetic-impurity scattering rate $\gamma/\Delta_0=\pi\xi_0/2\ell_{\rm imp}$ and (b) Dynes $\Gamma$ parameter.  
   }\label{fig15}
\end{figure}

Fig.~\ref{fig15} (a) shows $J_d$ as functions of the nonmagnetic-impurity scattering rate $\gamma/\Delta_0=\pi\xi_0/2\ell_{\rm imp}$; $J_d$ decreases as $\gamma$ increases (see also Fig.~\ref{fig13}). 
At $\gamma/\Delta_0 \gtrsim 1$, we have $J_d \propto \gamma^{-1/2}$, 
consistent with the dirty-limit result~\cite{1963_Maki, 2012_Clem_Kogan, 2020_Kubo_1, 2020_Kubo_2}:  
\begin{eqnarray}
J_d(\gamma \gg \Delta_0, \Gamma, T) 
=C(\Gamma,T) \frac{H_{c0}}{\lambda_{0, {\rm dirty}}} \propto \frac{1}{\sqrt{\gamma}}.
\end{eqnarray}
Here, $\lambda_{0, dirty} = \sqrt{\hbar \rho_n/\pi\mu_0\Delta_0}\propto \sqrt{\gamma}$ is the zero-temperature penetration depth of the idealized ($\Gamma\to 0$) dirty-limit BCS superconductor for the zero-current state (see also Sec.~\ref{section_lambda}), and the coefficient $C(\Gamma,T)$ can be calculated from the microscopic theory, e.g., $C(0,0)=0.595$ and $C(0,T)$ are calculated some decades ago~\cite{1963_Maki, 1980_Kupriyanov}; $C(\Gamma,T)$ is calculated in Refs.~\cite{2020_Kubo_1, 2020_Kubo_2}.
Fig.~\ref{fig15} (b) shows $J_d$ as functions of Dynes $\Gamma$ for different impurity concentrations; 
$J_d$ is significantly suppressed at $\Gamma/\Delta_0 \gtrsim 0.1$, where the pair-breaking scattering almost destroys the superconductivity.

\section{Discussion} 

The effects of Dynes $\Gamma$ and nonmagnetic-impurity scattering-rate $\gamma \propto \xi_0/\ell_{\rm imp}$ on various physical quantities relevant to superconducting devices have been investigated. 
Here, we summarize our results and discuss their implications for superconducting devices.

\subsection{Kinetic inductance} 

In Sec.~\ref{section_lambda_Lk}, we calculated the penetration depth $\lambda$ and the kinetic inductivity $L_k$ taking the effects of $\Gamma$ and $\gamma$ into account and showed that $\lambda$ and $L_k$ are monotonic increasing functions of $\Gamma$ and $\gamma$ (see Figs.~\ref{fig3} and \ref{fig4}). 
Eq.~(\ref{penetration_depth_zeroT_1}) can evaluate $\lambda$ for a nearly ideal ($\Gamma/\Delta_0 \ll 1$) and clean ($\gamma/\Delta_0 \ll 1$) BCS superconductor. 
Eqs.~(\ref{penetration_depth_zeroT_2})-(\ref{penetration_depth_dirty_idealBCS}) give other convenient formulas for $\lambda$. 
Combining these formulas with Eqs.~(\ref{LK_general}) and (\ref{Lk_dirty}), 
we get the analytical formulas for $L_k$. 
Note that Eq.~(\ref{penetration_depth_dirty_idealBCS}) is widely used for analyses of experimental data of superconducting devices, but it is applicable only to the idealized ($\Gamma\to 0$) dirty-limit ($\gamma/\Delta_0 =\pi\xi_0/\ell_{\rm imp}\gg 1$) superconductor. 
To analyze a wide range of materials, Eqs.~(\ref{penetration_depth_zeroT_1})-(\ref{Lk_dirty}) or the numerical results (Figs.~\ref{fig3} and \ref{fig4}) would be helpful.

The reset time of SSPD is known to be limited by the kinetic inductance~\cite{2006_Kerman}, and therefore nanowires with smaller kinetic inductances have been developed. 
On the other hand, a high kinetic inductance has also attracted attention in the quantum-circuit community; the so-called superinductor makes fluxonium qubit immune to charge fluctuations (see e.g., Ref.~\cite{2019_Niepce}). 
According to our results, not only $\gamma$ (impurity concentration) but also $\Gamma$ increase $L_k$. 
The effect of $\Gamma$ on $L_k$ would not be negligible. 
However, $\Gamma$ in the material of SSPD, KID, and other quantum circuit is usually not measured. 
Measurements of $\Gamma$ of circuit material using tunneling spectroscopy or recently proposed method to extract $\Gamma$ from complex conductivity, which will be discussed in Sec.~\ref{discussion_coherence_peak}, would give insight into a method to engineer $L_k$.

\subsection{$T$ dependence of $\sigma_1$ and coherence peak} \label{discussion_coherence_peak}

In Sec.~\ref{section_coherence_peak}, we studied the effects of $\Gamma$ and $\gamma$ on the $T$ dependence of $\sigma_1$ and the coherence peak. 
The height of the coherence peak decreases as $\omega$ or $\ell_{\rm imp} (\propto \gamma^{-1})$ increases, 
consistent with the previous theoretical study~\cite{1991_Marsiglio, 2021_Herman} and experiment~\cite{2008_Steinberg}. 
Note that the peak height decreases as the material gets clean. 
Also, a finite $\Gamma$ reduces the peak height, 
consistent with the previous studies for the dirty-limit superconductor~\cite{2017_Herman, 2020_Kubo_1}.

The recent finding of Ref.~\cite{2021_Herman} was also reproduced: 
the $\sigma_1(T)$ curve for a clean superconductor does not depend on $\Gamma$ and $\gamma$ separately but depends on the ratio $\Gamma/\gamma$ [see Fig.~\ref{fig5} (d)]. 
Also, it is possible to extract $\Gamma$ and $\gamma$ by fitting experimental data with the theory as proposed in Ref.~\cite{2021_Herman}, 
in which they evaluated $\Gamma$ and $\gamma$ using the recent experimental data of SRF resonant cavities~\cite{2021_Bafia}. 
This method would apply to KID and other resonators for quantum technologies.
Now, $\Gamma$ of resonator material can be easily obtained via measurements of $\sigma_1$, 
which are routinely performed in laboratories. 
Using experimentally determined $\Gamma$, it becomes possible to compare experimental data of devices with the more realistic theories~\cite{2017_Gurevich_Kubo, 2019_Kubo_Gurevich, 2020_Kubo_1, 2020_Kubo_2} rather than the simplest formulas such as Eqs.~(\ref{penetration_depth_dirty_idealBCS}) and (\ref{MB_ideal_1}).

\subsection{$\sigma_1$ and $R_s$ at a very low $T$} \label{discussion_very_low_T}

In Sec.~\ref{section_residual_sigma1} and \ref{section_Rs}, 
we calculated the subgap-state-induced residual-dissipative-conductivity $\lim_{T\to 0}\sigma_1$ and residual surface resistance $R_{\rm res} = \lim_{T\to 0} R_s$,  
which originates from a finite density of subgap states in the vicinity of the Fermi level (see Figs.~\ref{fig6} and \ref{fig9}). 
Eqs.~(\ref{sigma1_zeroT}) and (\ref{sigma1_zeroT_dirty}) are formulas for $\lim_{T\to 0}\sigma_1$. 
Also, we derived the formulas for $R_{\rm res}$, Eqs.~(\ref{Rres_approx}) and (\ref{Rres_approx_dirty}), 
which give good approximations of the numerical results of $R_{\rm res}$ [see Fig.~\ref{fig9} (b)].

These results can be tested by experiments: measure the temperature dependence of $\sigma_1$, extract $\Gamma$ and $\gamma$ from $\sigma_1(T)$ using Refs.~\cite{2021_Herman, 2021_Bafia}, and cool a resonator down until $Q$ becomes independent of $T$. 
However, as shown in Eq.~(\ref{Q_Gamma_TLS}), factors other than the subgap-state-induced $R_{\rm res}$ can also contribute to $Q$. 
Such contributions should be reduced to a level ignorable as compared with $Q_{\Gamma}^{-1}=R_{\rm res}/G$, which is typically $< 10^{-10}$ (see Sec.~\ref{section_Rs}). 
It would be difficult for a 2D resonator to detect $Q_{\Gamma}^{-1}$, in which $Q^{-1}_{\rm TLS}$ always overwhelms $Q_{\Gamma}^{-1}$. 
High-$Q$ SRF cavities~\cite{2020_Romanenko} may be helpful for this purpose.

\subsection{$\sigma_1$ and $R_s$ at a moderately low $T$} \label{discussion_moderately_low_T}

In Sec.~\ref{lowF_lowT} and \ref{section_Rs}, we studied $\sigma_1$ and $R_s$ at moderately-low-temperatures, $\hbar \omega \ll k_B T \ll \Delta_0$.

It was found that $\sigma_1$ of a dirty superconductor exhibits the well-known frequency dependence $\propto \ln(1/\omega)$, while that of a nearly-ideal ($\Gamma/\Delta_0 \ll 1$) clean ($\gamma/\Delta_0=\pi\xi_0/2\ell_{\rm imp} \ll1$) superconductor exhibits $\omega^{-2}$ dependence (see Fig.~\ref{fig7}),  
resulting in the frequency-independent surface resistance $R_s$, shown in Fig.~\ref{fig10}. 
For instance, $R_s$ of a nearly-ideal ($\Gamma/\Delta_0=0.001$) clean-limit ($\gamma/\Delta_0=0.001$) superconducting resonator made from a large-$\lambda/\xi$ material (e.g., ${\rm Nb_3 Sn}$, NbTiN) is frequency-independent between a few GHz and several tens of GHz (e.g., $3.7\,{\rm GHz} \lesssim f \lesssim 75\,{\rm GHz}$ for ${\rm Nb_3 Sn}$). 
Roughly speaking, the red curve in Fig.~\ref{fig10} (a) corresponds to a dirty ${\rm Nb_3 Sn}$ cavity operated at $T\sim 4\,{\rm K}$. 
When the resonant frequency is $f \sim 1\,{\rm GHz}$, which corresponds to $\hbar\omega/\Delta_0 = 10^{-3}$, its surface resistance is $R_s \sim {\rm n\Omega}$ from Fig.~\ref{fig10} (a), consistent with experiments. 
The frequency-independent $R_s$ is realized when the material is nearly ideal and clean (the black curve). 
$R_s$ of the black curve is one order of magnitude larger than the red curve at $\hbar\omega/\Delta_0 = 10^{-3}$ ($\sim 1\,{\rm GHz}$) but smaller than the red curve at $\hbar\omega/\Delta_0 \gtrsim 10^{-2}$ ($\sim 10\,{\rm GHz}$), 
where $R_s\sim {\rm \mu \Omega}$. 
The red curve increases in proportion to $\omega^2$, while the black curve remains $R_s\sim {\rm \mu \Omega}$ even at $\hbar\omega/\Delta_0 \sim 0.1$ ($\sim 100\,{\rm GHz}$). 
Hence, if there is a demand for a compact cavity with the resonant frequency $\sim$ a few tens GHz, 
it should be made from a nearly-ideal ($\Gamma/\Delta_0 \ll 1$) clean-limit ($\gamma/\Delta_0$) type-II superconductor to minimize $R_s$.

Also, it was found that $\sigma_1$ and $R_s$ are non-monotonic functions of $\Gamma$, 
and there exists the optimum value $\Gamma_*$ which minimizes $\sigma_1$ and $R_s$.
For the dirty limit, $\sigma_1(\Gamma_*)$ and $R_s(\Gamma_*)$ are a few tens of $\%$ smaller than those of the idealized dirty-limit BCS superconductor, 
consistent with the previous studies~\cite{2017_Gurevich_Kubo, 2019_Kubo_Gurevich}. 
On the other hand, for the clean limit, $\sigma_1(\Gamma_*)$ and $R_s(\Gamma_*)$ are one order of magnitude smaller than those of the idealized clean-limit BCS superconductor (see Figs.~\ref{fig8} and \ref{fig11}). 
It has been well-known that the optimum impurity-scattering rate $\gamma/\Delta_0 \sim \xi_0/\ell_{\rm imp} \sim 1$ minimizes $R_s$. 
Such an $R_s$ minimized by the optimum $\gamma$ can be further reduced by the optimum $\Gamma$  [see Fig.~\ref{fig11}]. 
These results suggest that we can significantly reduce $R_s$ by tuning $\Gamma$. 
While the physics and materials mechanisms behind $\Gamma$ are not well understood, 
comparison of experimentally determined $\Gamma$ (e.g., using tunneling spectroscopy~\cite{2003_Zasa} or complex conductivity measurement~\cite{2021_Herman, 2021_Bafia}) and various materials treatments can give useful information on how to engineer $\Gamma$.

\subsection{Depairing current density} 

In Sec.~\ref{section_Jd}, we studied the depairing current density $J_d$. 
As shown in Fig.~\ref{fig15}, $J_d$ is a monotonically decreasing function of $\Gamma$ and $\gamma \propto \ell_{\rm imp}^{-1}$. 
Hence, we need both $\Gamma$ and $\gamma$ to know $J_d$ of the materials of devices or cables. 
As repeatedly mentioned above, tunneling spectroscopy can determine $\Gamma$, and the $T$ dependence of $\sigma_1$~\cite{2021_Herman, 2021_Bafia} can determine $\Gamma$ and $\gamma$. 
Combining these measurements of $\Gamma$ and $\gamma$ with those of $J_d$, 
such as Refs.~\cite{1982_Romijn, 2004_Rusanov}, we can test the theory. 
Also, combining various materials treatments and the $\Gamma$ measurement can give beneficial information on ameliorating $J_d$, 
which is practically important to examine the quality of superconducting cable and a thin film laminated on the inner surface of a superconducting heterostructure cavity.


%


\end{document}